\documentclass[10pt,reqno]{ofj}

\usepackage{tensorNotation}
\usepackage{finiteVolumeNotation}

\usepackage{setspace}
\usepackage{lineno}
\usepackage{color}
\usepackage[dvipsnames,svgnames,x11names]{xcolor}
\usepackage[hang,flushmargin]{footmisc}
\usepackage{orcidlink}
\usepackage{cite}
\usepackage{xspace} 

\usepackage[]{hyperref}

\newcommand{\authorcontributions}[1]{%
\vspace{6pt}\noindent{\fontsize{9}{11.2}\selectfont\textbf{Author Contributions:} {#1}\par}}

\let\paragraph\undefined

\copyrightinfo{2024}{The Authors. This work is an open access article under the {\color{blue}{\href{https://creativecommons.org/licenses/by-sa/4.0/}{CC BY-SA 4.0}}} license}
\newcommand{\OF}[0]{OpenFOAM\textsuperscript{\textregistered}\xspace}

\OpenFOAMversions{\OF 8}

\usepackage[per-mode=symbol]{siunitx}
\usepackage{threeparttable}
\usepackage{subcaption}
\usepackage{enumerate}
\usepackage{comment}
\usepackage{csvsimple} 
\usepackage{tikz}
\usetikzlibrary{shapes.geometric, arrows, positioning, arrows.meta, backgrounds, fit, calc}
\tikzstyle{arrow} = [thick,->,>=latex]
\tikzset{every picture/.style={font issue=\small,line width=0.75pt},
	font issue/.style={execute at begin picture={#1\selectfont}},
	base/.style = {rectangle, minimum width=4cm, minimum height=0.7cm, text centered},
	block/.style = {base, rounded corners, draw=black, text width=4.2cm},
	ghost/.style = {rectangle, minimum width=0cm, minimum height=0cm},
}
\usepackage{pgfplots}
\pgfplotsset{compat=1.18}
\pgfplotsset{legend style={legend cell align=left}}
\pgfplotstableset{col sep=comma}
\definecolor{gray70}{rgb}{0.3, 0.3, 0.3}

\newcommand{\diff}{\ensuremath{\mathrm{d}}}
\newcommand{\Diff}{\ensuremath{\mathrm{D}}}
\newcommand{\tddt}[1]{\ensuremath{\frac{\diff #1}{\diff t}}} 
\newcommand{\intV}[1]{\int_V #1 \diff V}
\newcommand{\intS}[1]{\int_S #1 \diff\vec{S}}
\newcommand{\intVt}[1]{\int_{V(t)} #1 \diff V}
\newcommand{\intSt}[1]{\int_{S(t)} #1 \cdot \vec{n} \diff S}
\newcommand{\mdprod}{\cdot}
\newcommand{\mddprod}{:}
\newcommand{\mdiv}{\nabla\mdprod}
\newcommand{\abs}[1]{\left\lvert#1\right\rvert}
\newcommand{\norm}[1]{\left\lVert#1\right\rVert}
\newcommand{\normE}[1]{\norm{#1}_2}
\newcommand{\corr}[1]{\operatorname{corr}\left(#1\right)}
\newcommand{\mU}{\ensuremath{\vec{u}}} 
\newcommand{\mgU}{\ensuremath{\nabla\mU}} 
\newcommand{\mD}{\ensuremath{\vec{D}}} 
\newcommand{\mdevD}{\ensuremath{\dev\mD}} 
\newcommand{\shearStress}{\ensuremath{\gvec{\tau}}} 
\newcommand{\md}{\ensuremath{\vec{d}}} 
\newcommand{\thk}{k} 
\newcommand{\vol}{v} 
\newcommand{\phiv}{\gvec{\phi}} 
\newcommand{\varphiv}{\gvec{\varphi}} 
\newcommand{\A}{\ensuremath{A}} 
\newcommand{\Hu}{\ensuremath{\vec{H}}} 
\newcommand{\thv}{\gvec{\theta}}
\newcommand{\Sv}{\vec{S}} 
\newcommand{\pTaitA}{C_1} 
\newcommand{\pTaitB}{C_2} 

\begin{document}


\title[A Two-Phase Flow Solver with Variable Liquid Compressibility and Temperature Equation]{A Two-Phase Flow Solver with Variable Liquid Compressibility and Temperature Equation for Partitioned Simulation of Elastohydrodynamic Lubrication}




\author[N. Delaissé]{Nicolas Delaissé$^{1,*}$\orcidlink{0000-0002-3241-6487}}
\address{$^1$Department of Electromechanical, Systems and Metal Engineering, Ghent University, Sint-Pietersnieuwstraat 41, 9000 Gent, Belgium}
\email{Nicolas.Delaisse@UGent.be}

\author[P. Havaej]{Peyman Havaej$^{2}$\orcidlink{0000-0001-7221-8891}}
\address{$^2$Department of Electromechanical, Systems and Metal Engineering, Ghent University, Technologiepark Zwijnaarde 46, 9052 Zwijnaarde, Belgium}

\author[D. Fauconnier]{Dieter Fauconnier$^{2,3}$\orcidlink{0000-0002-0257-4687}}
\address{$^3$Flanders Make @ UGent -- Core Lab MIRO}

\author[J. Degroote]{Joris Degroote$^{1,3}$\orcidlink{0000-0003-4225-1791}}


\begin{abstract}
	This paper presents a new solver developed in \OF for the modeling of lubricant in the narrow gap between two  surfaces inducing hydrodynamic pressures up to few gigapascal.
	Cavitation is modeled using the homogeneous equilibrium model.
	The mechanical and thermodynamic constitutive behavior of the lubricant is accurately captured by inclusion of compressibility, lubricant rheology and thermal effects.
	Different constitutive models can be selected at run time, through the adoption of the modular approach of \OF.
	By combining the lubricant solver with a structural solver using a coupling tool, elastohydrodynamically lubricated contacts can be accurately simulated in a partitioned way.
	The solution approach is validated and examples with different slip conditions are included.
	
	The benefit for the \OF community of this work is the creation of a new solver for lubricant flow in challenging conditions and at the same the illustration of combining \OF solvers with other open-source software packages.
\end{abstract}

\date{\today}

\dedicatory{}

\maketitle



\section{Introduction}
\label{sec:introduction}

Lubrication is essential for all electromechanical machinery with parts in relative motion.
It is indispensable for many of the basic components in drivetrains, such as bearings, gears and cams.
Its primary goal is to separate the surfaces of a loaded contact by a lubricant film, to reduce wear and to guarantee an acceptable lifetime of the involved components.
At the same time, it minimizes noise and vibration, reduces friction and evacuates heat from the contact.

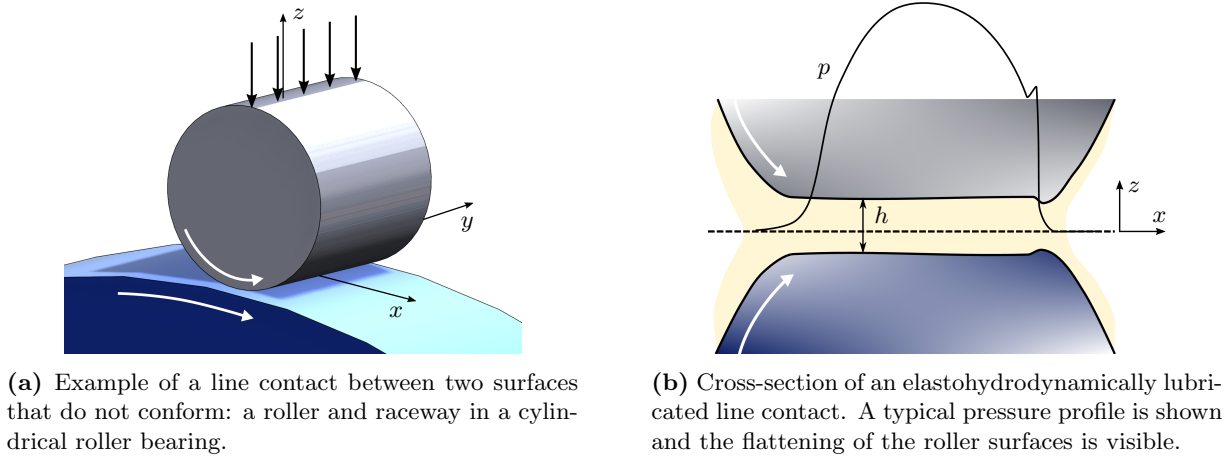
\begin{figure}[htbp]
	\begin{subfigure}[t]{.47\linewidth}
		\centering
		\begin{tikzpicture}[inner sep=0pt,
			every node/.style={align=center}]
			\node (img) at (0,0) {\includegraphics[width=.8\textwidth]{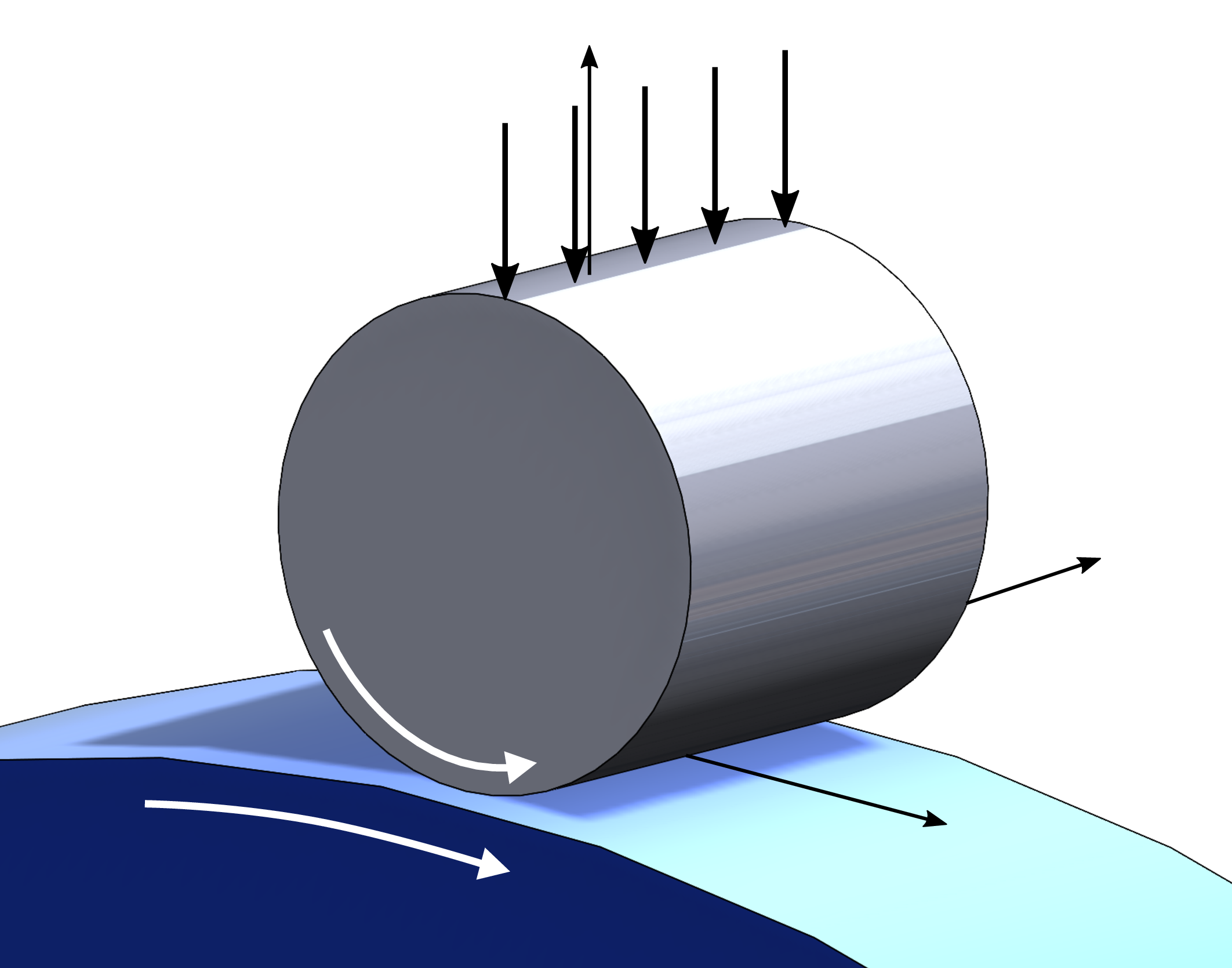}};
			\begin{scope}[shift=(img.south west), 
				x={(img.south east)},y={(img.north west)}] 
				\node at (.725, .12) {$x$};
				\node at (.88, .365) {$y$};
				\node at (.513, .95) {$z$};
			\end{scope}
		\end{tikzpicture}
		\caption{Example of a line contact between two surfaces that do not conform: a roller and raceway in a cylindrical roller bearing.}
		\label{fig:roller3d}
	\end{subfigure}
	\hfill
	\begin{subfigure}[t]{.47\linewidth}
		\centering
		\begin{tikzpicture}[inner sep=0pt,
			every node/.style={align=center}]
			\node (img) at (0,0) {\includegraphics[width=.8\textwidth]{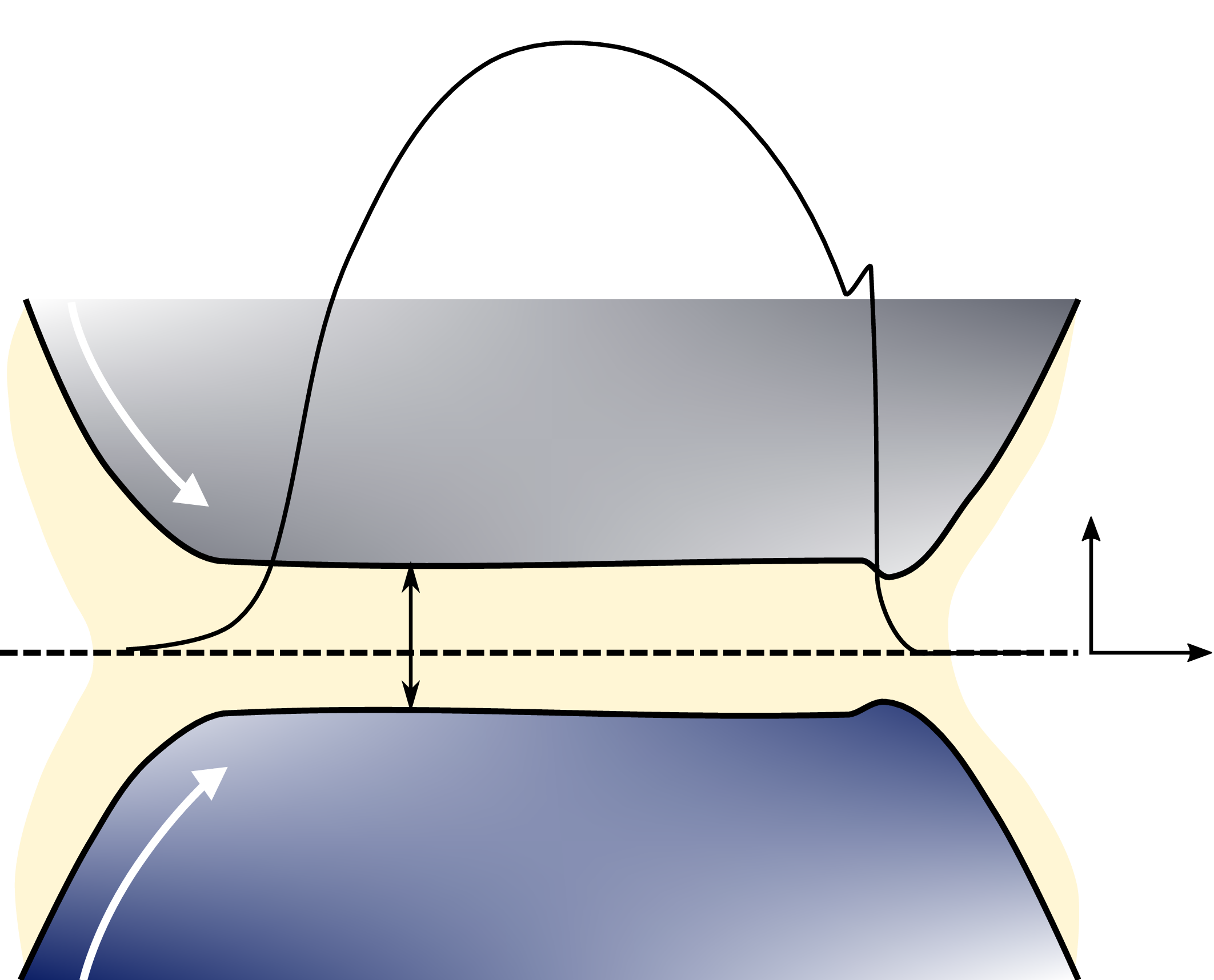}};
			\begin{scope}[shift=(img.south west), 
				x={(img.south east)},y={(img.north west)}] 
				\node at (.255, .77) {$p$};
				\node at (.38, .38) {$h$};
				
				\node at (.985, .375) {$x$};
				\node at (.93, .457) {$z$};
			\end{scope}
		\end{tikzpicture}
		\caption{Cross-section of an elastohydrodynamically lubricated line contact. A typical pressure profile is shown and the flattening of the roller surfaces is visible.}
		\label{fig:roller2d}
	\end{subfigure}
	\caption{Illustration of elastohydrodynamic lubrication.}
	\label{fig:roller}
\end{figure}

The degree of separation serves as a criterion to discern different lubrication regimes.
For example, for hydrodynamic lubrication, the surfaces are fully separated, and the load is in its entirety carried by the hydrodynamic pressure.
A special case of this regime occurs in the contact between surfaces that do not conform to each other, for example, in rolling element bearings, cams and gears.
In these so-called non-conformal contacts, the load is carried by a small area and the hydrodynamic pressure rises locally with typically four orders of magnitude, leading to a considerable increase in viscosity and elastic deformation of the solid surfaces, often an order of magnitude larger then the film thickness.
This phenomenon is called \emph{elastohydrodynamic lubrication} (EHL) and is advantageous because high load can be transferred with minimal friction.
The deformation and hydrodynamic pressure profile are sketched in \autoref{fig:roller}.

The loads in rolling element bearings, gears and cams are typically transferred through a number of non-conformal elastohydrodynamically lubricated line or point contacts.
Today's trend towards higher power density and better performing drive trains also pushes these components to their limit regarding: (1) being able to predict and prevent wear and failure in order to minimize outage, (2) limiting noise, vibration and harshness (NVH) and (3) reducing friction and as such the power consumption.

Above aspects underline the importance of understanding the behavior of EHL contacts under varying conditions.
This is all but an easy task due to the very high local pressures, typically exceeding \qty{1}{\giga\pascal} up to \qty{3}{\giga\pascal}, and very thin lubricant films, in the order of a few \qty{100}{\nano\metre}.
Analytical formulae to predict the film thickness in the contact exist \cite{Grubin1949, Dowson1995, Moes1992, Hamrock2004}, but they are based on very restrictive assumptions, e.g., exponential piezo-viscous relation, and the parameters in these equations are obtained through curve-fitting, resulting in limited accuracy.
Experimentally, it is challenging to measure quantities such as film thickness and pressure in non-conformal contacts without interfering with them, due to the small contact area, the inaccessible and opaque nature of the contact and the fast moving surfaces, in addition to the elastic deformation and extreme pressures.
Examples include ball-on-disk experiments \cite{Stachowiak2004} and film thickness measurements based on electrical resistance or capacitance \cite{Manjunath2024, Jablonka2012}.

Consequently, much research on EHL contacts is numerical in nature.
Since the shape and thickness of the lubricant film in the contact follows from interaction between the hydrodynamic pressure build-up and the elastic deformation of the solid surfaces, the phenomenon is a strongly coupled fluid-structure interaction problem.
Historically, the problem has been modeled using a monolithic solver combining the Reynolds and Boussinesq equations for the lubricant flow and solid deformation respectively.
Although the Reynolds-Boussinesq approach has certainly proved its usefulness, the necessary assumptions unavoidably result in inaccuracies, for example, through the negligence of inertial effects in the Reynolds equations, relevant at the inlet regions, and the inexact Boussinesq approximation for large deformations \cite{Scurria2021}.
More recently, high-fidelity CFD and CSM solvers have been coupled together in a partitioned approach \cite{Hajishafiee2017, Singh2021}.

Besides the challenges already mentioned, the modeling of the lubricant behavior itself is not straightforward.
In literature, different models have been proposed to capture compressibility, piezoviscosity (the steep increase of viscosity with pressure), shear thinning (the decrease of viscosity under shear strain) and thermal effects.

This work presents a new CFD lubricant solver developed in the \OF toolbox.
A compressible pressure-based approach is chosen since there are regions (away from the contact) with near constant density.
The current state-of-the-art includes cavitation through a homogeneous equilibrium model (HEM), which is already implemented in \OF in the existing solver \emph{cavitatingFoam}.
However, this solver only accounts for a constant compressibility $\psi$ of both the liquid and vapor phase.
While this approach is sufficient to model the vapor phase, it is not suitable for the liquid phase due to the large pressure variations.
Moreover, no thermal behavior is included.
The implementation of the lubricant models necessitates the introduction of a variable density and the inclusion of the temperature equation formulated above.

The object-oriented nature and structure of the \OF toolbox allows to benefit from the top-level syntax for the implementation of the new solver and to adopt a modular approach, where lubricant models can be used interchangeably.
The result is a new CFD solver and accompanying lubricant models that allow to accurately model lubricant flow and eventually, after coupling with a structural solver, allow to predict the phenomena occurring in EHL contacts.

The remainder of this work is structured as follows.
The next section details the governing equations and their implementation within \OF.
Thereafter, \autoref{sec:test_case} discusses the use of the solver in a partitioned EHL simulation and presents a test case.
Section \ref{sec:results} demonstrates the accuracy of the solver by validation with literature and provides some results for the test case, showing the effect and relevance of the variable compressibility and of the inclusion of temperature.
Finally, the conclusion is presented.

\section{Lubricant solver implementation}
\label{sec:solver}
To simulate lubricant flow with variable compressibility and temperature effects within \OF, a new solver is created.
This is achieved through the modification of \emph{cavitatingFoam}.
Before the required changes are discussed, the governing equations are presented.

\subsection{Governing equations for lubricant flow}
\label{sec:equations}
In this work, lubricant flow is modeled by the Navier-Stokes equations and equations asserting the conservation of mass and energy. 
Next to these \emph{flow equations}, cavitation is included through the use of a \emph{homogeneous equilibrium model} (HEM).
In addition to the equations solved for pressure $p$, velocity $\mU$ and temperature $T$, constitutive equations are required to express the dependence of other properties on these quantities and in this way model compressibility, piezoviscosity, shear thinning and the dependence of thermal parameters on pressure and temperature.
These \emph{lubricant models} exist in different forms in literature and the parameters they contain are tailored to specific lubricants.
The modular approach of \OF allows to switch and combine different lubricant models and readily change the parameters they use.

\subsubsection{Flow equations}
\label{sec:flow_equations}
The continuity, Navier-Stokes and energy equations are given in their convective form by\footnote{In this work, no body forces or external heat sources are considered.}
\begin{subequations} \label{equ:convective}
	\begin{alignat}{2}
		&\DDt{\rho} &&= - \rho \mdiv\mU\label{equ:convective_continuity} \\
		\rho&\DDt{\mU} &&= \mdiv\stress\label{equ:convective_momentum}\\
		\rho&\DDt{E} &&= \mdiv(\stress\mdprod\mU) - \mdiv\vec{q}\label{equ:convective_energy},
	\end{alignat}
\end{subequations}
where on the left-hand side, the material derivative $\DDt{}\dots=\ddt\dots+\mU\mdprod\nabla\dots$ is used.

The energy equation, \autoref{equ:convective_energy}, is reformulated in terms of temperature $T$ by splitting the total energy $E$ in internal energy $e$ and mechanical energy $K=\frac{1}{2}\normE{\mU}^2$.
The identities 
\begin{equation}
	\rho \DDt{K}\equiv\rho\DDt{\mU}\mdprod\mU \qquad \text{and} \qquad
	\mdiv(\stress\mdprod\mU) \equiv (\mdiv\stress)\mdprod\mU + \stress\mddprod\mgU,
\end{equation}
allow to rewrite the mechanical energy contributions
\begin{equation}
	\rho \DDt{K} - \mdiv(\stress\mdprod\mU) = \rho\DDt{\mU}\mdprod\mU - (\mdiv\stress)\mdprod\mU - \stress\mddprod\mgU,
\end{equation}
which reduce to $-\stress\mddprod\mgU$ after taking the momentum equation, \autoref{equ:convective_momentum}, into account \cite{Greenshields2022}.
The resulting equation for the internal energy equation becomes
\begin{equation}
	\rho \DDt{e}=\stress\mddprod\mgU - \mdiv\vec{q}.
\end{equation}

Before writing this equation in terms of temperature, it is returned to the conservation form using the conservation form of the continuity equation, \autoref{equ:convective_continuity}, given by
\begin{equation}
	\ddt{\rho}+\mdiv(\rho\mU)=0.\label{equ:conservative_continuity}
\end{equation}
Furthermore, the fluid stress tensor is split in a pressure and shear stress contribution: $\stress=-p\I+\shearStress$.
For the pressure part holds ${-p\I\mddprod\mgU}\equiv {-p\mdiv\mU}$, and, because the viscous shear stress tensor $\shearStress$ is symmetric with zero-diagonal sum, the term $\shearStress\mddprod\nabla \mU$ can be reduced to $\shearStress\mddprod\mdevD$, with $\mdevD$ the deviatoric part of the deformation tensor $\mD=({\mgU}+\trans{\mgU})/2$.
The energy equation thus becomes
\begin{equation}
	\ddt{}(\rho e) + \mdiv(\rho e \mU)=-p\mdiv\mU + \shearStress\mddprod\mdevD - \mdiv\vec{q}.
\end{equation}
The right-hand side reveals that the change in internal energy is determined by compression work, viscous heating and thermal conductivity.

The final equation is obtained by introducing the specific enthalpy $h\equiv e+p/\rho$ and Fourier's law of thermal conduction $\vec{q}=-\thk\nabla T$, as follows
\begin{equation}
	\ddt{}(\rho h) + \mdiv(\rho h \mU)=\DDt{p} + \shearStress\mddprod\mdevD + \mdiv(\thk\nabla T),
	\label{equ:temperature_eq}
\end{equation}
where $\thk$ is the thermal conductivity of the lubricant.
What remains is to determine the dependence of $h$ on temperature, but in order to do so, the cavitation model must be explained first.

\subsubsection{Homogeneous equilibrium model}
In this work, it is assumed that cavitation occurs due to the vaporization of the lubricant, i.e., flash evaporation, and no dissolved gasses are considered.
The phenomenon is included using a homogeneous equilibrium model (HEM), which assumes that in every point of the continuum, the liquid and vapor phase are in mechanical and thermodynamic equilibrium, meaning they have equal velocity, pressure and temperature \cite{Stewart1984, Karrholm2007}.
With decreasing pressure, this model maintains a specified cavitation pressure $p_{sat}$ in the cavitated region by converting liquid into vapor \cite{Hartinger2008}.
The pressure can only drop below the cavitation pressure when all the liquid in that region has evaporated.
Conversely, the pressure can only increase above the cavitation pressure, once all vapor in the region is converted back into liquid.
While the cavitation model has some limitations, e.g., it does not account for surface tension, the corresponding pressure values are small compared to the maximal pressure in the contact, and therefore these limitations are not expected to largely impact the integral quantities such as load and film thickness \cite{Almqvist2002}.

The vapor and liquid volume fraction are denoted as $\alpha_v$ and $\alpha_l$, respectively.
The mixture density, dynamic viscosity, compressibility, thermal conductivity, specific heat capacity and enthalpy are expressed as weighted averages.
\begin{subequations}
	\label{equ:hem}
	\begin{alignat}{1}
		\rho &= \alpha_v\rho_{v} + \alpha_l\rho_{l} \label{equ:mixture_density} \\
		\mu &= \alpha_v\mu_{v} + \alpha_l\mu_{l} \label{equ:mixture_viscosity} \\
		\psi &= \alpha_v\psi_{v} + \alpha_l\psi_{l} \label{equ:mixture_compressibility} \\
		\thk &= \alpha_v\thk_v + \alpha_l\thk_l \label{equ:mixture_thermal_conductivity} \\
		\rho c_p &= \alpha_v\rho_{v}c_{p,v} + \alpha_l\rho_{l}c_{p,l} \label{equ:mixture_heat_capacity} \\
		\rho h &= \alpha_v\rho_{v}h_v + \alpha_l\rho_{l}h_l \label{equ:mixture_enthalpy}
	\end{alignat}
\end{subequations}
The compressibility $\psi$, defined as $\left.\dd{\rho}{p}\right\vert_s$, will be used in the next section and is related to the speed of sound $c=\sqrt{1/\psi}$.
Different models to determine the mixture compressibility exist, but the one used here (\autoref{equ:mixture_compressibility}) corresponds to the linear compressibility model.
The specific heat capacities $c_{p,v}$ and $c_{p,l}$ are used to link the enthalpy to temperature \cite{Moran2006}, as follows 
\begin{subequations}
	\label{equ:enthalpy}
	\begin{align}
		h_v &= h_{l,R} + h_{vap}+c_{p,v}(T-T_R) \\
		h_l &= h_{l,R} + c_{p,l}(T-T_R)+\int_{0}^{p}\frac{1}{\rho_l}(1-T\beta)\diff p. \label{equ:enthalpy_liquid}
	\end{align}
\end{subequations}
In this equation $h_{l,R}$ is the enthalpy at reference temperature $T_R$ and zero pressure, $h_{vap}$ is the enthalpy of vaporization and $\beta$ is the volume expansivity given by $\left.-\frac{1}{\rho}\dd{\rho}{T}\right\vert_p$.

Since the pressure and temperature of the vapor phase only deviate slightly from the cavitation pressure and reference temperature (about \qty{3}{\pascal} and \qty{1}{\kelvin}), the vapor parameters $\mu_{v}$, $\psi_{v}$, $\thk_v$ and $c_{p,v}$ are considered constant.
This is not the case for the liquid parameters and their relation to the solution variables is subject of the next section.
Finally, the vapor density $\rho_v$ is expressed using the constant compressibility $\psi_{v}$
, as
\begin{equation}
	\rho_v=\psi_v p.
\end{equation}

\subsubsection{Lubricant models}
\label{sec:lubricant_models}
The previous sections described the governing equations to be solved for pressure, the velocity components and temperature.
Other variables are related to these through equations of state, called lubricant models.

There is no consensus in literature on which lubricant models are most suited, but the importance of their choice has been highlighted by multiple authors \cite{Bair1993, Bair2000, Tosic2019, Havaej2023}.
A frequently used combination \cite{Almqvist2002, Hartinger2008, Hajishafiee2017, Singh2021} is the Dowson-Higginson expression for compressibility \cite{Dowson1977}, the piezoviscous model of Roelands \cite{Roelands1966} (in some cases the adaptation by Houpert \cite{Houpert1985}) and if applicable the Ree-Eyring \cite{Evans1986} shear thinning equation.

Others argue \cite{Bair2019, Zolper2020, Bjoerling2014} for the use of the Doolittle model for the pressure-viscosity relation in combination with the Tait equation for compressibility and the Carreau model for shear thinning.
It is this combination that will be detailed below.
However, it is important to note that other lubricant models can be opted for by simply changing the library used by the \OF solver.

As compressibility equation, expressing $\rho_l$ in function of $p$ and $T$, the Tait equation is used \cite{Zolper2020}.
Based on the free-volume framework, this equation provides an expression for the specific volume $\vol_l=1/\rho_l$, as follows\footnote{The subscript $l$ referring to the liquid lubricant has been omitted from the specific volume for brevity.}
\begin{subequations}
	\label{equ:tait}
	\begin{alignat}{1}
		\frac{\vol}{\vol_0} &= 1-\frac{1}{1+K'_0}\ln\left[1+\frac{p}{K_0}\left(1+K'_0\right)\right] \\
		\frac{\vol_0}{\vol_{R}} &= 1+a_\vol(T-T_R) \\
		K_0 &= K_{00}e^{-\beta_K T}
	\end{alignat}
\end{subequations}
Here, $\vol_0$ is the specific volume at atmospheric pressure and only depends on $T$.
Its reference value at $T_R$ is denoted by $\vol_{R}$ and its reciprocal $\rho_{R}$.
Further, while $K_0$ depends on temperature, $K'_0$, $K_{00}$, $a_\vol$ and $\beta_K$ are material constants. 
By performing mathematical operations on the Tait equation, \autoref{equ:tait}, the volume expansivity $\beta$, liquid compressibility $\psi_l$ and integral term in \autoref{equ:enthalpy_liquid} can be calculated, see \autoref{sec:tait_derived}.

The expression for the lubricant viscosity $\mu_l$ is given by the Doolittle equation, which is also based on the liquid's free volume and combines well with the Tait equation \cite{Zolper2020}.\footnote{The subscript $l$ referring to the liquid lubricant has been omitted from the viscosity.}
\begin{subequations}
	\begin{alignat}{1}
		\ln\left(\frac{\mu}{\mu_{R}}\right) &= B\left(\frac{\vol_\infty}{\vol-\vol_\infty}-\frac{\vol_{\infty R}}{\vol_{R}-\vol_{\infty R}}\right) \\
		\frac{\vol_\infty}{\vol_{\infty R}} &= 1+a_\infty(T-T_R) \\
		R_0 &= \frac{\vol_{\infty R}}{\vol_{R}}
	\end{alignat}
\end{subequations}
Herein, $\vol_\infty$ represents the occupied volume, which only depends on temperature.
Its reference value at $T_R$ is analogously denoted by $\vol_{\infty R}$.
This parameter is linked to $\vol_{R}$, through the ratio $R_0$.
The reference viscosity at $T_R$ and atmospheric pressure is denoted $\mu_{R}$, and the parameters $B$ and $a_\infty$ are material specific constants.

Furthermore, the viscosity often also depends on the shear rate.
This non-Newtonian behavior is included with the Carreau shear thinning model \cite{Bair2002}.
To differentiate between the value of viscosity before and after accounting for shear thinning, the latter is denoted by $\eta$.
Nevertheless, it is the value after shear thinning, i.e., $\eta$, that has to be used as $\mu_l$ in \autoref{equ:mixture_viscosity}.
\begin{subequations}
	\begin{alignat}{1}
		\eta &= \mu\left[1+\left(\frac{\mu\dot{\gamma}}{G}\right)^2\right]^{\frac{n-1}{2}} \\
		G &= G_{R} \frac{T}{T_R}\frac{\vol_{R}}{\vol} \\
		G_{R} &= \frac{\mu_{R}}{\lambda_{R}}
	\end{alignat}
\end{subequations}
In this equation, $\dot{\gamma}$ is the strain rate, given by
\begin{equation}
	\dot{\gamma} = \sqrt{\vphantom{\mD}2\mD:\mD} \qquad \text{with} \qquad \mD = \frac{\nabla\mU+\trans{\nabla\mU}}{2},
\end{equation}
corresponding to the square root of twice the second main invariant of the deformation tensor $\mD$ \cite{Baur2007, Gurevich1964}. 
Further, $G$ is the effective shear modulus and $G_{R}$ its value at $T_R$ and atmospheric pressure. The constant $\lambda_{R}$ is the relaxation time at the same conditions, and $n$ is yet another material constant.

For sufficiently high shear stress $\tau=\eta\dot{\gamma}$ relative to the pressure $p$, a phenomenon called shear localization has been observed \cite{Gray1998}, resulting in a limitation on the value of the shear stress \cite{Bair2002}.
This limitation is modeled as follows
\begin{equation}
	\tau = \min\left(\eta\dot{\gamma},\Lambda p\right) \qquad \text{for} \qquad p > \qty{1}{\mega\pascal},
\end{equation}
where $\Lambda$ is the limiting shear stress pressure coefficient.
With the corrected value of $\tau$, a corrected value for $\eta=\tau/\dot{\gamma}$ is calculated.
Care is required to assure that this correction is only applied for sufficiently high pressure, e.g., \qty{1}{\mega\pascal}.

Lastly, the dependence of thermal conductivity $\thk_l$ and heat capacity $c_{p,l}$ of the liquid lubricant on the thermodynamic state \cite{Bjoerling2014} is modeled according to 
\begin{subequations}
	\begin{alignat}{1}
		\thk_l &= C_k\kappa^{-s} \\
		\kappa &= \left(\frac{\vol}{\vol_{R}}\right)\left[1+K\left(\frac{T}{T_R}\right)\left(\frac{\vol}{\vol_{R}}\right)^q\right]
	\end{alignat}
\end{subequations}
and
\begin{equation}
	\label{equ:heat_capacity}
	c_{p,l} = \vol \left(C_0 +m\left(\frac{T}{T_R}\right)\left(\frac{\vol}{\vol_{R}}\right)^{-3}\right).
\end{equation}
In these equations, $\kappa$ is a thermal conductivity scaling factor, while $C_k$, $s$, $K$, $q$, $C_0$ and $m$ are material constants.\footnote{Other sources use $-4$ as exponent instead of $-3$ in \autoref{equ:heat_capacity}, e.g., \cite{Habchi2010,Havaej2023}.}

\subsection{Adaptions with respect to \emph{cavitatingFoam}}
\label{sec:implementation}

To solve the flow equations including lubricant models from the previous section, the \OF solver \emph{cavitatingFoam} is modified.
This solver, is part of the standard \OF installation and includes an HEM model accounting for a constant compressibility $\psi$ of both the liquid and vapor phase.
While this approach is sufficient to model the vapor phase, it is not for the liquid phase, as explained in \autoref{sec:flow_equations}.
The implementation of the lubricant models necessitates the introduction of a variable compressibility and the inclusion of the temperature equation formulated above.
A similar approach was followed by Hartinger \cite{Hartinger2007} and Havaej et al. \cite{Havaej2023a}, yet there are important differences, as will be explained further.

Before proceeding, it is important to note that, in the finite volume method, the conservation laws are expressed for each cell and integrated over the cell volume $V$, after which they are transformed to a surface integral over the cell faces $S$, using Gauss's theorem \cite{Ferziger2002}.
While all equations are thus integral equations, \OF uses differential notation for conciseness and increased readability, and the same approach will be followed here \cite{Weller1998, Greenshields2020, Greenshields2022}.
For instance, in case of convection the following conversion applies.
\begin{equation}
	\label{equ:openfoam_flux}
	\text{convection: }\mdiv(\mU\thv)\quad\Rightarrow\intV{\mdiv(\mU\thv)} = \intS{}\cdot(\mU\thv) \approx \sum_f \Sv_f\cdot\mU_f\thv_f = \sum_f \phi_f\thv_f,
\end{equation}
where the subscript $f$ refers to the quantity evaluated at the faces, $\Sv_f$ is a face area vector orthogonal to the face and pointing outward from the cell, and the variable $\thv$ represents a general vector or scalar.
In this equation, the convective flux $\phi_f=\Sv_f\cdot\mU_f$ is introduced.
In the following, the bold variant will also be used in the differential notation $\mdiv(\phiv\thv)$.

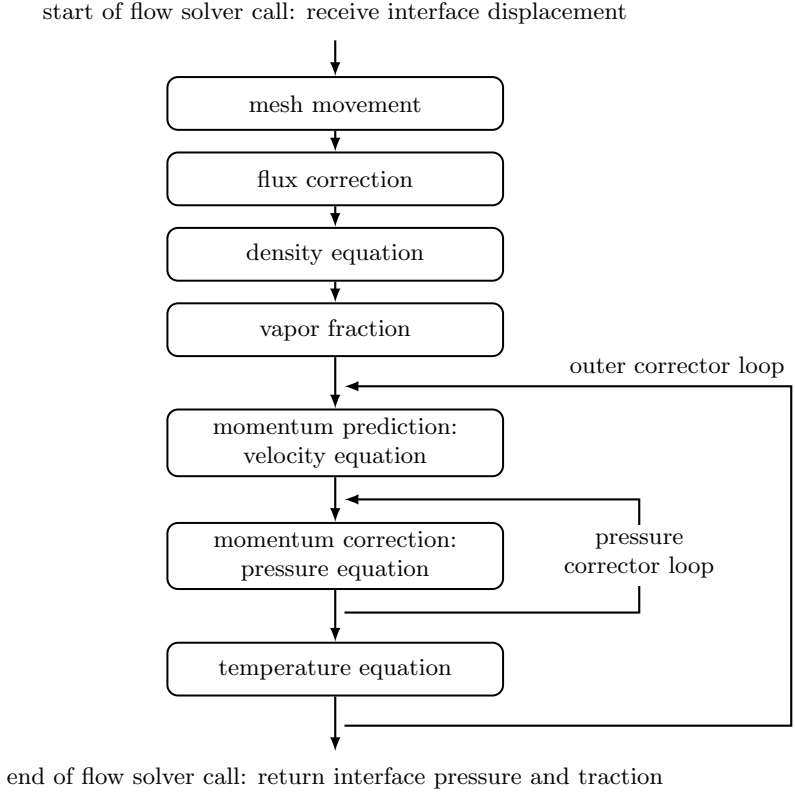
\begin{figure}[htbp]
	\centering
	\begin{tikzpicture}[node distance=1cm,
		every node/.style={fill=white, align=center}]
		\node (start) [base] {start of flow solver call: receive interface displacement};
		\node (mesh) [block, below of=start, yshift=-.2cm] {mesh movement};
		\node (flux) [block, below of=mesh] {flux correction};
		\node (density) [block, below of=flux] {density equation};
		\node (vapor) [block, below of=density] {vapor fraction};
		\node (bfpred) [ghost, below of=vapor, yshift=.25cm] {};
		\node (prediction) [block, below of=bfpred, yshift=.25cm] {momentum prediction: \\velocity equation};
		\node (bfcorr) [ghost, below of=prediction, yshift=.25cm] {};
		\node (correction) [block, below of=bfcorr, yshift=.25cm] {momentum correction: \\pressure equation};
		\node (afcorr) [ghost, below of=correction, yshift=.25cm] {};
		\node (temperature) [block, below of=afcorr, yshift=.25cm] {temperature equation};
		\node (aftemp) [ghost, below of=temperature, yshift=.25cm] {};
		\node (end) [base, below of=aftemp, yshift=.3cm] {end of flow solver call: return interface pressure and traction};
		
		\draw [arrow] (start) -- (mesh);
		\draw [arrow] (mesh) -- (flux);
		\draw [arrow] (flux) -- (density);
		\draw [arrow] (density) -- (vapor);
		\draw [arrow] (vapor) -- (prediction);
		\draw [arrow] (prediction) -- (correction);
		\draw [arrow] (correction) -- (temperature);
		\draw [arrow] (temperature) -- (end);
		
		\draw [arrow] (afcorr) -- ++(4,0) -- node[text width=2cm]{pressure corrector loop} ++(0,1.5) -- (bfcorr);
		\draw [arrow] (aftemp) -- ++(6,0) -- node[text width=3cm, yshift=2.5cm, xshift=-1.5cm, fill=none]{outer corrector loop} ++(0,4.5) |- (bfpred);
	\end{tikzpicture}
	\caption{Flowchart illustrating the PIMPLE loop of the flow solver for use in a partitioned FSI approach to simulate EHL.}
	\label{fig:pimple}
\end{figure}

The solver uses the so-called PIMPLE loop to achieve the pressure-velocity coupling in a segregated approach.
An overview of the main blocks is given in \autoref{fig:pimple}.
The steps are shortly described below with emphasis on what has been changed compared to \emph{cavitatingFoam}.
In the subsequent sections, this lubricant solver will be used in an iterative scheme together with a structural solver to simulate an elastohydrodynamically lubricated contact, which is a fluid-structure interaction phenomenon.
Here, however, the focus is on a single calculation of the flow solver in this iterative scheme with the structural solver, and thus with a deforming fluid domain.

\subsubsection{Mesh movement}
When the flow solver is called, the mesh is updated first, based on the displacement of the fluid-structure interface. In this case, this is achieved by solving the cell-centered Laplacian for displacement $\md$, given by
\begin{equation}
	\laplacian \md = 0,
\end{equation}
but other mesh motion techniques are equally possible.
Since the mesh moved, the flux $\phi_f=\mU_f\cdot\Sv_f$ has to be recalculated, and this is achieved with the face values of the velocity mapped to the new mesh position. 

\subsubsection{Flux correction}
Due to the mapping of the velocity field, the conservation of mass is not guaranteed.
Therefore, the flux is corrected by solving the following equation for $p_{corr}$,
\begin{equation}
	\mdiv(\rho\phiv) - \mdiv\left(\frac{1}{A_{corr}} \nabla p_{corr}\right) = \xi,
\end{equation}
where $A_{corr}$ is a constant equal to one with unit \qty{1}{\per\second}, and $\xi$ is the value that the divergence of the corrected mass flux has to match, i.e., $\mdiv(\rho\phiv)$ stored at the end of the previous time step.
The name $p_{corr}$ is chosen because of the analogy with the pressure equation (see further), but this variable is unrelated to the pressure itself.
The corrected flux is obtained as
\begin{equation}
	\phiv - \frac{1}{\rho} \left(\frac{1}{A_{corr}} \nabla p_{corr}\right).
\end{equation}
Lastly, the flux is made relative to the mesh motion, as will be required for the convective term in the following steps.
The relative flux will be denoted as $\varphiv$ for clarity.
Details on the relative flux and the difference with the absolute flux are given in \autoref{sec:flux}.

\subsubsection{Density equation}
Next, the continuity equation, \autoref{equ:conservative_continuity}, is solved for $\rho$ using the relative flux $\varphiv$ as
\begin{equation}
	\ddt{\rho} + \mdiv (\varphiv \rho)=0. \label{equ:continuity}
\end{equation}

\subsubsection{Vapor fraction}
The newly obtained density is used to determine the vapor fraction by comparing its value to the liquid saturation density $\rho_{l,sat}=\rho(p_{sat},T)$ ($p_{sat}$ is considered constant) and the vapor saturation density $\rho_{v,sat}=\psi_v p_{sat}$, as follows
\begin{subequations}
	\begin{align}
		\alpha_v &= \max\left(\min\left(\frac{\rho_{l,sat}-\rho}{\rho_{l,sat}-\rho_{v,sat}}, 1\right), 0\right),
		\intertext{from which also the liquid fraction}
		\alpha_l &= 1-\alpha_v
	\end{align}
	\label{equ:updateFraction}
\end{subequations}
\kern-3pt is determined.
Furthermore, with these new values, the compressibility $\psi$ is updated with \autoref{equ:mixture_compressibility}. \\

The previous steps are only executed at the start of the solver call. The following steps, on the other hand, occur within a so-called outer corrector loop.

\subsubsection{Momentum prediction: velocity equation}
The first step of the pressure-velocity coupling loop is the momentum prediction, in which the conservative form of the momentum equation, \autoref{equ:convective_momentum}, is solved for the velocity as follows
\begin{equation}
	\ddt{}(\rho\mU) + \mdiv(\rho\varphiv\mU) -\mdiv \left(\mu\left[\trans{\nabla\mU}-\frac{2}{3}\tr(\trans{\nabla\mU})\I\,\right] \right) - \mdiv \left(\mu\nabla\mU\right)= -\nabla p.
\end{equation}
The notation of the diffusion term is chosen to mimic the computation in \OF.
For the next step, the left-hand side of this equation is split into $\A\mU$ and $-\Hu(\mU)$, where $\A$ only has diagonal elements and hence only one value per cell.
The off-diagonal elements of the coefficient matrix and the terms that not depend on velocity (excluding the pressure) are contained within $\Hu(\mU)$.

\subsubsection{Momentum correction: pressure equation}
With this split of the momentum equation, a momentum corrector is obtained \cite{Karrholm2008}, as follows
\begin{equation}
	\mU = \frac{\Hu(\mU)}{\A} - \frac{1}{\A}\nabla p.
	\label{equ:momentumCorrection}
\end{equation}
By substituting this in the continuity equation, \autoref{equ:continuity}, a pressure equation is obtained, given by
\begin{equation}
	\ddt{\rho} + \psi\corr{\ddt{p}} + \mdiv\left(\rho\frac{\Hu(\mU)}{\A}\right) - \mdiv\left(\rho\frac{1}{\A}\nabla p\right) =0.
	\label{equ:pressure_equation}
\end{equation}
Again, the resulting flux, given by the last two terms, has to be made relative first, see \autoref{sec:flux}.
The second term contains the corrector operator $\corr{\dots}$ which is used to capture the dependence of $\rho$ on $p$. This term disappears upon convergence.
The pressure equation can be solved for $p$ two or three times in a row. These repetitions are the so-called non-orthogonality corrector iterations.
After the calculation of $p$, the flux $\varphi_f$ is calculated as
\begin{equation}
	\varphi_f =  \Sv_f \cdot \left[ \left(\frac{\Hu(\mU)}{\A}\right)_f - \left(\frac{1}{\A}\right)_f\left(\nabla p\right)_f \right].
\end{equation}

The pressure obtained from this equation does not take into account cavitation, and therefore, the pressure is modified.
The following steps are followed:
\begin{enumerate}
	\item The values of pressure, liquid density and liquid compressibility are stored for later use, in the respective variables $p_{ref}$, $\rho_{l,ref}$ and $\psi_{l,ref}$.
	\item The pressure equation, \autoref{equ:pressure_equation}, is solved, as such enforcing conservation of mass.
	\item The liquid density $\rho_l$ is updated with the new pressure.
	\item The mixture density $\rho$ is determined with \autoref{equ:mixture_density}.
	\item The vapor and liquid fraction are updated with \autoref{equ:updateFraction}, and the mixture compressibility is updated with \autoref{equ:mixture_compressibility}.
	\item Now, the pressure is modified by using a linearized relation between $p$ and $\rho$, given by
	\begin{equation}
		\rho \approx \alpha_v\psi_v p + \alpha_l \left(\rho_{l,ref}+\psi_{l,ref}(p-p_{ref})\right).
	\end{equation}
	Rearranging for $p$ results in
	\begin{equation}
		p = \frac{\rho - \alpha_l \rho_{l,ref} + \alpha_l \psi_{l,ref} p_{ref}}{\alpha_v\psi_v  + \alpha_l \psi_{l,ref}}
	\end{equation}
	The corrected pressure will be $p_{sat}$ everywhere where the vapor fraction is between zero and one.
	\item Next, the liquid density $\rho_l$ and compressibility $\psi_l$ are updated with the new pressure value.
	\item Finally, the mixture density is calculated once more using \autoref{equ:mixture_density}.
\end{enumerate}
At this point, the continuity errors are evaluated by solving the density equation, \autoref{equ:continuity}, and comparing its solution to the thermodynamic $\rho$ obtained above.
By integrating the magnitude of the difference over the entire domain and dividing by the total mass, a so-called local error is obtained that, in this work, has a typical order of magnitude of $\qty{1e-11}{}$.
Lastly the velocity is corrected with the updated pressure using \autoref{equ:momentumCorrection}. 

In the PIMPLE loop the momentum correction is repeated several times, updating $\A$ and $\Hu(\mU)$ with the new velocity.

\subsubsection{Temperature equation}
This step starts with updating the liquid thermal conductivity and heat capacity according to \autoref{equ:mixture_thermal_conductivity} and \autoref{equ:mixture_heat_capacity}, respectively.
Then, the liquid and vapor enthalpy are determined and subsequently used to calculate the product of mixture enthalpy and density as in \autoref{equ:mixture_enthalpy}.

With these values, the temperature equation \autoref{equ:temperature_eq} is solved as follows
\begin{equation}
	\begin{split}
		&\ddt{}(\rho h) + \corr{\ddt{}(\rho c_p T)} 
		+ \mdiv\left(\varphiv \rho h\right) + \corr{\mdiv\left(\varphiv \rho c_p T\right)} \\
		&\qquad=\DDt{p} + \shearStress:\mdevD + \mdiv\left(\thk\nabla T\right).
	\end{split}
\end{equation}
Similarly to the pressure equation, the correction terms include the dependence of $h$ on $T$ and disappear upon convergence.
A note on the calculation of $\Diff{p}/\Diff{t}$ is provided in \autoref{sec:material_derivative}.
This equation is solved for $T$, after which the liquid density $\rho_l$, compressibility $\psi_l$ and saturation density $\rho_{l,sat}$ are recalculated.

Lastly, the liquid viscosity $\mu_l$ is reevaluated with the new values of $p$ and $T$, and eventually the mixture viscosity is determined according to \autoref{equ:mixture_viscosity}.
In order to do this correctly, a compressible variant of the two-phase-mixture model has been created in \OF.\\

The corrector loop is executed until the initial residual of both the velocity equation and the pressure equation (in the first pressure corrector iteration) are below their respective tolerance.

\section{Test case}
\label{sec:test_case}
To illustrate the new lubricant solver, an elastohydrodynamically lubricated line contact is simulated.
Such a contact occurs for example in cylindrical roller bearings, gears or cams.
By neglecting end-effects, the problem can be rendered two-dimensional, representing an infinite line contact.


\subsection{Case description}
\label{sec:case}
\begin{figure}[htbp]
	\centering
	\begin{tikzpicture}[inner sep=0pt,
		every node/.style={align=center}]
		\node (img) at (0,0) {\includegraphics[width=1\textwidth]{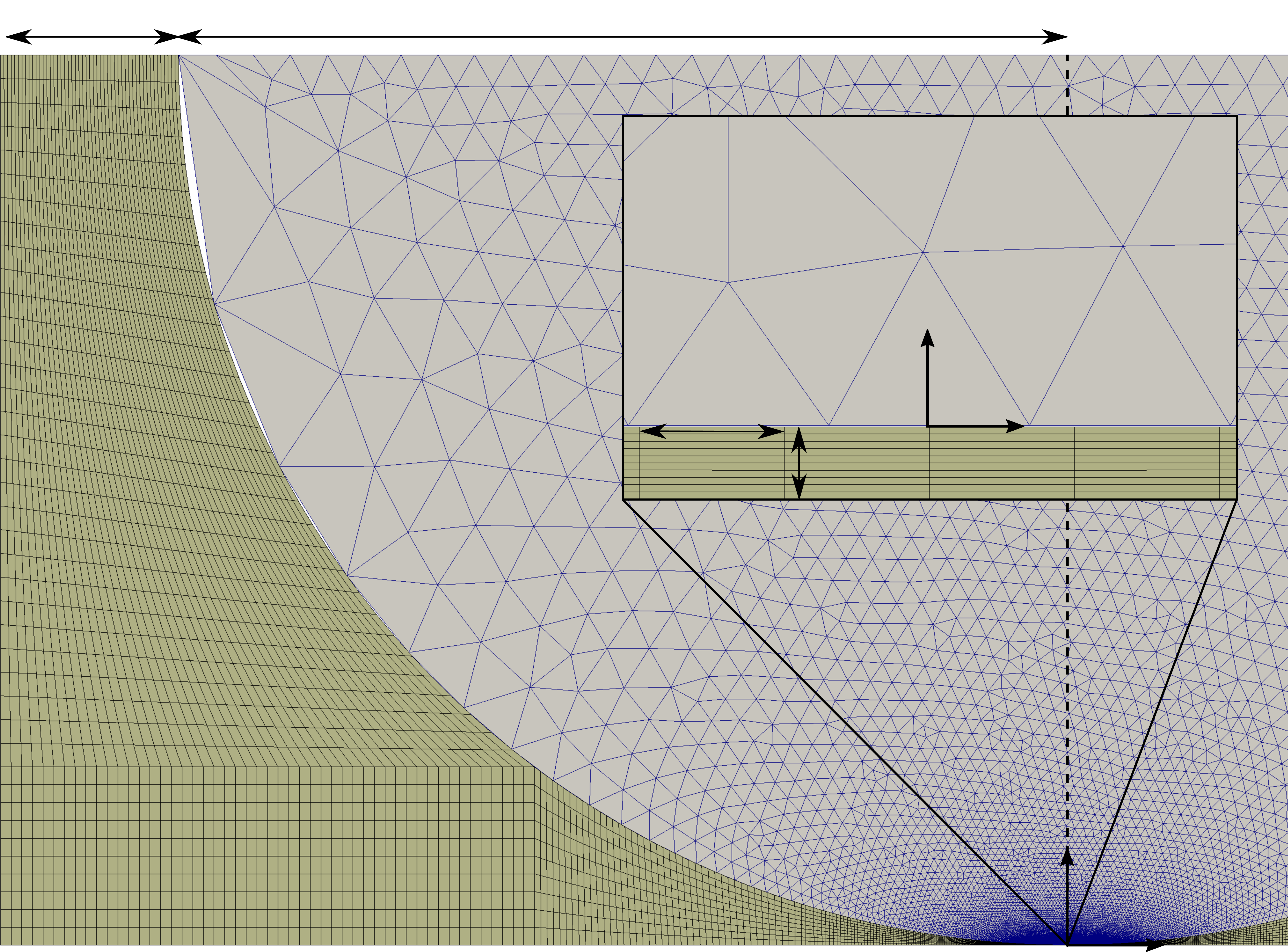}};
		\begin{scope}[shift=(img.south west), 
			x={(img.south east)},y={(img.north west)}] 
			\node at (.072, 1) [anchor=north] {$\qty{0.002}{\meter}$};
			\node at (.48, 1) [anchor=north] {$R = \qty{0.01}{\meter}$};
			
			\node at (.55, .58) {$\qty{500}{\nano\meter}$};
			\node at (.675, .515) {$\qty{250}{\nano\meter}$};
			
			\node at (.737, .642) {$z$};
			\node at (.785, .58) {$x$};
			
			\node at (.81, .1) {$z$};
			\node at (.893, .04) {$x$};
		\end{scope}
	\end{tikzpicture}
	\caption{Left part of the fluid mesh (yellow-green) and structural mesh (gray). Both are symmetric with respect to the $z$-axis.}
	\label{fig:mesh}
\end{figure}
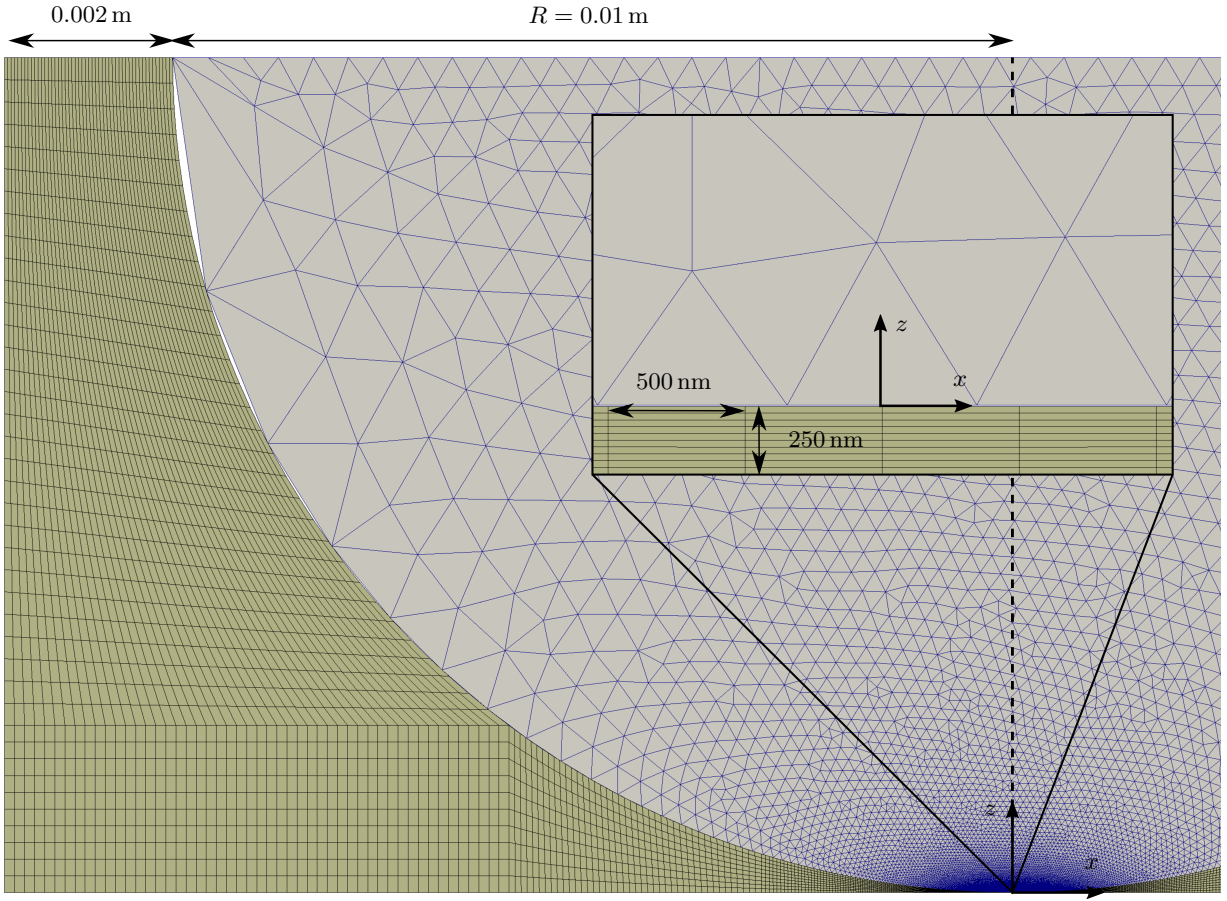

The case considered consists of a flexible roller on a flat rigid plate.
Any line contact between two elastic bodies can be reduced to such a contact \cite{Dowson1977}.
The left part of the resulting geometry, which is symmetric around the $z$-axis, is shown in \autoref{fig:mesh}.

The radius $R$ equals \qty{10}{\milli\meter}.
The entrainment speed $u$ is \qty{2.5}{\meter\per\second} and is defined as the average of the two rolling velocities $u_1$ and $u_2$ of the upper and lower contacting surfaces, respectively,
\begin{equation}
	u = \frac{u_1+u_2}{2}.
\end{equation}

In case of pure rolling, both surfaces have the same velocity.
In reality, however, slip may occur.
This is characterized by a sliding speed $u_s=\abs{u_1-u_2}$ and the slip-to-roll ratio (SRR), given by
\begin{equation}
	\text{SRR} = \frac{u_s}{u}= \frac{2\abs{u_1-u_2}}{u_1+u_2}. \label{equ:slip}
\end{equation}

\subsection{Structural solver and coupling tool}
Besides the lubricant flow solver, a structural solver is also required, since EHL is a fluid-structure interaction phenomenon.
In this work, the structural side is modeled with finite elements using the open-source package Kratos Multiphysics, specifically the Structural Mechanics Application in version 9.1 \cite{Dadvand2010}.
The coupling between them is achieved with the open-source coupling tool CoCoNuT \cite{Delaisse2023} (\href{https://github.com/pyfsi/coconut/}{GitHub repository pyfsi/coconut}), which treats the solvers as black boxes and calls them when needed.
The solid solver and coupling tool can be used out-of-the-box, without modifications.


\begin{figure}[htbp]
	\centering
	\begin{tikzpicture}[inner sep=0pt,
		every node/.style={align=center}]
		\node (img) at (0,0) {\includegraphics[width=\textwidth]{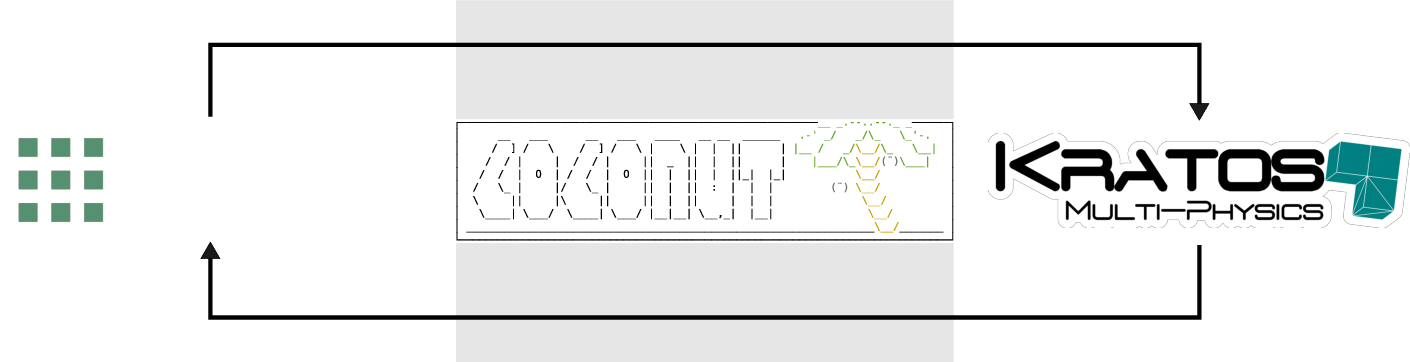}};
		\begin{scope}[shift=(img.south west), 
			x={(img.south east)},y={(img.north west)}] 
			\node at (.09, 0.5) [anchor=west] {\LARGE\OF};
			\node at (.5, 0.85) [anchor=north] {pressure and traction};
			\node at (.5, 0.15) [anchor=south] {displacement};
		\end{scope}
	\end{tikzpicture}
	\caption{Partitioned coupling in each time step between the flow solver (\OF) and the structural solver (Kratos Multiphysics) with the coupling tool CoCoNuT.
	}
	\label{fig:partitioned_coupling}
\end{figure}
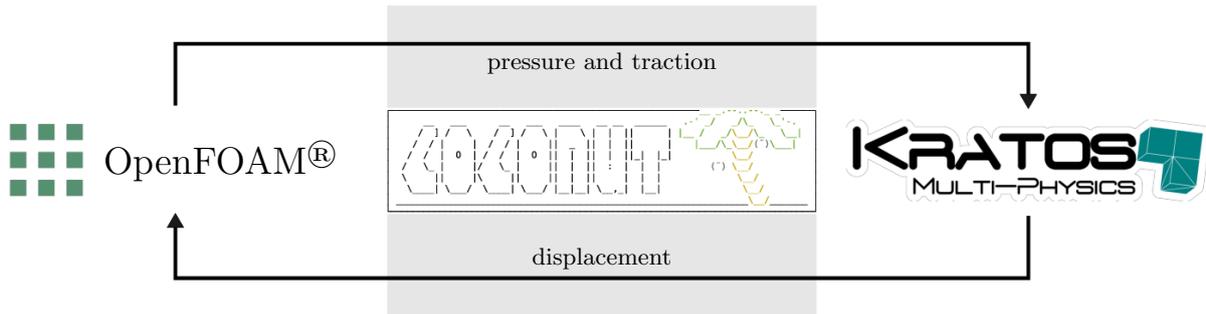
A schematic illustration of the partitioned coupling of the flow and structural solver is given in \autoref{fig:partitioned_coupling}.
By iterating between the two solvers within each time step, also called implicit coupling, it is assured that displacement, pressure and traction data on the common interface are the same in both solver up to a certain tolerance.
This iterative scheme is stabilized and accelerated using a coupling algorithm within CoCoNuT.
In this case a quasi-Newton algorithm is used, more specifically IQN-ILSM \cite{Delaisse2022} with reuse of 50 time steps and without filtering.
Interpolation of interface data between the fluid and structural grid is achieved with radial basis mapping, and the initial value of the interface displacement for the next time step is determined through quadratic extrapolation.
The coupling loop in a time step is considered converged when the Euclidean norm of the coupling residual, which is defined as the difference between the input and output displacement of the flow and structural solver, respectively, becomes smaller than \qty{1e-10}{\meter}.

\subsection{Discretization}
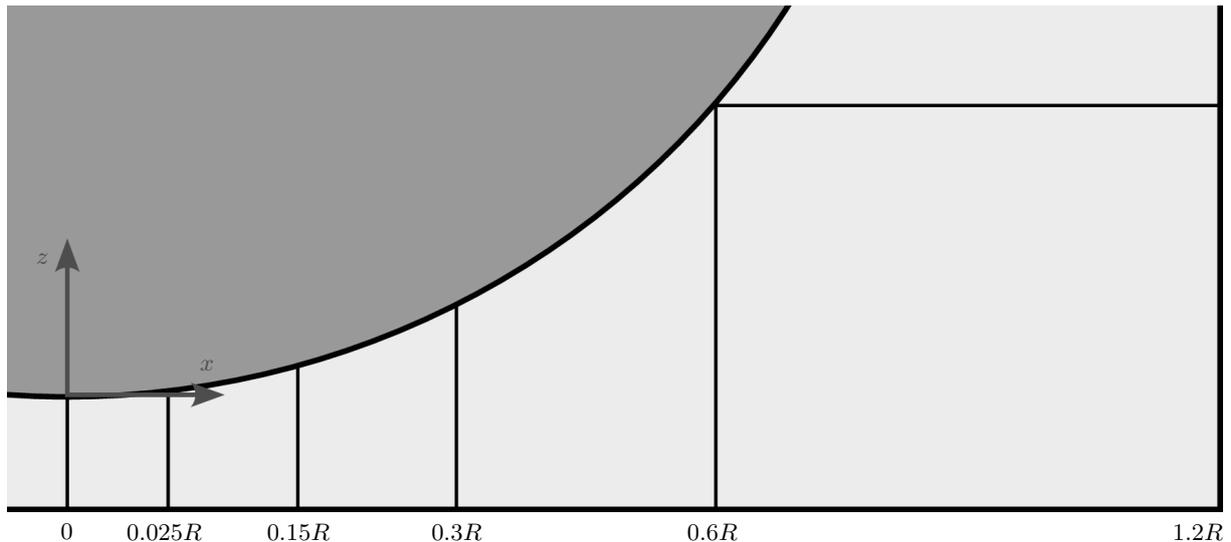
\begin{figure}[htbp]
	\centering
	\begin{tikzpicture}[inner sep=0pt,
		every node/.style={align=center}]
		\node (img) at (0,0) {\includegraphics[width=1.0\textwidth]{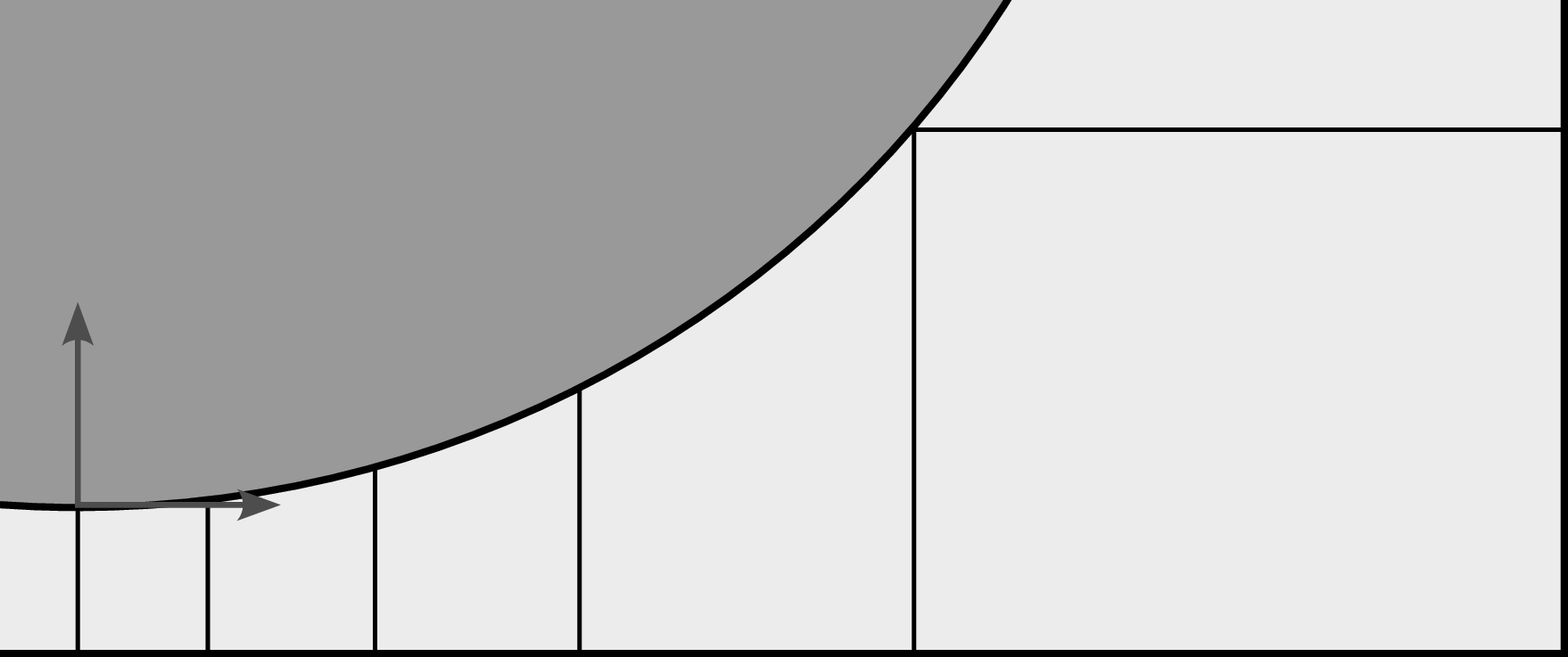}};
		\begin{scope}[shift=(img.south west), 
			x={(img.south east)},y={(img.north west)}] 
			\node at (.05, -0.02) [anchor=north] {$0$};
			\node at (.13, -0.02) [anchor=north] {$0.025R$};
			\node at (.24, -0.02) [anchor=north] {$0.15R$};
			\node at (.37, -0.02) [anchor=north] {$0.3R$};
			\node at (.58, -0.02) [anchor=north] {$0.6R$};
			\node at (1, -0.02) [anchor=north east] {$1.2R$};
			\node at (.165, .29) [text=gray70] {\small $\displaystyle x$};
			\node at (.03, .5) [text=gray70] {\small $\displaystyle z$};
		\end{scope}
	\end{tikzpicture}
	\caption{The fluid mesh is divided into zones to obtain a proper meshing of the contact region. The right-bottom part is shown and the $x$-position of the zones is indicated relative to the roller radius (not to scale).
	}
	\label{fig:zones}
\end{figure}

\autoref{fig:mesh} shows the fluid and structural mesh.
The lubricant domain is meshed with a structured grid of \qty{20 100}{} cells.
The minimum cell size in the $x$-direction is \qty{5e-7}{\meter}. This value was also selected by To{\v{s}}i{\'{c}} et al. \cite{Tosic2019} after performing a mesh verification test, however, with other discretization techniques.
In the height direction, there are $10$ cells, following literature \cite{Tosic2019, Hartinger2007}, and the initial minimal film thickness is \qty{2.5e-7}{\meter}.
In order to obtain proper meshing of the contact region, it is divided into several zones as shown in \autoref{fig:zones}.
Both the central and outer zones are uniformly meshed, and the grading and number of cells in the intermediate zones have been chosen so that the transition is smooth.
With respect to time discretization, a first order, implicit scheme is used, namely backward Euler.
For the discretization of the convective terms in the continuity, momentum and temperature equation, a first order upwind scheme is used.
Central discretization is used for the diffusive terms.

The structural domain is half a circle, meshed with an unstructured grid with \qty{32 786}{} first-order, triangular, plane strain elements as shown in \autoref{fig:mesh}.
As with the fluid mesh, a grading is applied so that the resolution is highest in the contact region.
Time discretization is achieved with the implicit, second order Bossak method.
Both the flow and structural solver use a time step size of $\qty{1e-8}{\second}$.

\subsection{Solver tolerances}
In the flow solver, a (normalized) tolerance value of \qty{1e-7}{} is used for both pressure and velocity and the maximal number of outer corrector iterations is set to \qty{100}{}.
Three pressure corrector iterations are executed in this work and, since the mesh is of sufficient orthogonal quality, no non-orthogonality corrector iterations are applied here, see also \autoref{sec:implementation}.
The velocity equations are relaxed with a factor $0.5$, and no relaxation factor is applied to the pressure or temperature.

The structural solver employs an absolute residual tolerance equal to \qty{1e-7}{\newton}.

\subsection{Boundary conditions}
For the fluid domain with compressible two-phase flow, boundary conditions need to be specified for the velocity, pressure, temperature and density.
For the pressure, the zero gradient condition is chosen for the walls, and the total pressure condition is applied on the other boundaries, which means that $p$ is equal to the prescribed pressure $p_0$ in case of outflow and to $p_0 - \rho_f \normE{\mU}/2$ in case of inflow.
The total pressure used here is atmospheric pressure, i.e., \qty{1e5}{\pascal}.

For the density, the same zero gradient condition is used on the walls, but an inlet-outlet condition is used everywhere else.
This condition specifies a fixed value, equal to $\rho_{l,R}$, for inflow, and a zero gradient condition for outflow.

This is also used for the temperature on the same boundaries with a fixed value equal to $T_{R}$.
On the walls a fixed temperature is specified, since it is assumed that the roller has a constant temperature.
This way no thermal equation has to be solved for the structure, nor are coupling conditions required for temperature and heat flux.
This simplifying assumption is supported by both the time scale of the thermal transport in the roller, which is typically larger than the mechanical and hydrodynamic time scales, and the fact that, in reality, the contact point on the roller continuously changes due to its rotation.
Other works use for example a Carslaw-Jaeger boundary condition \cite{Havaej2023a} to account for the accumulation of heat in the roller.

Lastly, for the velocity, an inlet-outlet-velocity condition is applied on the open boundaries.
This prescribes a zero gradient condition for the normal component and an inlet-outlet condition for the tangential component, for which the prescribed value is zero.
For top and bottom wall, new boundary conditions are implemented that add the velocity due to the motion of the surfaces and the imposed tangential velocity.
For the bottom wall this velocity is prescribed directly, while for the roller it is derived from an applied rotation.

Also the solution of the deformation of the roller requires boundary conditions.
While the top surface is fixed, a distributed fluid load is applied on the bottom surface of the roller.

Note that both the mesh and boundary conditions are completely symmetric, with the exception of the velocities of the plane and roller imposed in the flow solver, which are in this case in the positive $x$-direction.

\subsection{Model parameters}
Further, the parameters used for the solid and lubricant modeling are summarized in \autoref{tab:parameters}.
Their definition and the formulation of the corresponding lubricant models are given in \autoref{sec:lubricant_models}.

The lubricant is squalane (SQL), which is also the prevalent choice in literature, mainly because of the availability of data for this lubricant, e.g., density and viscosity in function of pressure, temperature and -- in case of viscosity -- shear rate.
The Tait, Doolittle and Carreau parameters in \autoref{tab:parameters} have been adopted from Bair \cite{Bair2006}.
Parameters regarding thermal conductivity and heat capacity originate from Björling et al. \cite{Bjoerling2014}.

\begin{table}[htbp]
	\caption{
		Material parameters of the equivalent roller and the lubricant squalane.}
	\label{tab:parameters}
	\centering
	\begin{threeparttable}
		\begin{tabular}{lll}
			\toprule
			Model & Parameter & Value \\
			\midrule
			Solid
			& $\rho_{s}$ & \qty{8750}{\kilogram\per\meter\cubed} \\
			& $E_s$ & \qty{105}{\giga\pascal} \tnote{\textdagger}\\
			& $\nu_{s}$ & \qty{0.3}{} \\
			\midrule
			Lubricant vapor
			& $\mu_v$ & \qty{8.97e-6}{\newton\second\per\meter\squared} \\
			& $\psi_v$ & \qty{5.76e-6}{\second\squared\per\meter\squared} \\
			& $h_{vap}$ & \qty{287e3}{\joule\per\kg} \\
			& $c_{p,v}$ & \qty{1800}{\joule\per\kilogram\per\kelvin} \\
			& $\thk_{v}$ & \qty{0.025}{\joule\per\meter\per\second\per\kelvin} \\
			& $p_{sat}$ & \qty{5000}{\pascal} \\
			\noalign{\smallskip}
			Lubricant liquid
			& $T_{R}$ & \qty{313.15}{\kelvin} \\
			\noalign{\smallskip}
			\qquad Tait 
			& $\rho_{l,R}$ & \qty{794.6}{\kilogram\per\meter\cubed} \\
			& $\beta_K$ & \qty{6.232e-3}{1\per\kelvin} \\
			& $a_\vol$ & \qty{8.36e-4}{1\per\kelvin} \\
			& $K_{00}$ & \qty{8.658e9}{\newton\per\meter\squared} \\
			& $K'_0$ & \qty{11.74}{} \\
			\noalign{\smallskip}
			\qquad Doolittle 
			& $\mu_{R}$ & \qty{0.0157}{\newton\second\per\meter\squared} \\
			& $a_\infty$ & \qty{-7.273e-4}{1\per\kelvin} \\
			& $B$ & \qty{4.71}{} \\
			& $R_0$ & \qty{0.6568}{} \\
			\noalign{\smallskip}
			\qquad Carreau 
			& $\Lambda$ & \qty{0.075}{} \\
			& $\lambda_{R}$ & \qty{2.2622e-9}{\second} \\
			& $n$ & \qty{0.463}{} \\
			\noalign{\smallskip}
			\qquad Thermal conductivity 
			& $K$ & \qty{-0.115}{} \\
			& $C_k$ & \qty{0.074}{\joule\per\meter\per\second\per\kelvin} \\
			& $q$ & \qty{2}{} \\
			& $s$ & \qty{4.5}{} \\
			\noalign{\smallskip}
			\qquad Heat capacity 
			& $C_0$ & \qty{9.4e5}{\joule\per\meter\cubed\per\kelvin} \\
			& $m$ & \qty{6.2e5}{\joule\per\meter\cubed\per\kelvin} \\
			\bottomrule
		\end{tabular}
		\begin{tablenotes}\footnotesize
			\item [\textdagger] This work uses the so-called \emph{equivalent geometry}, representing the contact between two elastic surfaces as the contact between an elastic surface and a rigid plane \cite{Dowson1977}. As a consequence, the modulus of elasticity has to be adapted and is in this case half of the original value for steel.
		\end{tablenotes}
	\end{threeparttable}
\end{table}

\subsection{Laminar flow}
Finally, this section ends with a note on the typical Reynolds number in this problem.
Choosing the initial film thickness as characteristic length $L$, the Reynolds number becomes
\begin{equation}
	\mathrm{Re} = \frac{\rho_{R} u L}{\mu_{R}} \approx 0.0316,
\end{equation}
where the reference values from \autoref{tab:parameters} are used for density and viscosity.
Considering that the viscosity will increase significantly when the pressure increases, e.g., a value of \qty{20}{\pascal\second} is not uncommon, it can be concluded that the flow is laminar, such that no turbulence has to modeled.
Note that this holds for most lubrication problems.

\section{Results}
\label{sec:results}
In this section, the newly developed solver is validated by using it in an EHL simulation as described in \autoref{sec:test_case} and comparing the results with literature.
After the validation, some simulation results with different slip values are presented.

\subsection{Prescribed bottom plane motion}
\label{sec:plane_motion}
First, it is explained how the contact is loaded in a transient simulation by moving the bottom plane.
This is necessary since in the final geometry the bottom plane has moved about six times the initial gap between the roller and the plane, which would lead to a distorted fluid mesh if done at once.
This approach is followed in the following two sections.
The bottom rigid plane will be moved up in a prescribed way, as depicted in \autoref{fig:plate_displacement}.
To obtain a smooth motion, constant velocity intervals are connected by a sigmoid curve, at intermediate velocity jumps, as well as at the start of the simulation.
The resulting displacement is obtained through numerical integration.
As the time step size is $\qty{1e-8}{\second}$, \qty{30 000}{} time steps are required.
\begin{figure}[htbp]
	\begin{minipage}{\textwidth}
		\centering
		\begin{tikzpicture}
			\begin{axis}[
				width=0.9\textwidth,
				height=0.35\linewidth,
				xlabel=Time ($\qty{}{\second}$),
				ylabel=Vertical displacement of bottom plane ($\qty{}{\meter}$),
				]
				\addplot [blue] table[x=time(s),y=displacement(m),col sep=comma]{srr0_output/displacement_plate_input_smaller.csv};
				\coordinate (lb) at (axis cs:1.4733e-4,1.48e-6);
				\coordinate (rt) at (axis cs:1.54e-4,1.52e-6);
				\coordinate (rb) at (axis cs:1.54e-4,1.48e-6);
				\coordinate (spy point) at (axis cs:2.25e-4,0.75e-6);
			\end{axis}
			\node (sp) at (spy point)
			{
				\begin{tikzpicture}[trim axis left,trim axis right]
					\begin{axis}[
						width=4cm, height=4cm,
						xtick=\empty, ytick=\empty, minor tick num=0,
						xmin=1.4733e-4,xmax=1.54e-4,
						ymin=1.48e-6,ymax=1.52e-6,
						]
						\addplot [blue,thick] table[x=time(s),y=displacement(m),col sep=comma]{srr0_output/displacement_plate_input_zoom.csv};
					\end{axis}
				\end{tikzpicture}%
			};
			\draw [-,thin] (sp)--(rb);
			\draw [thin] (lb) rectangle (rt);
		\end{tikzpicture}
		\caption{Prescribed vertical displacement of the bottom rigid plane.
			The zoomed in detail shows that the motion is smooth, which is achieved by using a sigmoid curve when changing the velocity to a new value.
		}
		\label{fig:plate_displacement}
	\end{minipage}
	\begin{minipage}{\textwidth}
		\centering
		\begin{tikzpicture}
			\begin{axis}[
				width=0.9\textwidth,
				height=0.35\linewidth,
				xlabel=Time ($\qty{}{\second}$),
				ylabel=Load ($\qty{}{\newton\per\meter}$),
				cycle list={{blue},	{red, dashed},{brown!60!black, dashdotted},},
				]
				\addplot table[x=time(s),y=load(N/m),col sep=comma]{srr0_output/load_contact_plate_output_smaller.csv};
				\draw [color=gray,dashed] ({axis cs:1.5e-4,\pgfkeysvalueof{/pgfplots/ymin}}) -- ({axis cs:1.5e-4,\pgfkeysvalueof{/pgfplots/ymax}});
			\end{axis}
		\end{tikzpicture}
		\caption{Contact load on bottom plane in function of time, obtained by integrating the relative pressure for $\abs{x}<\qty{5e-4}{\meter}$.
			The gray dashed line indicates the point where the plane stops moving.
		}
		\label{fig:plate_load}
	\end{minipage}
\end{figure}

At the start of the simulation, the lubricant pressure in the gap between the two solids increases, as a consequence of entrainment of lubricant by the motion of the surfaces.
Consequently, the thickness of the film will become larger as well.
This effect would even occur for a fixed plane.
Here, however, the bottom plane is gradually moved much higher than the original roller position (the initial gap is \qty{0.25}{\micro\meter}), such that the pressure keeps increasing and a higher load is obtained, as depicted in \autoref{fig:plate_load}. 
Comparing this figure with \autoref{fig:plate_displacement} reveals that the increase in load is approximately proportional to the upward displacement of the plane.

\begin{figure}[htbp]
	\centering
	\begin{tikzpicture}
		\begin{axis}[
			width=0.9\textwidth,
			height=0.65\linewidth,
			xlabel=Time ($\qty{}{\second}$),
			ylabel=Film thickness ($\qty{}{\meter}$),
			cycle list={{blue},	{red, dashed},{brown!60!black, dashdotted},},
			]
			\addplot table[x=time(s),y=hc(m),col sep=comma]{srr0_output/hc_hmin_roller_input_smaller.csv};
			\addlegendentry{$h_c$}
			\addplot table[x=time(s),y=hmin(m),col sep=comma]{srr0_output/hc_hmin_roller_input_smaller.csv};
			\addlegendentry{$h_{min}$}
			\draw [color=gray,dashed] ({axis cs:1.5e-4,\pgfkeysvalueof{/pgfplots/ymin}}) -- ({axis cs:1.5e-4,\pgfkeysvalueof{/pgfplots/ymax}});
		\end{axis}
	\end{tikzpicture}
	\caption{Evolution of central and minimal film thickness in time due to the pressure build-up at the start of the simulation and the upward motion of the bottom plane.
		The gray dashed line indicates the point where the plane stops moving.
	}
	\label{fig:hc_hmin}
\end{figure}
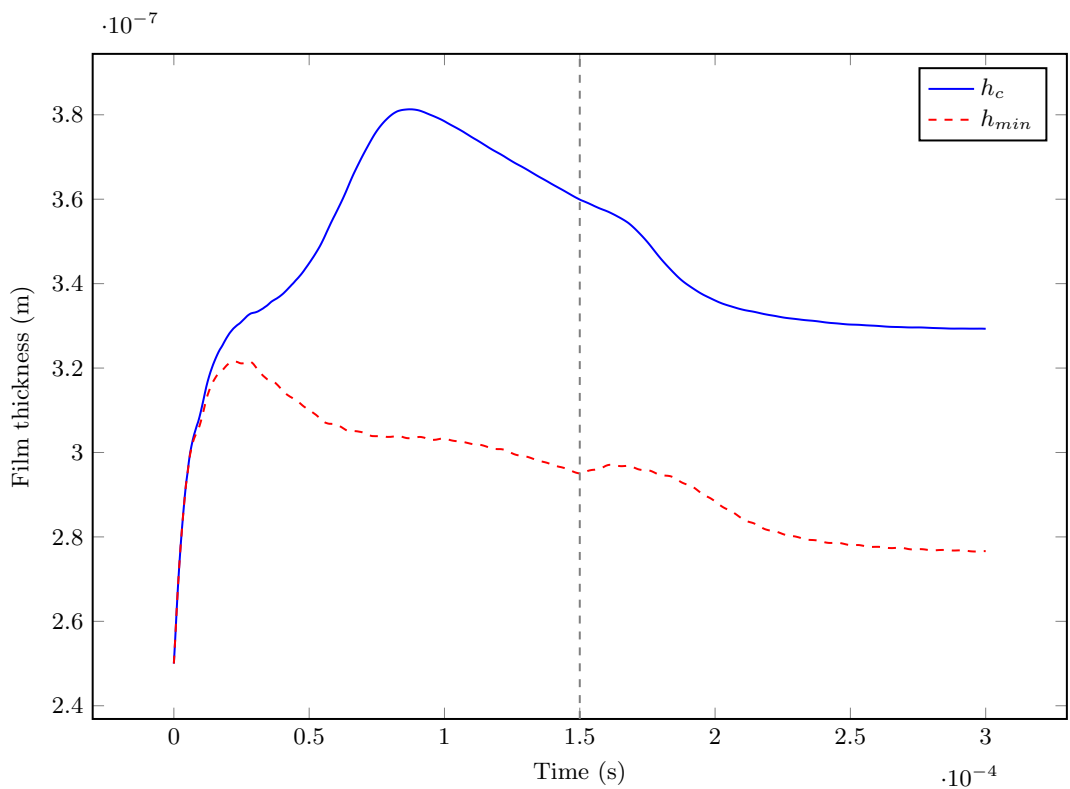

Further, \autoref{fig:hc_hmin} shows the time evolution of the central and minimal thickness, $h_c$ and $h_{min}$, which are the film thickness at the center of the contact ($x=\qty{0}{m}$) and the minimal value of the film thickness, respectively.
Initially, these two often used quantities coincide since the roller has not yet been deformed.
Moreover they grow, indicating that the deformation caused by the pressure build-up outweighs the upward motion of the plane.
At two different times, both will begin to decrease, so then the upward motion of the plate is greater than the deformation of the roller due to the pressure increase.
The continued decrease of the film thickness after the plane has been stopped, shows the presence of inertia effects: the decrease in film thickness is no longer due to the upward moving plane, but due to reducing deformation of the roller.
A steady state is only reached after holding the plane in its final position for approximately \qty{1.5e-4}{\second} more.

\subsection{Validation}
In this section the solution framework is validated.
As explained in the introduction, experimental data is difficult to obtain, and therefore the simulation results are compared against other numerical solutions in literature.
For these cases, the nodes on the bottom plane where allowed to move in the $x$-direction, i.e., a slip boundary condition for the mesh motion.
Furthermore, the calculation of the mesh motion was always performed on the original mesh to avoid mesh distortion problems.

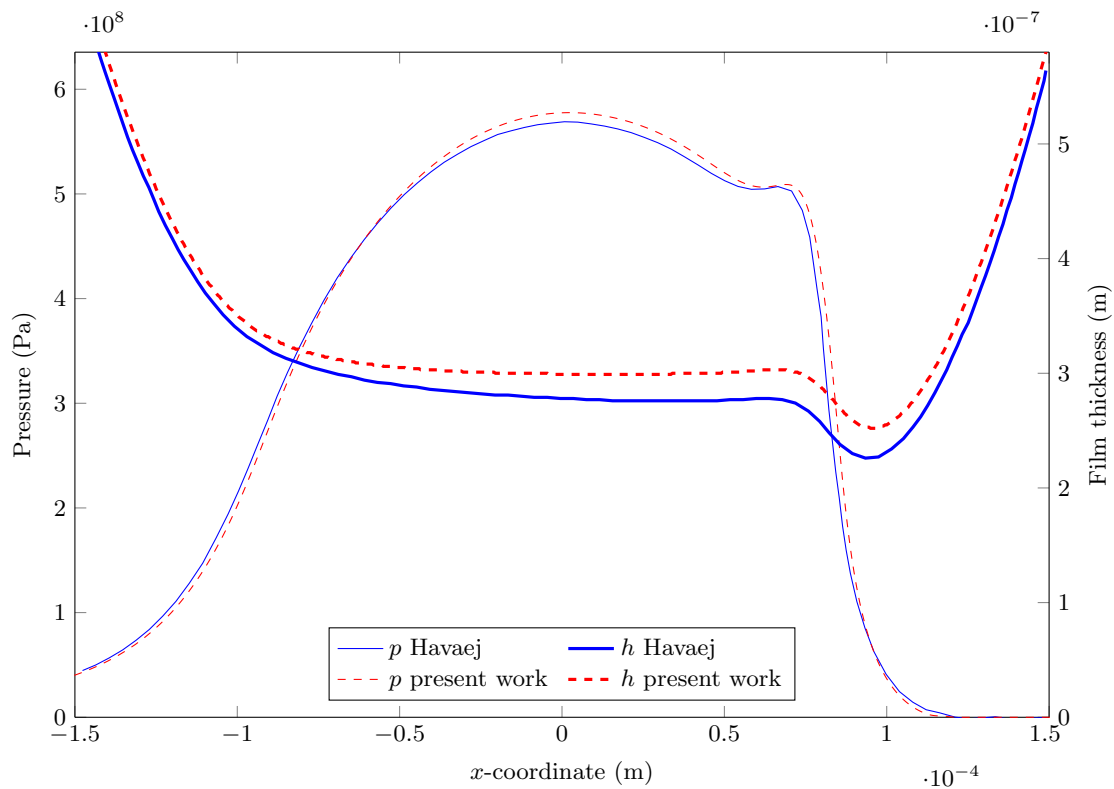
\begin{figure}[htbp]
	\centering
	\begin{tikzpicture}
		\pgfplotsset{set layers}
		\begin{axis}[
			width=0.9\linewidth,
			height=0.65\linewidth,
			xmin=-1.5e-4,xmax=1.5e-4,
			xtick={-1.5e-4, -1e-4, -0.5e-4, 0, 0.5e-4, 1e-4},
			extra x tick style={tick label style={xshift=-1.5pt}},
			extra x ticks={1.5e-4},
			extra x tick labels={$1.5$},
			axis y line*=left,
			ymin=0,
			xlabel=$x$-coordinate (\qty{}{\meter}),
			ylabel=Pressure (\qty{}{\pascal}),
			cycle list={{blue},	{red, dashed},{brown!60!black, dashdotted},},
			]
			\addplot table[x=havaej_xp(m),y=havaej_p(Pa),col sep=comma]{validation/validation.csv};
			\addplot table[x=x_coord(m),y=pressure(Pa),col sep=comma]{havaej100_output/pressure_plate_output.csv};
		\end{axis}
		\begin{axis}[
			width=0.9\textwidth,
			height=0.65\linewidth,
			xmin=-1.5e-4,xmax=1.5e-4,
			axis y line*=right,
			axis x line=none,
			ymin=0,ymax=5.8e-7,
			ylabel=Film thickness (\qty{}{\meter}),
			cycle list={
				{blue},
				{blue, very thick},
				{red, dashed},	
				{red, dashed, very thick},	
				{brown!60!black, dashdotted},
				{brown!60!black, dashdotted, very thick},
			},
			legend columns=2, 
			legend style={
				at={(0.5,0.025)},
				anchor=south,
				legend cell align=left,
				/tikz/column 2/.style={column sep=5pt}
			},
			]
			\addplot coordinates {(0,0)};
			\addlegendentry{$p$ Havaej}
			\addplot table[x=havaej_xh(m),y=havaej_h(m),col sep=comma]{validation/validation.csv};
			\addlegendentry{$h$ Havaej}
			\addplot coordinates {(0,0)};
			\addlegendentry{$p$ present work}
			\addplot table[x=x_coord(m),y=h(m),col sep=comma]{havaej100_output/h_smaller.csv};
			\addlegendentry{$h$ present work}
		\end{axis}
	\end{tikzpicture}
	\caption{Validation by comparison of pressure and film thickness profiles for a thermal sliding contact (SRR $1$) with a load of \qty{100}{\kilo\newton\per\meter} simulated by Havaej et al. \cite{Havaej2023a}.
	}
	\label{fig:validation100}
\end{figure}

First, a thermal simulation is considered with SSR $1$ and a load of \qty{100}{\kilo\newton\per\meter}.
This case was simulated by Havaej et al. \cite{Havaej2023a} using almost identical lubricant models and parameters.
However, they also included an energy equation for the solid and conjugate heat transfer on the fluid-structure interface, as well as a Carslaw-Jaeger boundary condition \cite{Kim2001a, Kim2001b} on the rigid plane to represent heat transfer to a moving surface.
These differences are assumed to explain the limited discrepancies observed in \autoref{fig:validation100}.

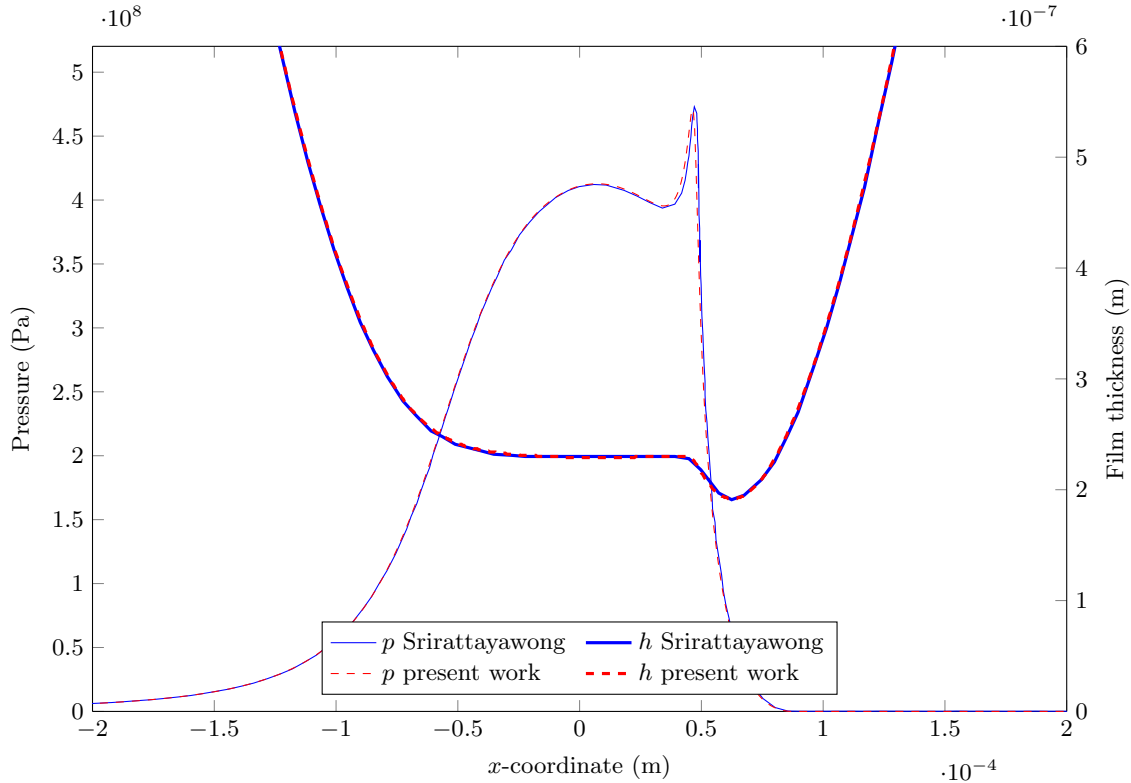
\begin{figure}[htbp]
	\centering
	\begin{tikzpicture}
		\pgfplotsset{set layers}
		\begin{axis}[
			width=0.9\linewidth,
			height=0.65\linewidth,
			xmin=-2e-4,xmax=2e-4,
			axis y line*=left,
			ymin=0,
			xlabel=$x$-coordinate (\qty{}{\meter}),
			ylabel=Pressure (\qty{}{\pascal}),
			cycle list={{blue},	{red, dashed},{brown!60!black, dashdotted},},
			]
			\addplot table[x=srirattayawong_xp(m),y=srirattayawong_p(Pa),col sep=comma]{validation/validation.csv};
			\addplot table[x=x_coord(m),y=pressure(Pa),col sep=comma]{tosic50_output/pressure_plate_output.csv};
		\end{axis}
		\begin{axis}[
			width=0.9\textwidth,
			height=0.65\linewidth,
			xmin=-2e-4,xmax=2e-4,
			ymin=0,ymax=6e-7,
			axis y line*=right,
			axis x line=none,
			ylabel=Film thickness (\qty{}{\meter}),
			cycle list={
				{blue},
				{blue, very thick},
				{red, dashed},	
				{red, dashed, very thick},	
				{brown!60!black, dashdotted},
				{brown!60!black, dashdotted, very thick},
			},
			legend columns=2, 
			legend style={
				at={(0.5,0.025)},
				anchor=south,
				legend cell align=left,
				/tikz/column 2/.style={column sep=5pt}
			},
			]
			\addplot coordinates {(0,0)};
			\addlegendentry{$p$ Srirattayawong}
			\addplot table[x=srirattayawong_xh(m),y=srirattayawong_h(m),col sep=comma]{validation/validation.csv};
			\addlegendentry{$h$ Srirattayawong}
			\addplot coordinates {(0,0)};
			\addlegendentry{$p$ present work}
			\addplot table[x=x_coord(m),y=h(m),col sep=comma]{tosic50_output/h_smaller.csv};
			\addlegendentry{$h$ present work}
		\end{axis}
	\end{tikzpicture}
	\caption{Validation by comparison of pressure and film thickness profiles for isothermal pure rolling (SRR $0$) with a load of \qty{50}{\kilo\newton\per\meter} independently simulated by To{\v{s}}i{\'{c}} et al. \cite{Tosic2019} and Srirattayawong \cite{Srirattayawong2014} (data of To{\v{s}}i{\'{c}} et al. not shown as it is similar).
	}
	\label{fig:validation50}
\end{figure}

Second, an isothermal pure rolling (SRR $0$) calculation is considered with a load of \qty{50}{\kilo\newton\per\meter}, independently simulated by To{\v{s}}i{\'{c}} et al. \cite{Tosic2019} and Srirattayawong \cite{Srirattayawong2014}.
Both used the Dowson compressibility and Roelands piezoviscosity models with oil as lubricant.
In order to obtain a fair comparison, the same lubricant models were implemented and identical parameters were used.
This is straightforward thanks to the modularity of the \OF framework.
Details can be found in their respective works.
\autoref{fig:validation50} shows the very closely matching film thickness and pressure profiles.

\subsection{Results with different slip ratios}
In this section, a selection of simulation results are presented for the plane motion detailed in \autoref{sec:plane_motion}.
Three different SRR values (\autoref{equ:slip}) are considered: $0$, $1$ and $2$.
In each case, the entrainment speed is \qty{2.5}{\meter\per\second}.

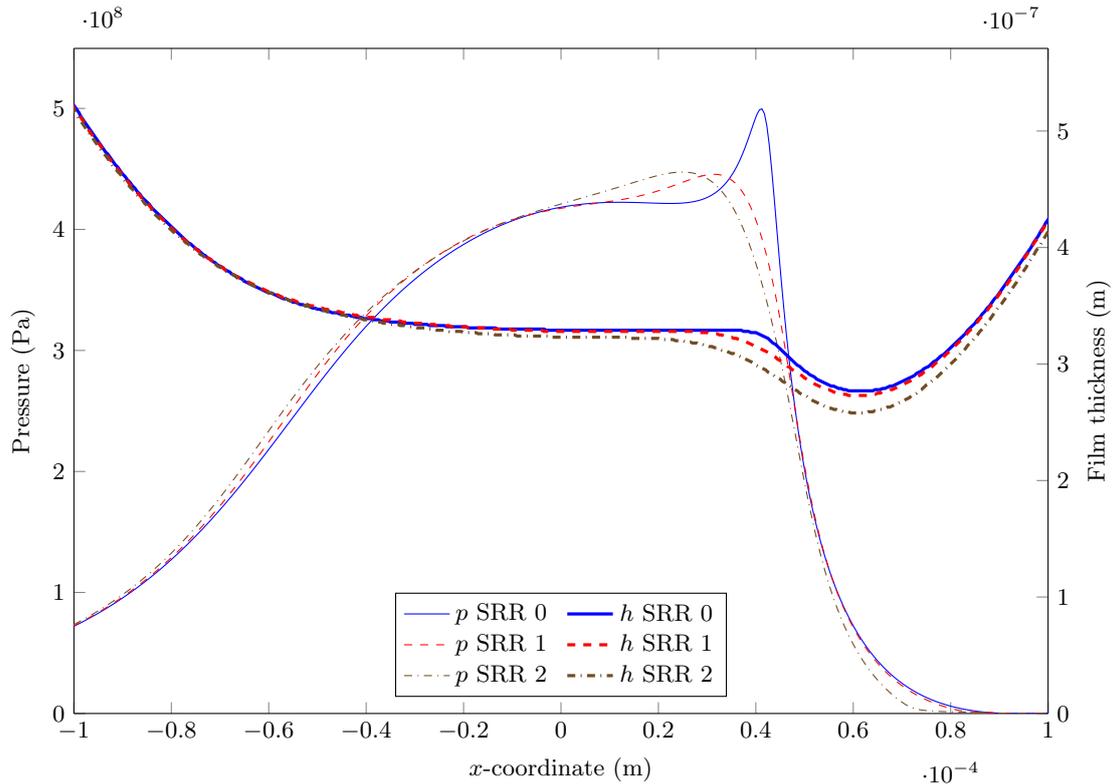
\begin{figure}[htbp]
	\centering
	\begin{tikzpicture}
		\pgfplotsset{set layers}
		\begin{axis}[
			width=0.9\linewidth,
			height=0.65\linewidth,
			xmin=-1e-4,xmax=1e-4,
			axis y line*=left,
			ymin=0,
			xlabel=$x$-coordinate (\qty{}{\meter}),
			ylabel=Pressure (\qty{}{\pascal}),
			cycle list={{blue},	{red, dashed},{brown!60!black, dashdotted},},
			]
			\addplot table[x=x_coord(m),y=pressure(Pa),col sep=comma]{srr0_output/pressure_plate_output.csv};
			\addplot table[x=x_coord(m),y=pressure(Pa),col sep=comma]{srr1_output/pressure_plate_output.csv};
			\addplot table[x=x_coord(m),y=pressure(Pa),col sep=comma]{srr2_output/pressure_plate_output.csv};
		\end{axis}
		\begin{axis}[
			width=0.9\textwidth,
			height=0.65\linewidth,
			xmin=-1e-4,xmax=1e-4,
			axis y line*=right,
			axis x line=none,
			ymin=0,
			ylabel=Film thickness (\qty{}{\meter}),
			cycle list={
				{blue},
				{blue, very thick},
				{red, dashed},	
				{red, dashed, very thick},	
				{brown!60!black, dashdotted},
				{brown!60!black, dashdotted, very thick},
			},
			legend columns=2, 
			legend style={
				at={(0.5,0.025)},
				anchor=south,
				legend cell align=left,
				/tikz/column 2/.style={column sep=5pt}
			},
			]
			\addplot coordinates {(0,0)};
			\addlegendentry{$p$ SRR $0$}
			\addplot table[x=x_coord(m),y=h(m),col sep=comma]{srr0_output/h_smaller.csv};
			\addlegendentry{$h$ SRR $0$}
			\addplot coordinates {(0,0)};
			\addlegendentry{$p$ SRR $1$}
			\addplot table[x=x_coord(m),y=h(m),col sep=comma]{srr1_output/h_smaller.csv};
			\addlegendentry{$h$ SRR $1$}
			\addplot coordinates {(0,0)};
			\addlegendentry{$p$ SRR $2$}
			\addplot table[x=x_coord(m),y=h(m),col sep=comma]{srr2_output/h_smaller.csv};
			\addlegendentry{$h$ SRR $2$}
		\end{axis}
	\end{tikzpicture}
	\caption{Pressure and film thickness profiles for different SRR values.}
	\label{fig:srr_p_h}
\end{figure}

\autoref{fig:srr_p_h} shows the different pressure profiles and film thicknesses.
Characteristic for pure rolling is the pressure peak, also known as the Petrusevich peak, just before the constriction, which is much less present in case of slip.
This peak is no numerical error, but a typical phenomenon occurring in EHL contacts that should be accurately captured.
Overall, the change in film thickness and pressure is limited.
In all cases, pressure does not vary significantly across the film, neither do the vapor fraction or density.
This is especially the case for pure rolling ($\text{SRR}=0$).
	
\autoref{tab:srr} summarizes the differences in load and film thicknesses, as well as in friction coefficient $C_f$, which is defined by
\begin{equation}
	C_f = - \frac{\int_S -\left(\shearStress \cdot \vec{n}\right)_x \diff S}{\int_S p_r \diff S} =  \frac{\int_S \left(\shearStress \cdot \vec{n}\right)_x \diff S}{w}.
\end{equation}
The subscript $x$ refers to the component in the $x$-direction, and the subscript $r$ to the relative pressure.
Further, $w$ is the normal pressure load, $S$ is the bottom plane surface, $\shearStress$ the viscous stress tensor, and $\vec{n}$ is a unit normal on $S$ pointing away from the fluid, i.e., the unit vector in the negative $y$-direction.\footnote{While it suffices to integrate the relative pressure over a smaller region, e.g., $\abs{x}<\qty{5e-4}{\meter}$, the traction has to be integrated over the entire surface to obtain an accurate value.}
The minus sign in the numerator is required to obtain the force on the wall, while the additional minus sign is added to obtain a positive value for the coefficient.
As expected, the friction is much higher in case of slip, which is a consequence of the higher shear stress.

\begin{table}[htbp]
	\caption{Simulation results for different values of SSR, where $u_1$ and $u_2$ are the roller and plane surface velocities, respectively.}
	\label{tab:srr}
	\centering
	\begin{tabular}{*{7}{l}}
		\toprule
		SRR & $u_1$ ($\qty{}{\meter\per\second}$) & $u_2$ ($\qty{}{\meter\per\second}$) & $w$ ($\qty{}{\newton\per\meter}$) & $h_c$ ($\qty{}{\micro\meter}$) & $h_{min}$ ($\qty{}{\micro\meter}$) & $C_f$ \\
		\midrule
		$0$ & $2.5$  & $2.5$  & \qty{52 086}{} & $0.329$ & $0.277$ & $0.00115$ \\
		$1$ & $1.25$ & $3.75$ & \qty{52 027}{} & $0.328$ & $0.273$ & $0.02693$ \\
		$2$ & $0$    & $5$    & \qty{51 771}{} & $0.323$ & $0.258$ & $0.03331$ \\
		\bottomrule
	\end{tabular}
\end{table}

\begin{figure}[p]
	\centering
	\def \srr {./srr0_output}
	\def \ybound {2.25651e-4}
	\captionsetup[subfigure]{skip=-1pt}
	\begin{subfigure}{0.95\textwidth}
		\begin{tikzpicture}[inner sep=0pt]
			\node (img) at (0,0) {\includegraphics[width=0.928\textwidth,trim={194 640 194 633},clip]{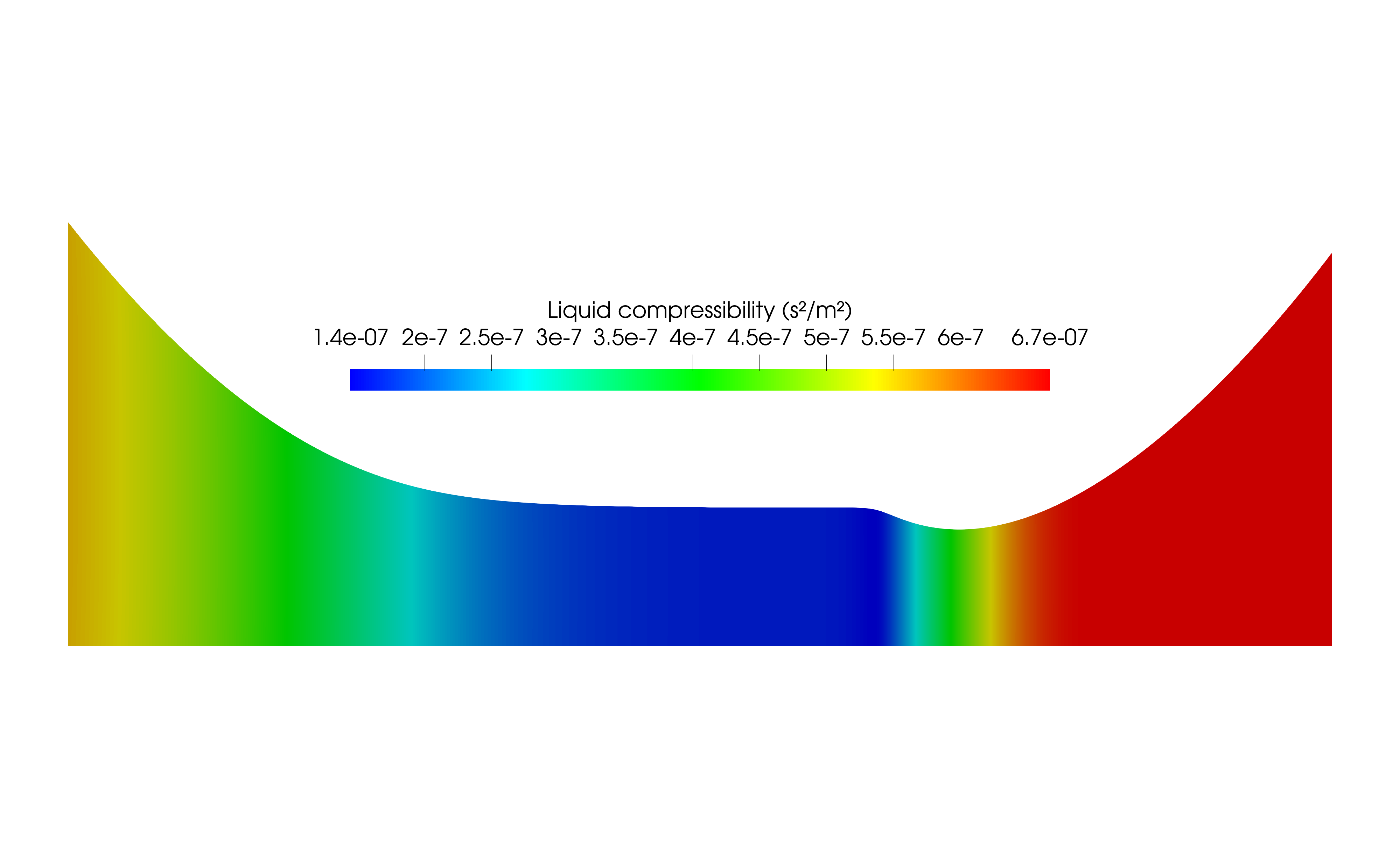}};
			\begin{scope}[shift=(img.south west), x={(img.south east)},y={(img.north west)}]
				\begin{axis}[
					width=(0.95*1.34*0.941*13.1cm),height=(0.95*1.34*0.953*5.44cm),
					xmin=-150, xmax=150, ymin=0, ymax=(\ybound-1.25e-4)/100*1e6,
					axis line style={thin},
					xtick={-150, -100, -50, 0, 50, 100},
					xticklabel style={anchor=north, yshift=-0.5ex, font=\scriptsize},
					extra x tick style={tick label style={xshift=-1.5pt}, font=\scriptsize},
					extra x ticks={150},
					ytick={0.25, 0.5, 0.75, 1},
					yticklabel style={anchor=east, xshift=-0.5ex, font=\scriptsize},
					extra y tick style={tick label style={yshift=0.5ex}, font=\scriptsize},
					extra y ticks={0},
					minor xtick={-125, -75, -25, 25, 75, 125},
					]
				\end{axis}
			\end{scope}
		\end{tikzpicture}
		\caption{Liquid compressibility.}
	\end{subfigure}
	
	\vspace{30pt}
	
	\begin{subfigure}{0.95\textwidth}
		\begin{tikzpicture}[inner sep=0pt]
			\node (img) at (0,0) {\includegraphics[width=0.928\textwidth,trim={194 640 194 633},clip]{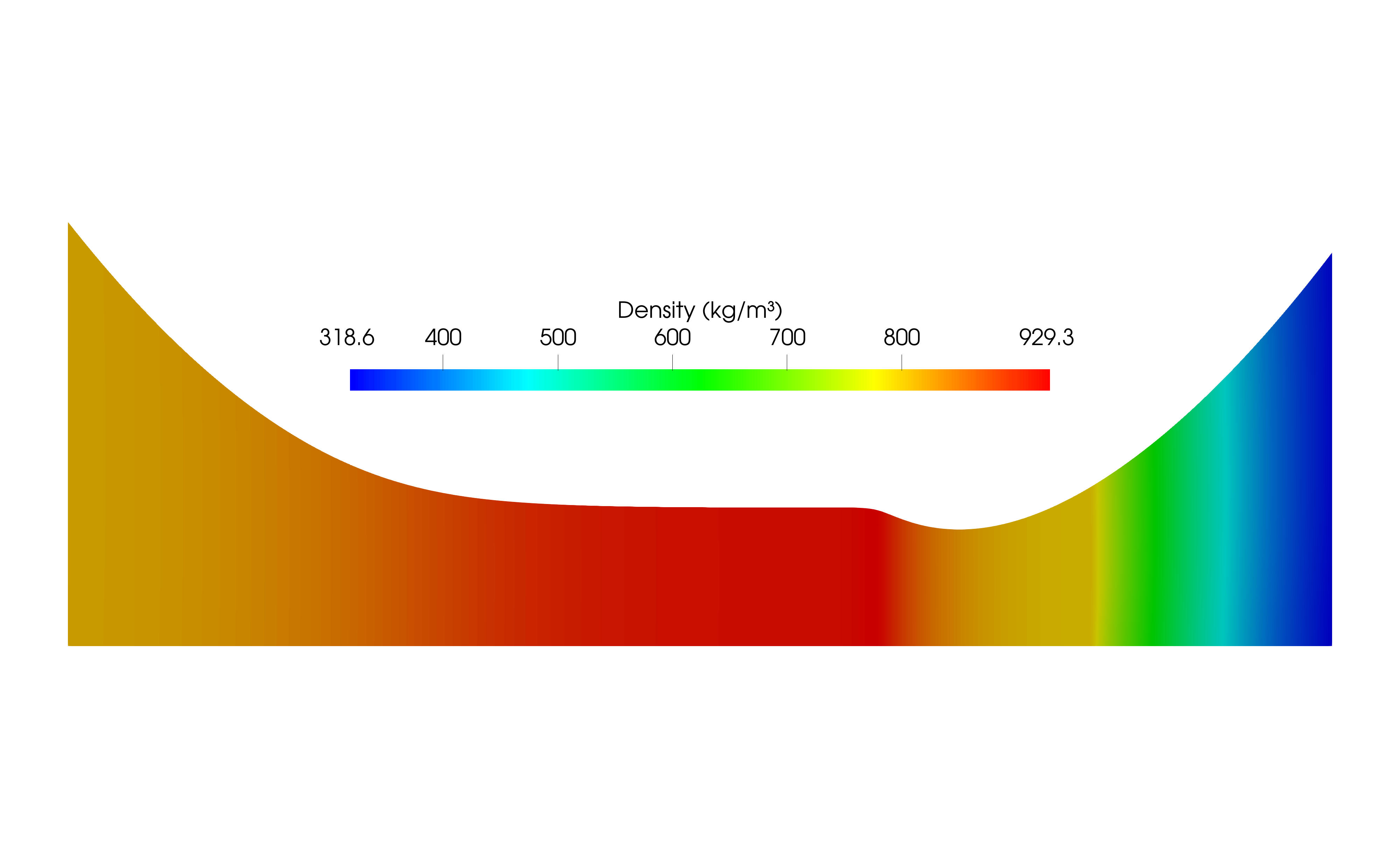}};
			\begin{scope}[shift=(img.south west), x={(img.south east)},y={(img.north west)}]
				\begin{axis}[
					width=(0.95*1.34*0.941*13.1cm),height=(0.95*1.34*0.953*5.44cm),
					xmin=-150, xmax=150, ymin=0, ymax=(\ybound-1.25e-4)/100*1e6,
					axis line style={thin},
					xtick={-150, -100, -50, 0, 50, 100},
					xticklabel style={anchor=north, yshift=-0.5ex, font=\scriptsize},
					extra x tick style={tick label style={xshift=-1.5pt}, font=\scriptsize},
					extra x ticks={150},
					ytick={0.25, 0.5, 0.75, 1},
					yticklabel style={anchor=east, xshift=-0.5ex, font=\scriptsize},
					extra y tick style={tick label style={yshift=0.5ex}, font=\scriptsize},
					extra y ticks={0},
					minor xtick={-125, -75, -25, 25, 75, 125},
					]
				\end{axis}
			\end{scope}
		\end{tikzpicture}
		\caption{Density.}
	\end{subfigure}
	
	\vspace{30pt}
	
	\begin{subfigure}{0.95\textwidth}
		\centering
		\begin{tikzpicture}[inner sep=0pt]
			\node (img) at (0,0) {\scalebox{1}[0.5]{\includegraphics[width=0.928\textwidth,trim={1095 121 1095 121},clip]{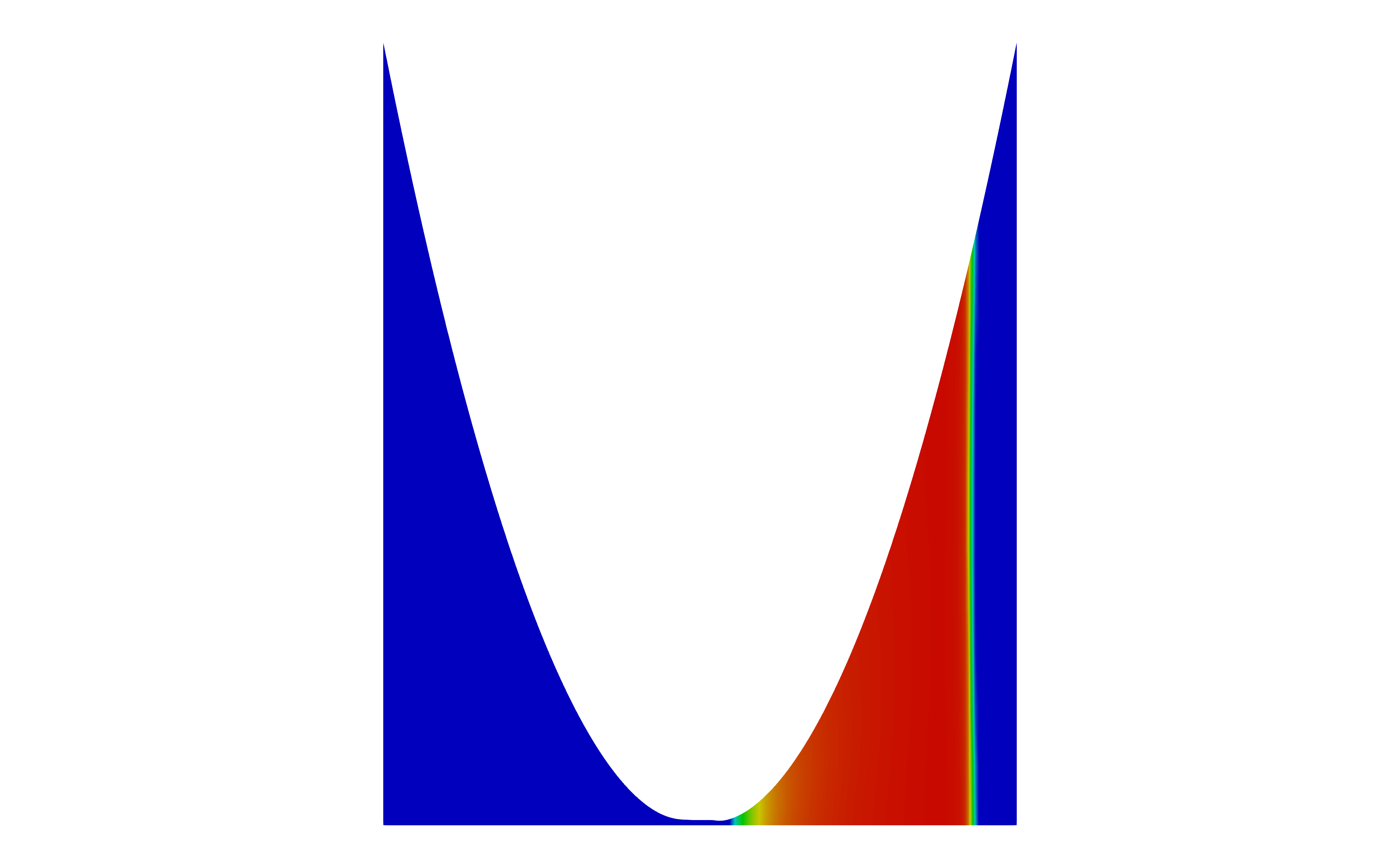}}}; 
			\node (leg) at (0,2.5) {\includegraphics[width=0.7\textwidth,trim={145.5 0 145.5 0},clip]{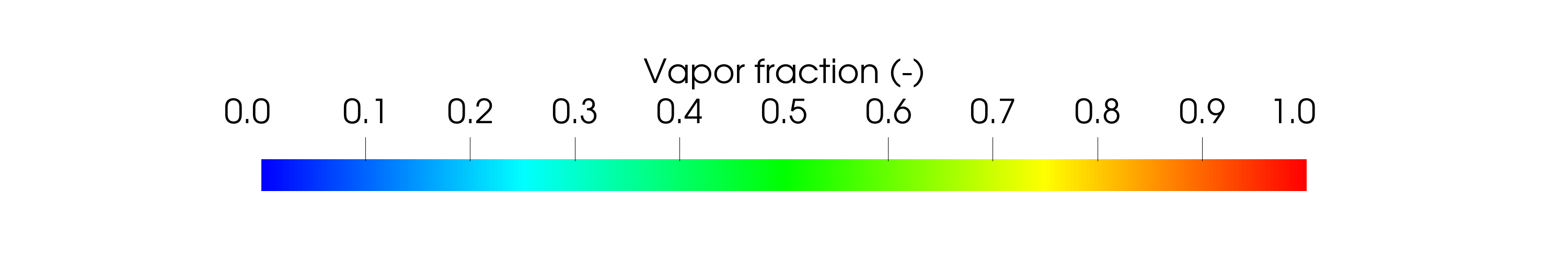}};
			\begin{scope}[shift=(img.south west), x={(img.south east)},y={(img.north west)}]
				\begin{axis}[
					width=(0.95*1.34*0.941*13.1cm),height=(1.035*1.34*0.5*14.85cm),
					xmin=-1000, xmax=1000, ymin=0, ymax=(2.53378e-3 - 6.24999e-5)/50*1e6,
					axis line style={thin},
					xtick={-1000, -500, 0, 500},
					xticklabel style={anchor=north, yshift=-0.5ex, font=\scriptsize},
					extra x tick style={tick label style={xshift=-1.5pt}, font=\scriptsize},
					extra x ticks={1000},
					ytick={10, 20, 30, 40, 50},
					yticklabel style={anchor=east, xshift=-0.5ex, font=\scriptsize},
					extra y tick style={tick label style={yshift=0.5ex}, font=\scriptsize},
					extra y ticks={0},
					minor xtick={-750, -250, 250, 750},
					minor ytick={5, 15, 25, 35, 45},
					]
				\end{axis}
			\end{scope}
		\end{tikzpicture}
	\caption{Vapor fraction.}
	\end{subfigure}
	\caption{Plots of liquid compressibility, density and vapor fraction for SRR $0$.
	The axis labels are given in \qty{}{\micro\meter}.  Note that the first two figures are scaled with a factor $100$ in the $y$-direction, while the last is scaled with a factor $25$.}
	\label{fig:srr0_extra}
\end{figure}

\begin{figure}[p]
	\centering
	\def \srr {./srr0_output}
	\def \ybound {2.25651e-4}
	\captionsetup[subfigure]{skip=-1pt}
	\begin{subfigure}{0.95\textwidth}
		\begin{tikzpicture}[inner sep=0pt]
			\node (img) at (0,0) {\includegraphics[width=0.928\textwidth,trim={194 640 194 633},clip]{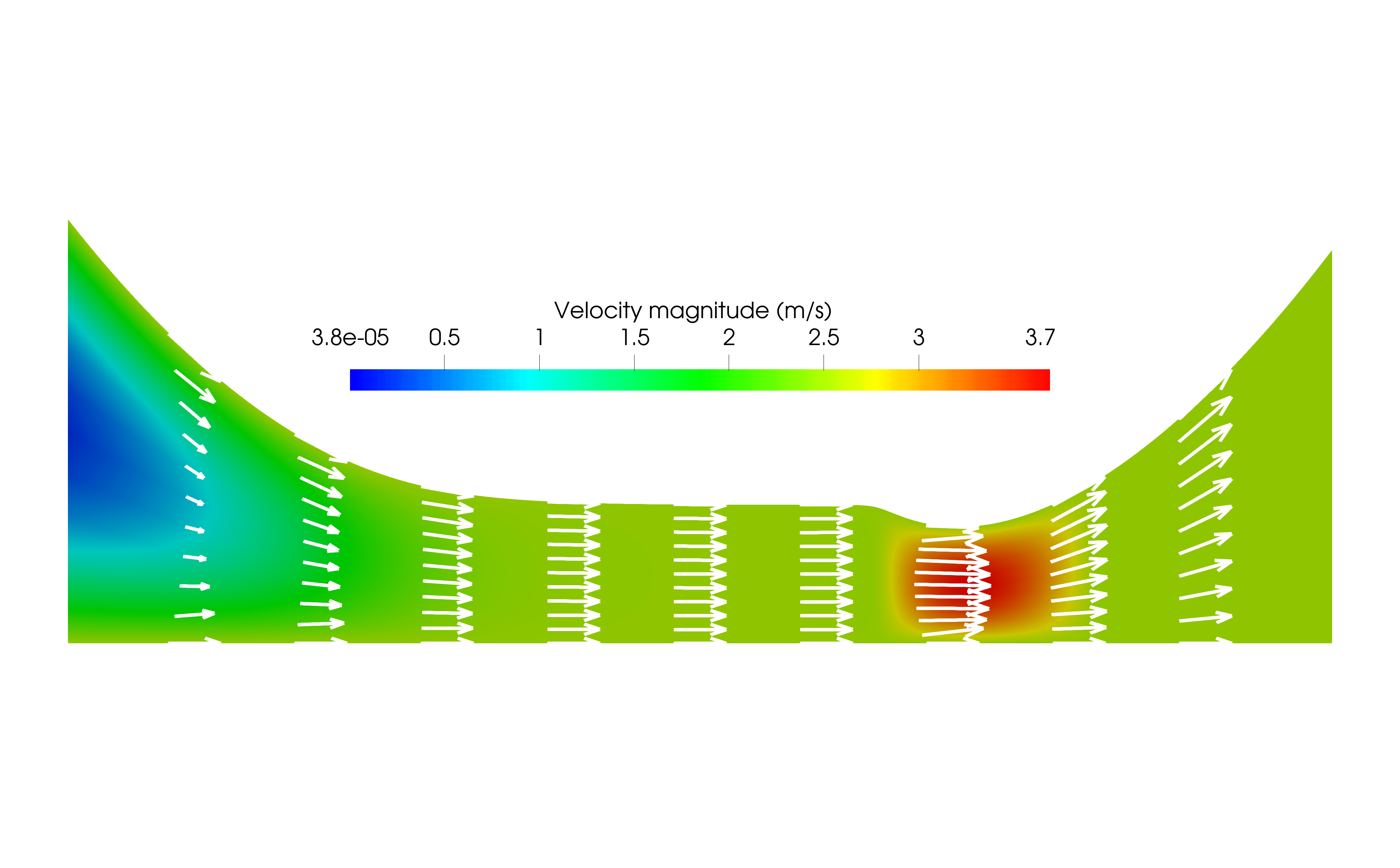}};
			\begin{scope}[shift=(img.south west), x={(img.south east)},y={(img.north west)}]
				\begin{axis}[
					width=(0.95*1.34*0.941*13.1cm),height=(0.95*1.34*0.953*5.44cm),
					xmin=-150, xmax=150, ymin=0, ymax=(\ybound-1.25e-4)/100*1e6,
					axis line style={thin},
					xtick={-150, -100, -50, 0, 50, 100},
					xticklabel style={anchor=north, yshift=-0.5ex, font=\scriptsize},
					extra x tick style={tick label style={xshift=-1.5pt}, font=\scriptsize},
					extra x ticks={150},
					ytick={0.25, 0.5, 0.75, 1},
					yticklabel style={anchor=east, xshift=-0.5ex, font=\scriptsize},
					extra y tick style={tick label style={yshift=0.5ex}, font=\scriptsize},
					extra y ticks={0},
					minor xtick={-125, -75, -25, 25, 75, 125},
					]
				\end{axis}
			\end{scope}
		\end{tikzpicture}
		\caption{Velocity.}
	\end{subfigure}
	
	\vspace{0pt}
	
	\begin{subfigure}{0.95\textwidth}
		\begin{tikzpicture}[inner sep=0pt]
			\node (img) at (0,0) {\includegraphics[width=0.928\textwidth,trim={194 633 194 633},clip]{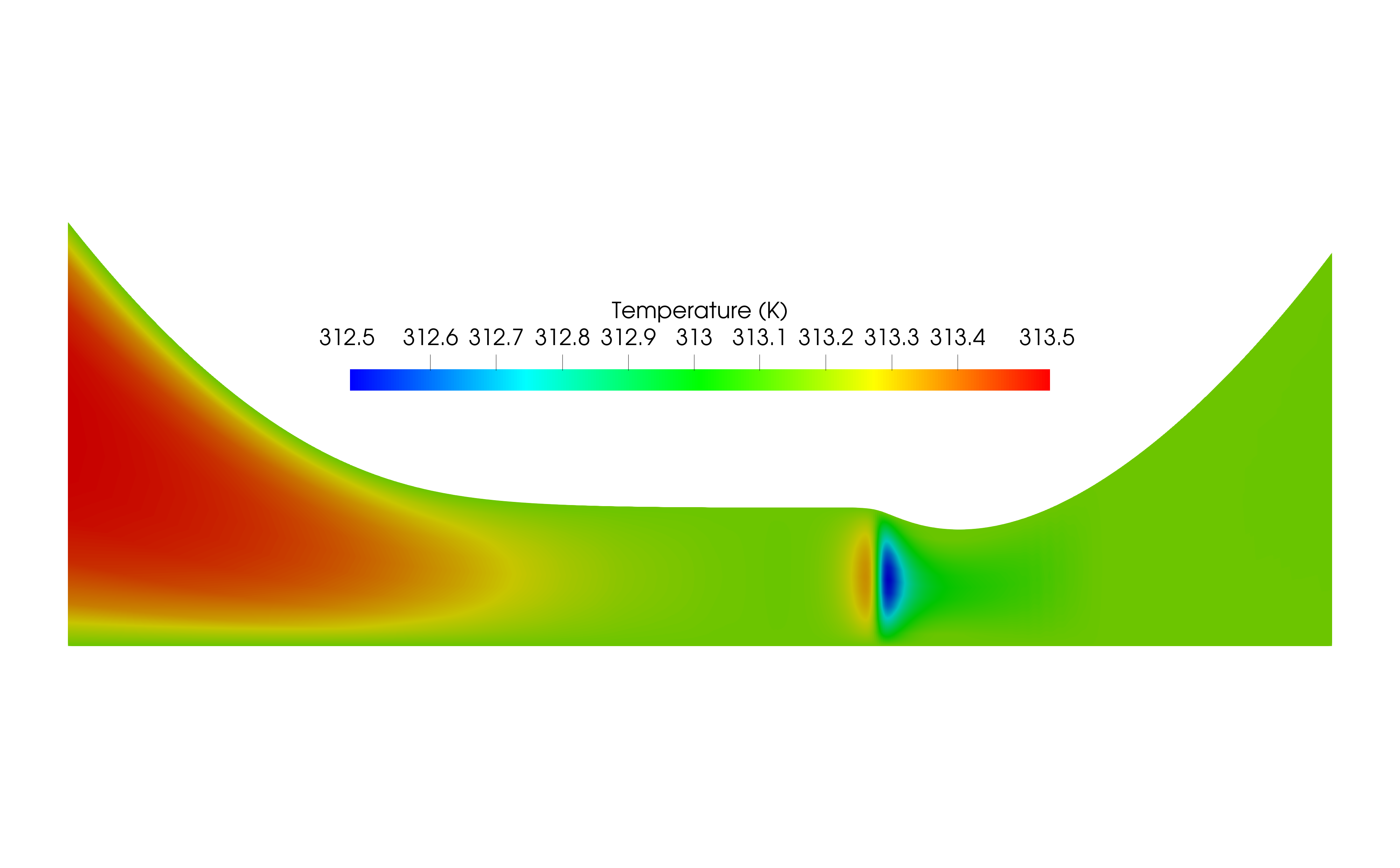}};
			\begin{scope}[shift=(img.south west), x={(img.south east)},y={(img.north west)}]
				\begin{axis}[
					width=(0.95*1.34*0.941*13.1cm),height=(0.95*1.34*0.953*5.44cm),
					xmin=-150, xmax=150, ymin=0, ymax=(\ybound-1.25e-4)/100*1e6,
					axis line style={thin},
					xtick={-150, -100, -50, 0, 50, 100},
					xticklabel style={anchor=north, yshift=-0.5ex, font=\scriptsize},
					extra x tick style={tick label style={xshift=-1.5pt}, font=\scriptsize},
					extra x ticks={150},
					ytick={0.25, 0.5, 0.75, 1},
					yticklabel style={anchor=east, xshift=-0.5ex, font=\scriptsize},
					extra y tick style={tick label style={yshift=0.5ex}, font=\scriptsize},
					extra y ticks={0},
					minor xtick={-125, -75, -25, 25, 75, 125},
					]
				\end{axis}
			\end{scope}
		\end{tikzpicture}
		\caption{Temperature.}
	\end{subfigure}
	
	\vspace{0pt}
	
	\begin{subfigure}{0.95\textwidth}
		\begin{tikzpicture}[inner sep=0pt]
			\node (img) at (0,0) {\includegraphics[width=0.928\textwidth,trim={194 633 194 633},clip]{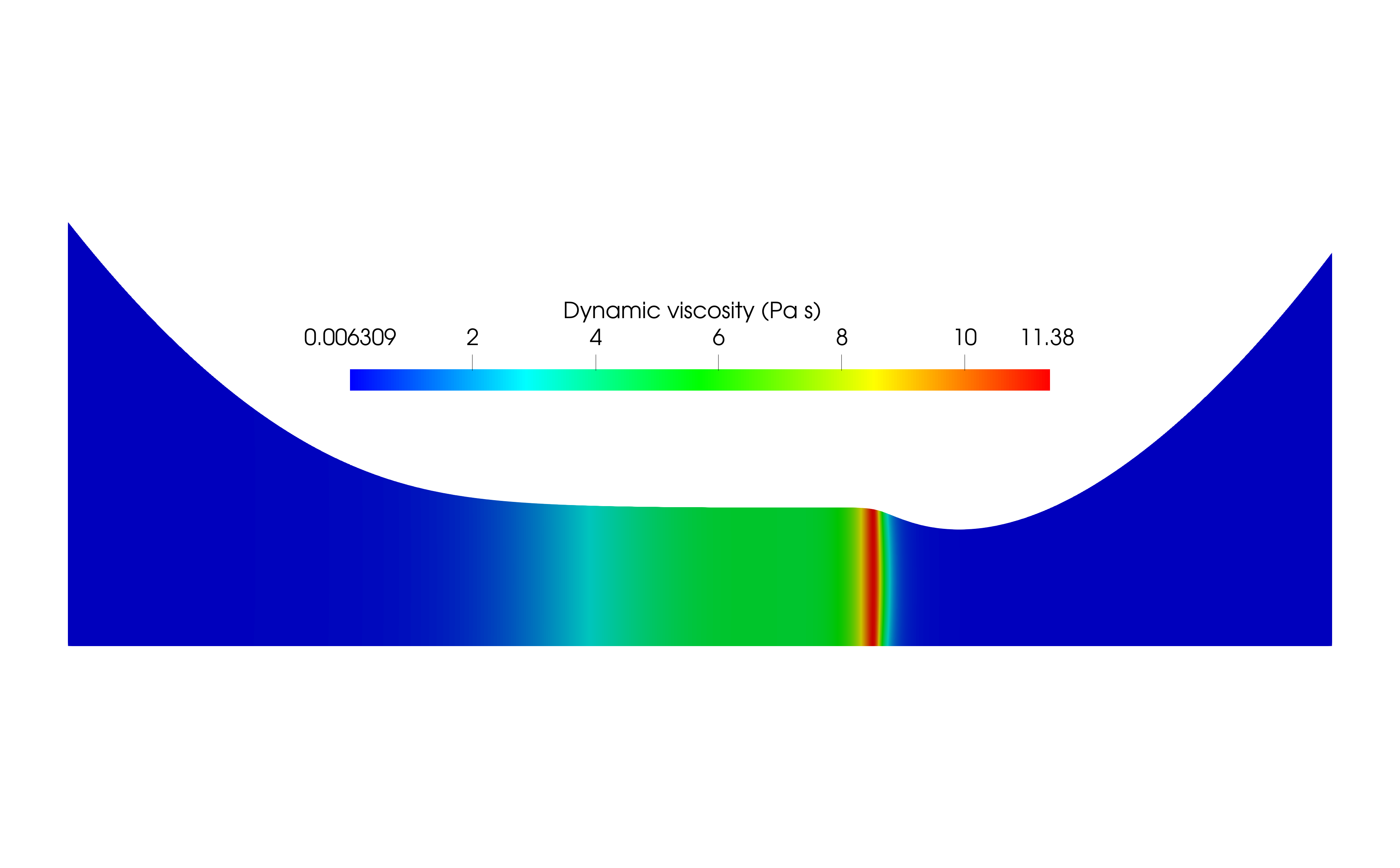}};
			\begin{scope}[shift=(img.south west), x={(img.south east)},y={(img.north west)}]
				\begin{axis}[
					width=(0.95*1.34*0.941*13.1cm),height=(0.95*1.34*0.953*5.44cm),
					xmin=-150, xmax=150, ymin=0, ymax=(\ybound-1.25e-4)/100*1e6,
					axis line style={thin},
					xtick={-150, -100, -50, 0, 50, 100},
					xticklabel style={anchor=north, yshift=-0.5ex, font=\scriptsize},
					extra x tick style={tick label style={xshift=-1.5pt}, font=\scriptsize},
					extra x ticks={150},
					ytick={0.25, 0.5, 0.75, 1},
					yticklabel style={anchor=east, xshift=-0.5ex, font=\scriptsize},
					extra y tick style={tick label style={yshift=0.5ex}, font=\scriptsize},
					extra y ticks={0},
					minor xtick={-125, -75, -25, 25, 75, 125},
					]
				\end{axis}
			\end{scope}
		\end{tikzpicture}
		\caption{Dynamic viscosity.}
	\end{subfigure}	
	
	\vspace{0pt}
	
	\begin{subfigure}{0.95\textwidth}
		\begin{tikzpicture}[inner sep=0pt]
			\node (img) at (0,0) {\includegraphics[width=0.928\textwidth,trim={194 633 194 633},clip]{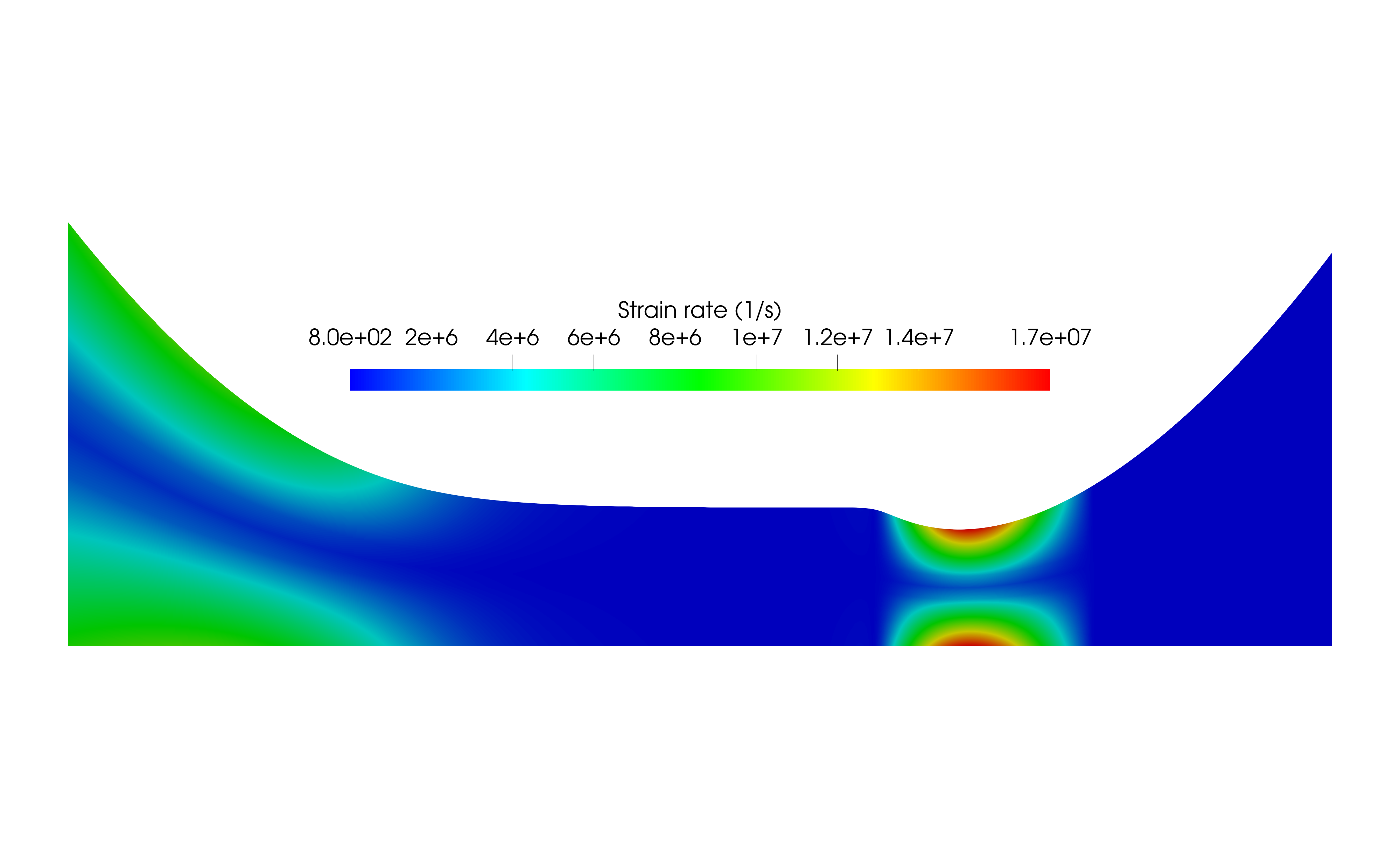}};
			\begin{scope}[shift=(img.south west), x={(img.south east)},y={(img.north west)}]
				\begin{axis}[
					width=(0.95*1.34*0.941*13.1cm),height=(0.95*1.34*0.953*5.44cm),
					xmin=-150, xmax=150, ymin=0, ymax=(\ybound-1.25e-4)/100*1e6,
					axis line style={thin},
					xtick={-150, -100, -50, 0, 50, 100},
					xticklabel style={anchor=north, yshift=-0.5ex, font=\scriptsize},
					extra x tick style={tick label style={xshift=-1.5pt}, font=\scriptsize},
					extra x ticks={150},
					ytick={0.25, 0.5, 0.75, 1},
					yticklabel style={anchor=east, xshift=-0.5ex, font=\scriptsize},
					extra y tick style={tick label style={yshift=0.5ex}, font=\scriptsize},
					extra y ticks={0},
					minor xtick={-125, -75, -25, 25, 75, 125},
					]
				\end{axis}
			\end{scope}
		\end{tikzpicture}
		\caption{Strain rate.}
	\end{subfigure}
	\caption{Plots of velocity, dynamic viscosity, temperature and strain rate for SRR $0$.
		The axis labels are given in \qty{}{\micro\meter}. Note that the figures are scaled with a factor $100$ in the $y$-direction.}
	\label{fig:srr0}
\end{figure}

\begin{figure}[p]
	\centering
	\def \srr {./srr1_output}
	\def \ybound {2.25443e-4}
	\captionsetup[subfigure]{skip=-1pt}
	\begin{subfigure}{0.95\textwidth}
		\begin{tikzpicture}[inner sep=0pt]
			\node (img) at (0,0) {\includegraphics[width=0.928\textwidth,trim={194 646 194 633},clip]{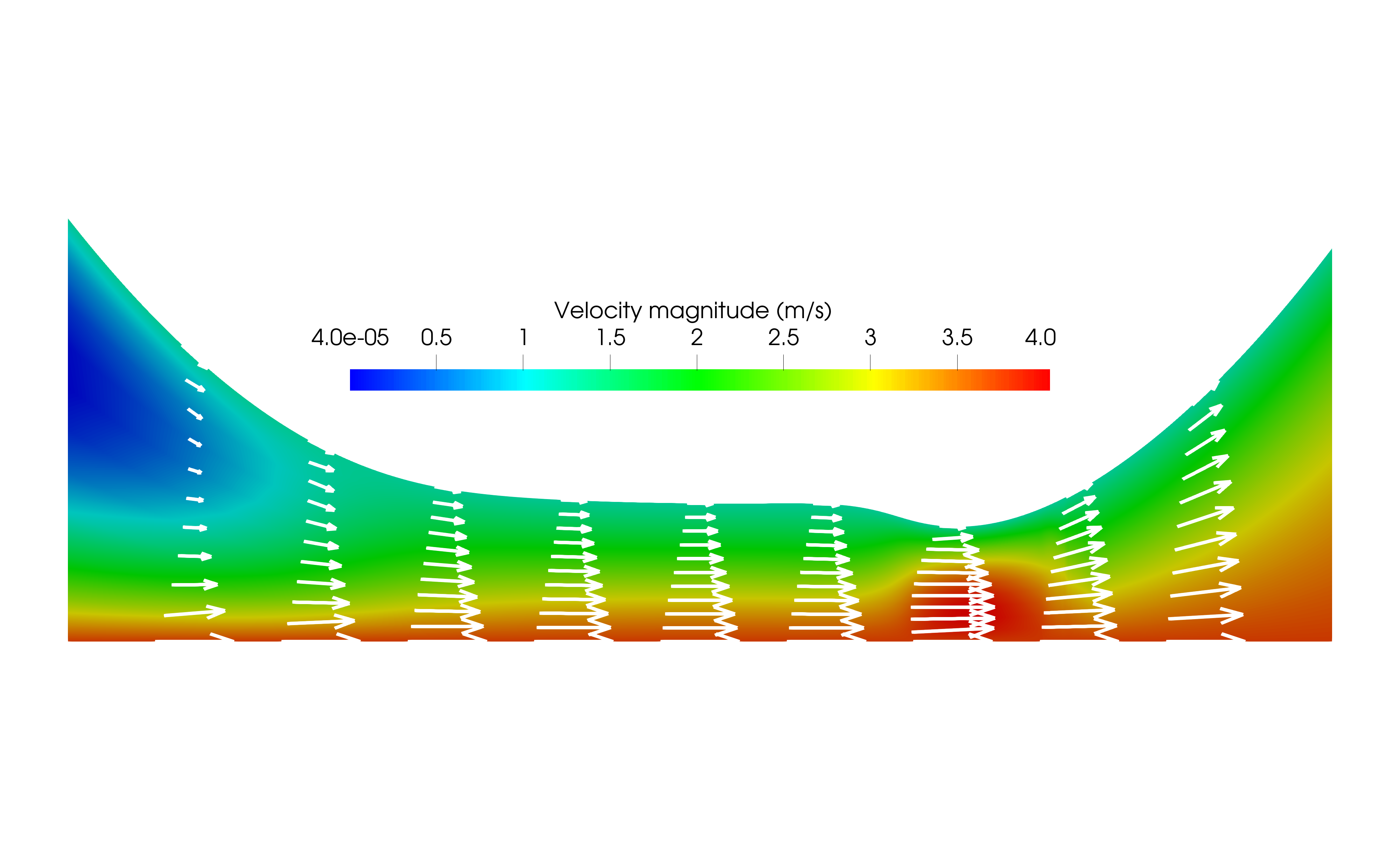}};
			\begin{scope}[shift=(img.south west), x={(img.south east)},y={(img.north west)}]
				\begin{axis}[
					width=(0.95*1.34*0.941*13.1cm),height=(0.95*1.34*0.953*5.44cm),
					xmin=-150, xmax=150, ymin=0, ymax=(\ybound-1.25e-4)/100*1e6,
					axis line style={thin},
					xtick={-150, -100, -50, 0, 50, 100},
					xticklabel style={anchor=north, yshift=-0.5ex, font=\scriptsize},
					extra x tick style={tick label style={xshift=-1.5pt}, font=\scriptsize},
					extra x ticks={150},
					ytick={0.25, 0.5, 0.75, 1},
					yticklabel style={anchor=east, xshift=-0.5ex, font=\scriptsize},
					extra y tick style={tick label style={yshift=0.5ex}, font=\scriptsize},
					extra y ticks={0},
					minor xtick={-125, -75, -25, 25, 75, 125},
					]
				\end{axis}
			\end{scope}
		\end{tikzpicture}
		\caption{Velocity.}
	\end{subfigure}
	
	\vspace{0pt}
	
	\begin{subfigure}{0.95\textwidth}
		\begin{tikzpicture}[inner sep=0pt]
			\node (img) at (0,0) {\includegraphics[width=0.928\textwidth,trim={194 633 194 633},clip]{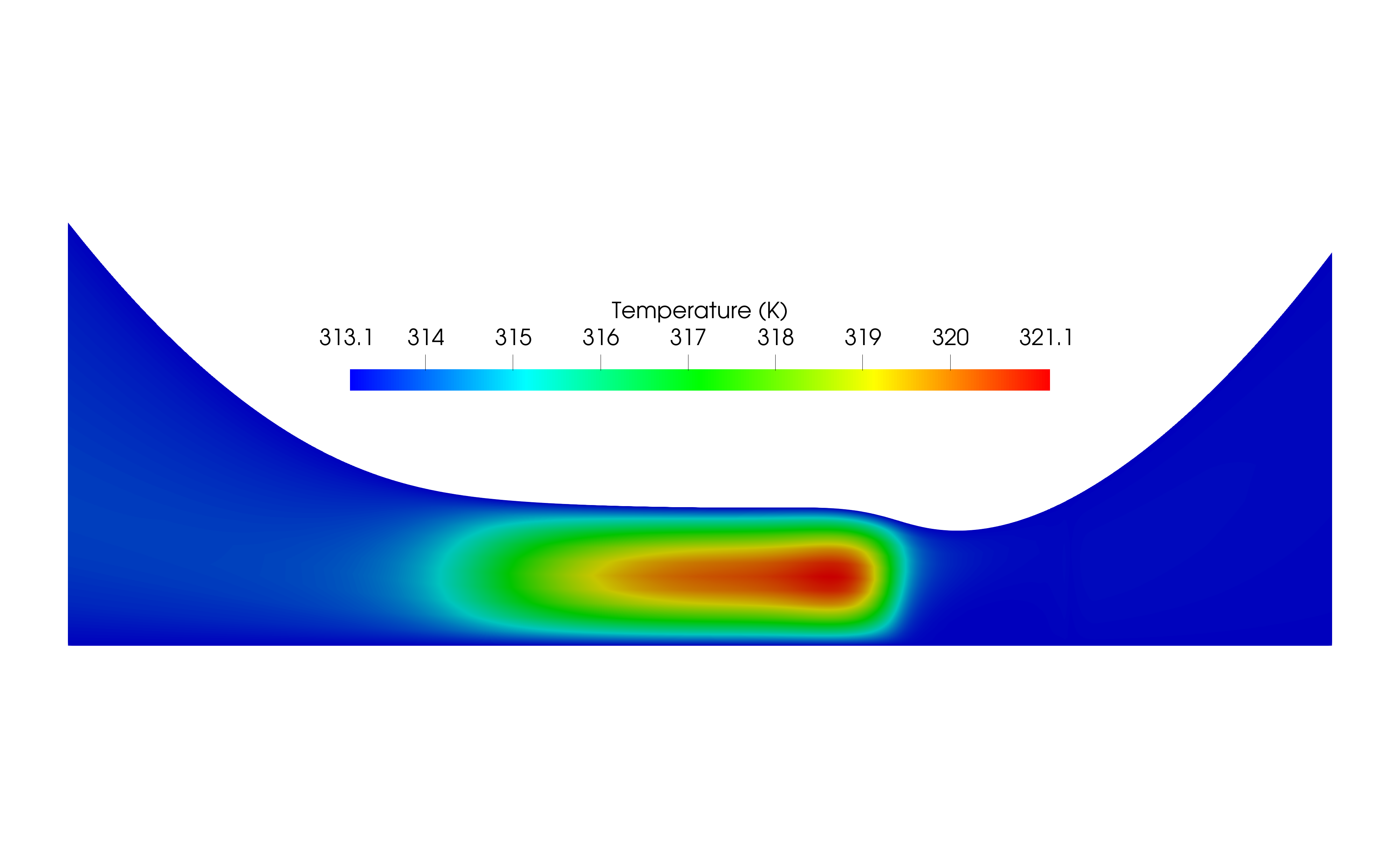}};
			\begin{scope}[shift=(img.south west), x={(img.south east)},y={(img.north west)}]
				\begin{axis}[
					width=(0.95*1.34*0.941*13.1cm),height=(0.95*1.34*0.953*5.44cm),
					xmin=-150, xmax=150, ymin=0, ymax=(\ybound-1.25e-4)/100*1e6,
					axis line style={thin},
					xtick={-150, -100, -50, 0, 50, 100},
					xticklabel style={anchor=north, yshift=-0.5ex, font=\scriptsize},
					extra x tick style={tick label style={xshift=-1.5pt}, font=\scriptsize},
					extra x ticks={150},
					ytick={0.25, 0.5, 0.75, 1},
					yticklabel style={anchor=east, xshift=-0.5ex, font=\scriptsize},
					extra y tick style={tick label style={yshift=0.5ex}, font=\scriptsize},
					extra y ticks={0},
					minor xtick={-125, -75, -25, 25, 75, 125},
					]
				\end{axis}
			\end{scope}
		\end{tikzpicture}
		\caption{Temperature.}
	\end{subfigure}
	
	\vspace{0pt}
	
	\begin{subfigure}{0.95\textwidth}
		\begin{tikzpicture}[inner sep=0pt]
			\node (img) at (0,0) {\includegraphics[width=0.928\textwidth,trim={194 633 194 633},clip]{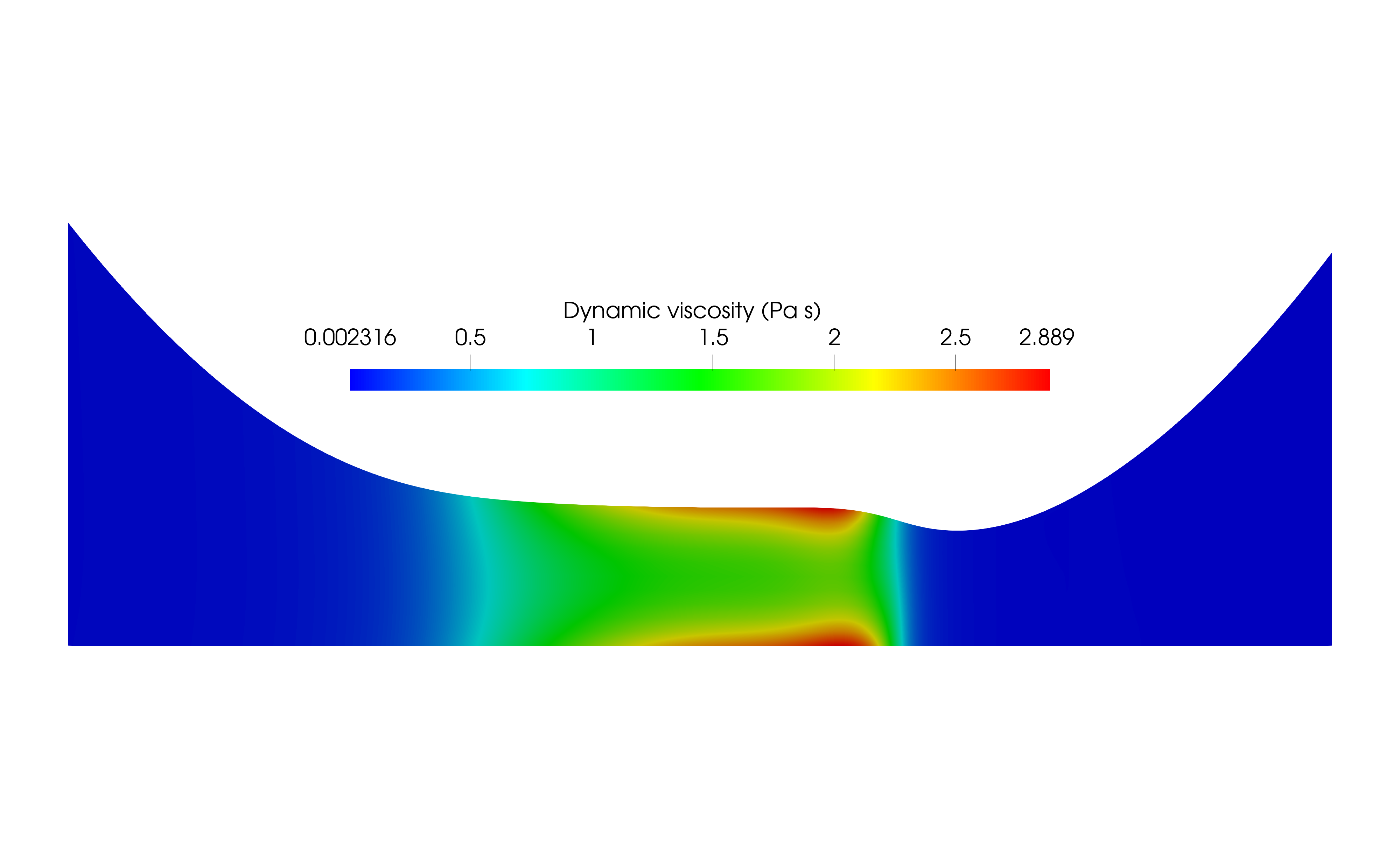}};
			\begin{scope}[shift=(img.south west), x={(img.south east)},y={(img.north west)}]
				\begin{axis}[
					width=(0.95*1.34*0.941*13.1cm),height=(0.95*1.34*0.953*5.44cm),
					xmin=-150, xmax=150, ymin=0, ymax=(\ybound-1.25e-4)/100*1e6,
					axis line style={thin},
					xtick={-150, -100, -50, 0, 50, 100},
					xticklabel style={anchor=north, yshift=-0.5ex, font=\scriptsize},
					extra x tick style={tick label style={xshift=-1.5pt}, font=\scriptsize},
					extra x ticks={150},
					ytick={0.25, 0.5, 0.75, 1},
					yticklabel style={anchor=east, xshift=-0.5ex, font=\scriptsize},
					extra y tick style={tick label style={yshift=0.5ex}, font=\scriptsize},
					extra y ticks={0},
					minor xtick={-125, -75, -25, 25, 75, 125},
					]
				\end{axis}
			\end{scope}
		\end{tikzpicture}
		\caption{Dynamic viscosity.}
	\end{subfigure}	
	
	\vspace{0pt}
	
	\begin{subfigure}{0.95\textwidth}
		\begin{tikzpicture}[inner sep=0pt]
			\node (img) at (0,0) {\includegraphics[width=0.928\textwidth,trim={194 633 194 633},clip]{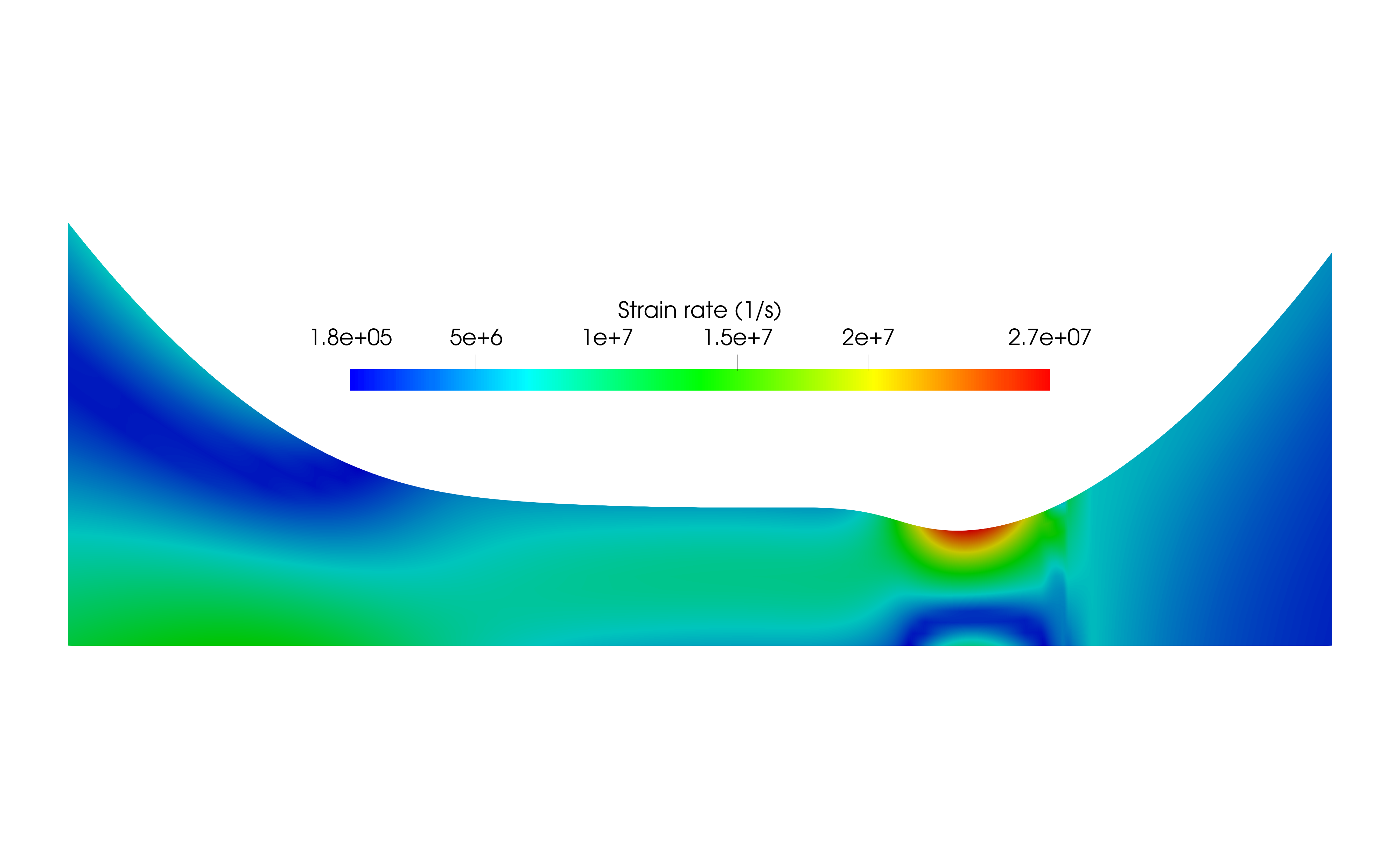}};
			\begin{scope}[shift=(img.south west), x={(img.south east)},y={(img.north west)}]
				\begin{axis}[
					width=(0.95*1.34*0.941*13.1cm),height=(0.95*1.34*0.953*5.44cm),
					xmin=-150, xmax=150, ymin=0, ymax=(\ybound-1.25e-4)/100*1e6,
					axis line style={thin},
					xtick={-150, -100, -50, 0, 50, 100},
					xticklabel style={anchor=north, yshift=-0.5ex, font=\scriptsize},
					extra x tick style={tick label style={xshift=-1.5pt}, font=\scriptsize},
					extra x ticks={150},
					ytick={0.25, 0.5, 0.75, 1},
					yticklabel style={anchor=east, xshift=-0.5ex, font=\scriptsize},
					extra y tick style={tick label style={yshift=0.5ex}, font=\scriptsize},
					extra y ticks={0},
					minor xtick={-125, -75, -25, 25, 75, 125},
					]
				\end{axis}
			\end{scope}
		\end{tikzpicture}
		\caption{Strain rate.}
	\end{subfigure}
	\caption{Plots of velocity, dynamic viscosity, temperature and strain rate for SRR $1$.
		The axis labels are given in \qty{}{\micro\meter}. Note that the figures are scaled with a factor $100$ in the $y$-direction.}
	\label{fig:srr1}
\end{figure}

\begin{figure}[p]
	\centering
	\def \srr {./srr2_output}
	\def \ybound {2.25097e-4}
	\captionsetup[subfigure]{skip=-1pt}
	\begin{subfigure}{0.95\textwidth}
		\begin{tikzpicture}[inner sep=0pt]
			\node (img) at (0,0) {\includegraphics[width=0.928\textwidth,trim={194 650 194 633},clip]{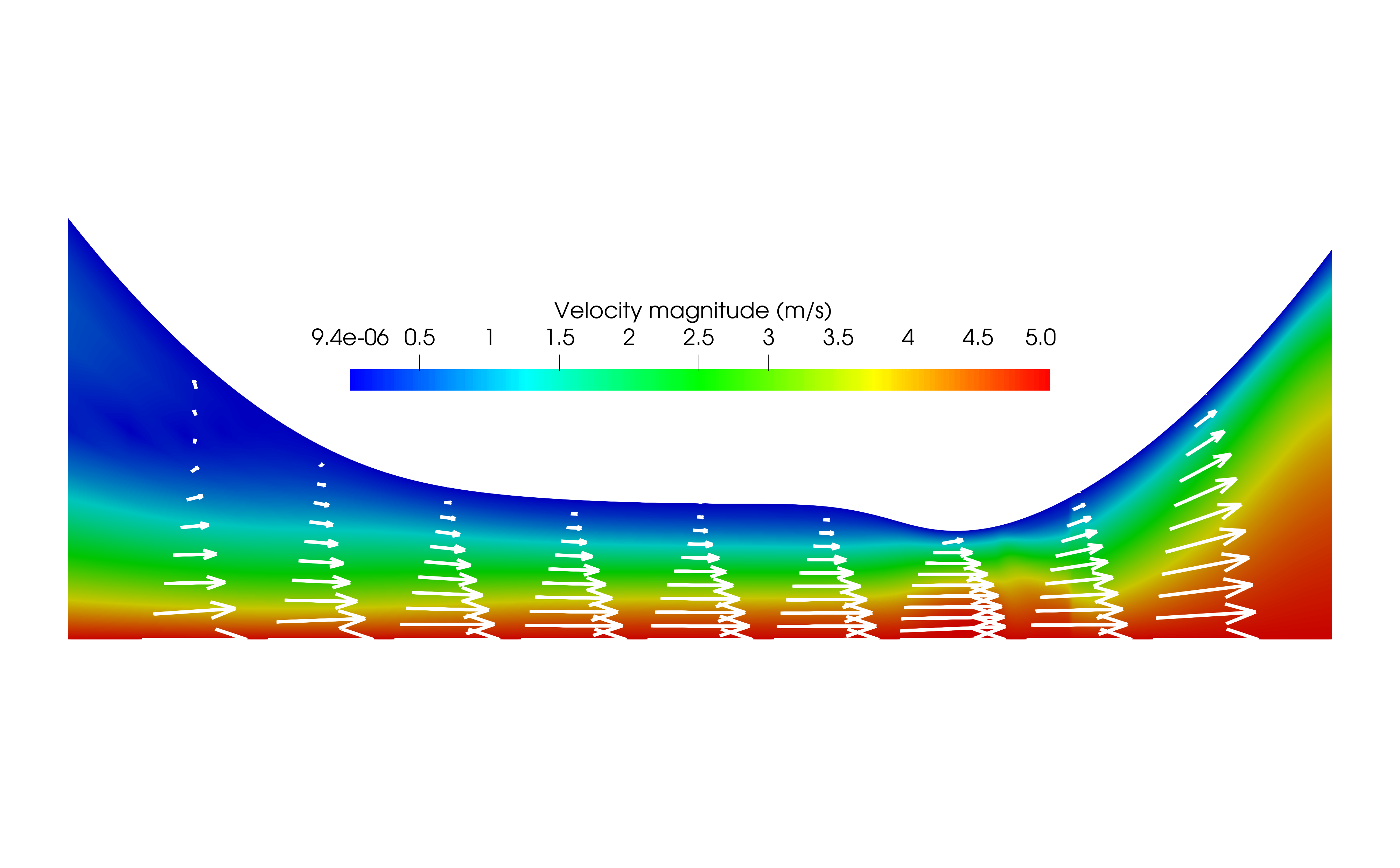}};
			\begin{scope}[shift=(img.south west), x={(img.south east)},y={(img.north west)}]
				\begin{axis}[
					width=0.95*(1.34*0.941*13.1cm),height=(0.95*1.34*0.953*5.44cm),
					xmin=-150, xmax=150, ymin=0, ymax=(\ybound-1.25e-4)/100*1e6,
					axis line style={thin},
					xtick={-150, -100, -50, 0, 50, 100},
					xticklabel style={anchor=north, yshift=-0.5ex, font=\scriptsize},
					extra x tick style={tick label style={xshift=-1.5pt}, font=\scriptsize},
					extra x ticks={150},
					ytick={0.25, 0.5, 0.75, 1},
					yticklabel style={anchor=east, xshift=-0.5ex, font=\scriptsize},
					extra y tick style={tick label style={yshift=0.5ex}, font=\scriptsize},
					extra y ticks={0},
					minor xtick={-125, -75, -25, 25, 75, 125},
					]
				\end{axis}
			\end{scope}
		\end{tikzpicture}
		\caption{Velocity.}
	\end{subfigure}
	
	\vspace{0pt}
	
	\begin{subfigure}{0.95\textwidth}
		\begin{tikzpicture}[inner sep=0pt]
			\node (img) at (0,0) {\includegraphics[width=0.928\textwidth,trim={194 633 194 633},clip]{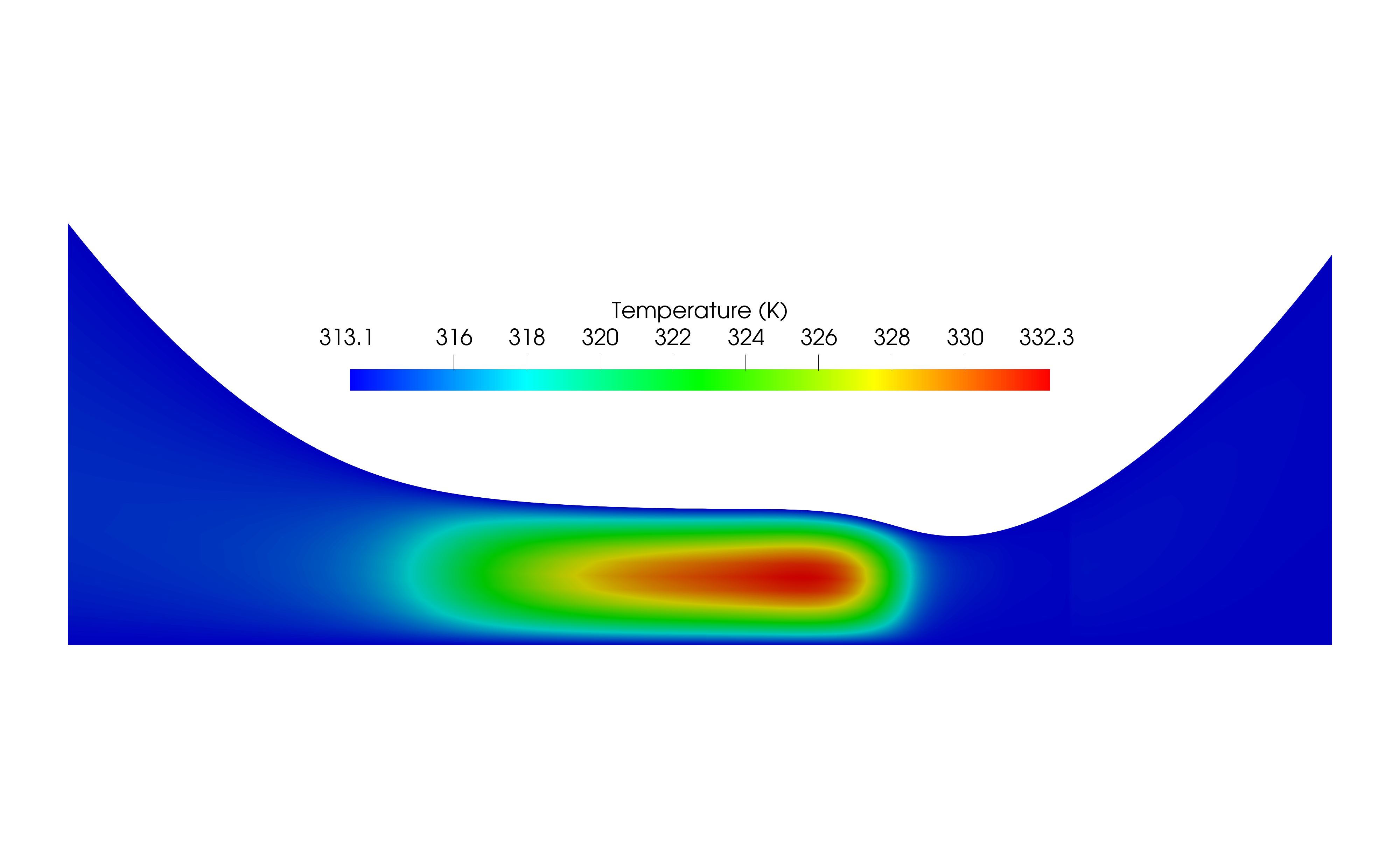}};
			\begin{scope}[shift=(img.south west), x={(img.south east)},y={(img.north west)}]
				\begin{axis}[
					width=(0.95*1.34*0.941*13.1cm),height=(0.95*1.34*0.953*5.44cm),
					xmin=-150, xmax=150, ymin=0, ymax=(\ybound-1.25e-4)/100*1e6,
					axis line style={thin},
					xtick={-150, -100, -50, 0, 50, 100},
					xticklabel style={anchor=north, yshift=-0.5ex, font=\scriptsize},
					extra x tick style={tick label style={xshift=-1.5pt}, font=\scriptsize},
					extra x ticks={150},
					ytick={0.25, 0.5, 0.75, 1},
					yticklabel style={anchor=east, xshift=-0.5ex, font=\scriptsize},
					extra y tick style={tick label style={yshift=0.5ex}, font=\scriptsize},
					extra y ticks={0},
					minor xtick={-125, -75, -25, 25, 75, 125},
					]
				\end{axis}
			\end{scope}
		\end{tikzpicture}
		\caption{Temperature.}
	\end{subfigure}
	
	\vspace{0pt}
	
	\begin{subfigure}{0.95\textwidth}
		\begin{tikzpicture}[inner sep=0pt]
			\node (img) at (0,0) {\includegraphics[width=0.928\textwidth,trim={194 633 194 633},clip]{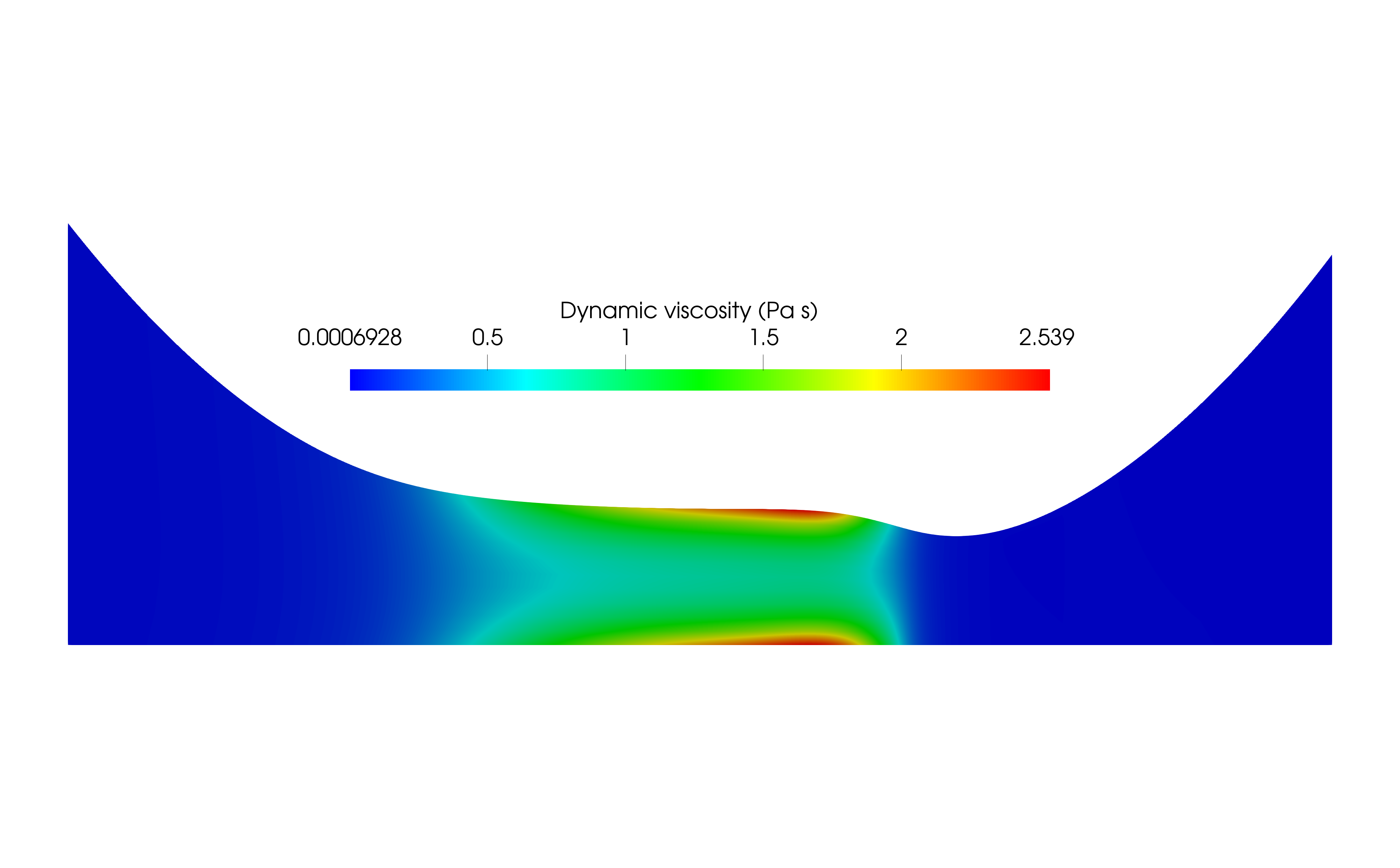}};
			\begin{scope}[shift=(img.south west), x={(img.south east)},y={(img.north west)}]
				\begin{axis}[
					width=(0.95*1.34*0.941*13.1cm),height=(0.95*1.34*0.953*5.44cm),
					xmin=-150, xmax=150, ymin=0, ymax=(\ybound-1.25e-4)/100*1e6,
					axis line style={thin},
					xtick={-150, -100, -50, 0, 50, 100},
					xticklabel style={anchor=north, yshift=-0.5ex, font=\scriptsize},
					extra x tick style={tick label style={xshift=-1.5pt}, font=\scriptsize},
					extra x ticks={150},
					ytick={0.25, 0.5, 0.75, 1},
					yticklabel style={anchor=east, xshift=-0.5ex, font=\scriptsize},
					extra y tick style={tick label style={yshift=0.5ex}, font=\scriptsize},
					extra y ticks={0},
					minor xtick={-125, -75, -25, 25, 75, 125},
					]
				\end{axis}
			\end{scope}
		\end{tikzpicture}
		\caption{Dynamic viscosity.}
	\end{subfigure}	
	
	\vspace{0pt}
	
	\begin{subfigure}{0.95\textwidth}
		\begin{tikzpicture}[inner sep=0pt]
			\node (img) at (0,0) {\includegraphics[width=0.928\textwidth,trim={194 633 194 633},clip]{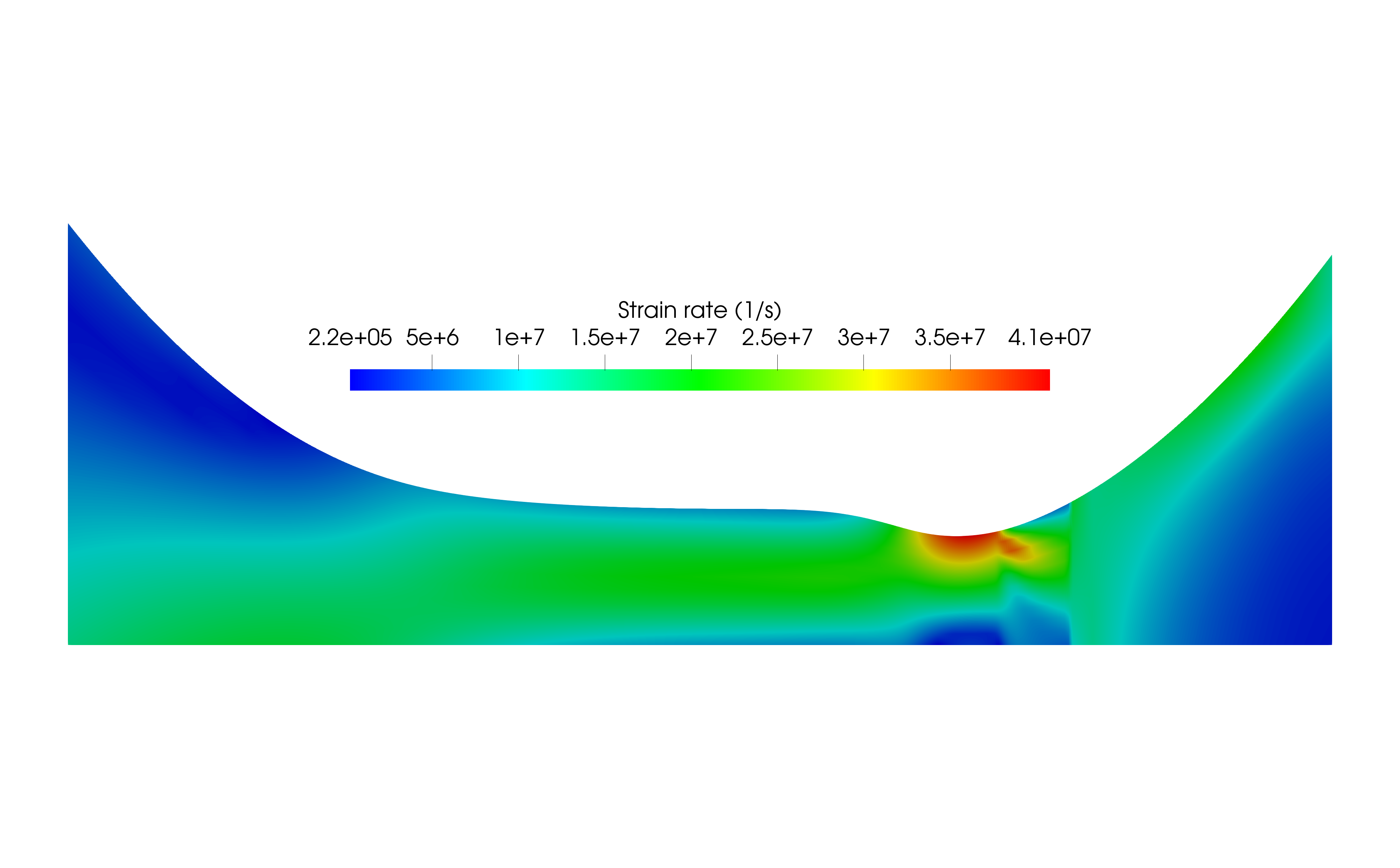}};
			\begin{scope}[shift=(img.south west), x={(img.south east)},y={(img.north west)}]
				\begin{axis}[
					width=(0.95*1.34*0.941*13.1cm),height=(0.95*1.34*0.953*5.44cm),
					xmin=-150, xmax=150, ymin=0, ymax=(\ybound-1.25e-4)/100*1e6,
					axis line style={thin},
					xtick={-150, -100, -50, 0, 50, 100},
					xticklabel style={anchor=north, yshift=-0.5ex, font=\scriptsize},
					extra x tick style={tick label style={xshift=-1.5pt}, font=\scriptsize},
					extra x ticks={150},
					ytick={0.25, 0.5, 0.75, 1},
					yticklabel style={anchor=east, xshift=-0.5ex, font=\scriptsize},
					extra y tick style={tick label style={yshift=0.5ex}, font=\scriptsize},
					extra y ticks={0},
					minor xtick={-125, -75, -25, 25, 75, 125},
					]
				\end{axis}
			\end{scope}
		\end{tikzpicture}
		\caption{Strain rate.}
	\end{subfigure}
	\caption{Plots of velocity, dynamic viscosity, temperature and strain rate for SRR $2$.
		The axis labels are given in \qty{}{\micro\meter}. Note that the figures are scaled with a factor $100$ in the $y$-direction.}
	\label{fig:srr2}
\end{figure}

\autoref{fig:srr0_extra} shows the variation of liquid compressibility and density, as well as the location of the vapor region in case of pure rolling. The maximum and minimum liquid compressibility differ by a factor of $7$, underlining the importance of including its variability.
The biggest effect of varying slip is seen in the velocity, viscosity, temperature and strain rate, which are all closely linked.
The corresponding plots are presented in \autoref{fig:srr0}, \autoref{fig:srr1} and \autoref{fig:srr2}.

For SRR $0$, the temperature is nearly constant, as almost no heat is generated by viscous effects, and the highest strain rate occurs at the constriction.

For SRR $1$, the velocity difference between the surfaces results in a Couette flow.
As a consequence, there is considerable shear thinning.
Due to the local increase in temperature (up to \qty{8}{\kelvin}) and accompanying drop in viscosity, a shear band arises in the middle of the contact.
These higher shear rates result in further shear thinning and decrease of viscosity.
Compared to SRR $0$, the viscosity is an order of magnitude lower.
The maximal shear rate occurs in the constriction at the top, slower surface, while at the bottom surface the shear rate is locally smaller, since the acceleration in the constriction cancels out the velocity gradient of the Couette flow.

For SRR $2$, similar conclusions can be made, although with more extreme effects.
The increased shear results in a temperature increase of \qty{29}{\kelvin} in the center of the contact, and the shear band is more pronounced, leading to even lower viscosity values.

\pagebreak
\section{Conclusion}
\label{sec:conclusion}
This work presents a solver for lubricant flow in small gaps and subjected to extreme pressures, developed within \OF 8. 
The basis for this solver is the standard solver \emph{cavitatingFoam}, to which variable liquid compressibility and a thermal equation have been added.
The hydrodynamic and thermal lubricant behavior is accurately captured by inclusion of cavitation, compressibility, piezoviscosity, shear thinning and models for the thermal conductivity and heat capacity.
The use of a library of lubricant models allow them to be interchangeably selected, benefiting from the modular \OF structure. 

Next to detailing the solver and its implementation, its use is showcased by simulating an EHL line contact.
To this end, the solver is coupled to the open-source structural solver Kratos Multiphysics, modeling the deforming surfaces in contact, with the help of the open-source coupling tool CoCoNuT.
Besides these, any other partitioned coupling code or structural solver could be chosen.
In these simulations, the rigid plane is moved upward according to a prescribed motion, resulting in a similarly increasing load.
The resulting solution setup is successfully validated against results found in literature.

Finally, the performance is illustrated for different slip conditions ranging from pure rolling to pure slip.
While for pure rolling (SRR $0$) the temperature increase and shear rate are limited, as expected, in case of slip (SRR $1$ and SRR $2$), an area of higher temperature and shear rate values is formed in the center of the contact.
This illustrates the necessity of including variable liquid compressibility and thermal effects, as opposed to the standard \emph{cavitatingFoam}, and shows the capabilities of this solver to be used in other EHL simulations with varying load or speed constraints
In this way, the effect of peak loads or speed reversal on lubricated contacts can for example be investigated in bearings.



\section*{Acknowledgements}

\noindent N. Delaissé gratefully acknowledges the funding received from the Research Foundation -- Flanders (FWO) as part of the SBO CONTACTLUB project (S006519N) and the funding received from the Ghent University Special Research Fund (BOF) (01P02523).

%
%
\authorcontributions{
Conceptualisation, N.D. and P.H.;
methodology, N.D. and P.H.;
software, N.D. and P.H.;
validation, N.D., P.H., J.D. and D.F.;
formal analysis, N.D.;
investigation, N.D. and P.H.;
resources, J.D. and D.F.;
data curation, J.D. and D.F.;
writing---original draft preparation, N.D.;
writing---review and editing, N.D., P.H., J.D. and D.F.;
visualisation, N.D.;
supervision, J.D. and D.F.;
project administration, J.D. and D.F.;
funding acquisition, J.D. and D.F.
All authors have read and agreed to the published version of the manuscript.
}


\appendix
\section{Additional notes on the lubricant solver}
\label{sec:ehladd}
In this section, some additional notes on the lubricant solver are presented.
First, several parameters are derived from the Tait compressibility equation.
Next, the concept of relative and absolute flux is explained in detail, and finally, the calculation of the total derivative is discussed.
The subscript $l$ referring to the liquid is omitted.

\subsection{Parameters derived from the Tait equation}
\label{sec:tait_derived}
The Tait equation, \autoref{equ:tait}, expresses the variation of the specific volume $\vol=1/\rho$ with pressure and temperature.
Several parameters can be derived from this relation.
For ease of notation, the constant
\begin{subequations}
	\begin{align}
		&\pTaitA = \frac{1}{1+K'_0} \\
		\intertext{and the pressure and (through $K_0$) temperature dependent variable}
		&\pTaitB = 1 + \frac{p}{\pTaitA K_0} = 1 + p\frac{1+K'_0}{K_0}
	\end{align}
	are introduced.
\end{subequations}

First, the liquid compressibility $\psi$ is given by
\begin{equation}
	\psi = \dd{\rho}{p} = -\frac{1}{\vol^2}\dd{\vol}{p}
	= \frac{1}{\vol_{0}K_0 \pTaitB \left(1-\pTaitA\ln{\pTaitB}\right)^2}.
\end{equation}

Second, the volumetric expansivity $\beta$ is given by
\begin{equation}
	\beta = \left.-\frac{1}{\rho}\dd{\rho}{T}\right\vert_p
	= \left.\frac{1}{\vol}\dd{\vol}{T}\right\vert_p
	= \frac{-\beta_K p}{K_0 \pTaitB \left(1-\pTaitA\ln{\pTaitB}\right)} + \frac{\vol_{0R}a_{\vol}}{\vol_0}.
\end{equation}
Note that the explicit value is not required by the solver, since the integral term in which it appears is more easily calculated by inserting the definition of the volume expansivity.

Finally, this term $\int_{0}^{p}\frac{1}{\rho}\left(1-T\beta\right)\diff p$ in the expression for liquid enthalpy $h_l$ (\autoref{equ:enthalpy_liquid}) is determined.
It represents the change of $h_l$ with $p$ and is given by
\begin{equation}
	\begin{split}
		\int_{0}^{p}\frac{1}{\rho}\left(1-T\beta\right)\diff p 
		&= \int_{0}^{p}\left(\vol-T \left.\dd{\vol}{T}\right\vert_p \right)\diff p \\
		&= \vol_{0R}\left(1-a_\vol T_R\right)\left(p-K_0 \pTaitA^2\left(1-\pTaitB+\pTaitB\ln{\pTaitB}\right)\right) \\
		&\qquad+\vol_0 \beta_K T \pTaitA (p-K_0 \pTaitA \ln{\pTaitB}),
	\end{split}
\end{equation}
where it is assumed that $p$ and $T$ are independent quantities.

\subsection{Relative and absolute flux}
\label{sec:flux}
In finite volume problems with dynamic mesh movement, the governing equations are integrated over a changing volume.
As explained by Ferziger et al. \cite{Ferziger2002}, this requires the use of Leibniz's rule for the time derivative term.
For example, integrating the continuity equation, \autoref{equ:conservative_continuity}, over the changing volume $V(t)$ results in
\begin{equation}
	\begin{split}
		&\intVt{\ddt{\rho}} + \intVt{\mdiv\left(\rho\mU\right)} \\
		&\qquad= \tddt{}\intVt{\rho} - \intSt{\rho \tddt{\vec{r}}} + \intVt{\mdiv\left(\rho \mU\right)},
	\end{split}
\end{equation}
where $S(t)$ is the moving boundary of $V(t)$ and $\vec{r}$ is the position vector.
After applying Gauss's theorem on the divergence term, as follows
\begin{equation}
	\tddt{}\intVt{\rho} - \intSt{\rho \tddt{\vec{r}}} + \intSt{\rho \mU},
\end{equation}
it becomes clear that this additional term can be considered as a modification of the transporting velocity, given by
\begin{equation}
	\label{equ:continuity_moving}
	\tddt{}\intVt{\rho} + \intSt{\rho \left(\mU-\mU_b\right)},
\end{equation}
where $\mU_b$ is the velocity of the integration boundary, i.e., the grid velocity.

In other words, when integrating the time derivative term in the conservation laws, \autoref{equ:convective}, in their conservative form over a changing volume, the transporting velocity in the convective term is the velocity relative to the grid motion.

Using the \OF differential notation for integral equations, the convection term in a moving grid becomes ${\mdiv\left(\left(\mU-\mU_b\right)\thv\right)}$, with $\thv$ a general vector or scalar.
Similar to \autoref{equ:openfoam_flux}, the flux notation can be introduced, albeit here, using the relative flux $\varphi=\Sv_f\cdot\left(\mU_f-\mU_{b,f}\right)$ as follows
\begin{equation}
	\begin{alignedat}{2}
		&\text{convection in a moving grid: }\quad\mdiv\left(\left(\mU-\mU_b\right)\thv\right) = \mdiv\left(\varphiv\thv\right) \\
		&\qquad\Rightarrow\intVt{\mdiv\left(\left(\mU-\mU_b\right)\thv\right)} = \int_{S(t)}\diff\Sv\cdot\left(\left(\mU-\mU_b\right)\thv\right)
			\approx \sum_f \Sv_f\cdot\left(\mU_f-\mU_{b,f}\right)\thv_f = \sum_f \varphi_f\thv_f.
	\end{alignedat}
\end{equation}
As before, the bold symbol $\varphiv$ is used in \OF differential notation.
Note that the order of the transporting velocity and $\thv$ is important, since the dot product of a tensor and a vector is not commutative.

Finally, care should be taken when determining the grid velocity $\mU_b$.
Simply using the face velocities leads to violating the space conservation law (SCL), also called geometric conservation law, given by
\begin{equation}
	\label{equ:space_conservation}
	\tddt{}\intVt{} = \intSt{\mU_b}.
\end{equation}
As a consequence, the relative flux $\varphi_f$ is in practice calculated as 
\begin{equation}
	\varphi_f = \Sv_f\cdot\mU_f - \Sv_f\cdot\mU_{b,f}= \Sv_f\cdot\mU_f - \frac{V_P - V_P^o}{\Delta t}= \phi_f - \frac{V_P - V_P^o}{\Delta t},
\end{equation}
in which $\phi_f$ is the absolute flux, the subscript $P$ denotes the value at the cell center, and the superscript $o$ refers to the old value.

Note that, in the special case of the incompressible flow on a moving grid, the continuity equation, \autoref{equ:continuity_moving}, combines with the SCL, \autoref{equ:space_conservation}, to give
\begin{equation}
	\intSt{\mU} = 0,
\end{equation}
or in \OF notation $\mdiv\mU=0$, which is the same as the continuity equation for a stationary grid.

\emph{\OF issue report.}
In version 8 of the \OF code released by \OF foundation, a bug related to the relative and absolute flux has been reported.
At the end of \texttt{pEqn.H} of the \emph{cavitatingFoam} solver, the absolute flux is required for the correction of the face velocity \texttt{Uf} instead of the relative one, see also the \href{https://bugs.openfoam.org/view.php?id=3825}{\OF issue report}.
This has been remedied from version $10$ onward.

\subsection{Calculating the material derivative}
\label{sec:material_derivative}
This section discusses the calculation of the material derivative, and the determination of $\Diff p / \Diff t$ will be used as example.
This material derivative can be rewritten as
\begin{equation}
	\DDt{p} \equiv \ddt{p} + \mU\cdot\nabla p = \ddt{p} + \mdiv\left(p \mU\right) - p \nabla \cdot \mU.
\end{equation}
The last form is called the conservation form and is preferred in finite volume calculations.
As explained in \autoref{sec:flux}, integration over a moving volume $V(t)$ results in
\begin{equation}
		\intVt{\DDt{p}} =\ddt{}\intVt{p} + \intSt{p\left(\mU-\mU_b\right)} - \intVt{p \nabla \cdot \mU}.
\end{equation}
In \OF notation, this becomes
\begin{equation}
	\DDt{p} = \ddt{p} + \mdiv\left(p\left(\mU-\mU_b\right)\right) - p \nabla \cdot \mU = \ddt{p} + \mdiv\left(p\varphiv\right) - p \mdiv \phiv.
\end{equation}
Note the use of both the relative flux $\varphiv$ and the absolute flux $\phiv$.

\bibliographystyle{IEEEtran}

\bibliography{Bibliography}

\begin{thebibliography}{10}
\providecommand{\url}[1]{#1}
\csname url@samestyle\endcsname
\providecommand{\newblock}{\relax}
\providecommand{\bibinfo}[2]{#2}
\providecommand{\BIBentrySTDinterwordspacing}{\spaceskip=0pt\relax}
\providecommand{\BIBentryALTinterwordstretchfactor}{4}
\providecommand{\BIBentryALTinterwordspacing}{\spaceskip=\fontdimen2\font plus
\BIBentryALTinterwordstretchfactor\fontdimen3\font minus
  \fontdimen4\font\relax}
\providecommand{\BIBforeignlanguage}[2]{{%
\expandafter\ifx\csname l@#1\endcsname\relax
\typeout{** WARNING: IEEEtran.bst: No hyphenation pattern has been}%
\typeout{** loaded for the language `#1'. Using the pattern for}%
\typeout{** the default language instead.}%
\else
\language=\csname l@#1\endcsname
\fi
#2}}
\providecommand{\BIBdecl}{\relax}
\BIBdecl

\bibitem{Grubin1949}
A.~N. Grubin and I.~E. Vinogradova, \emph{Book 30}.\hskip 1em plus 0.5em minus
  0.4em\relax Central Scientific Research Institute for Technology and
  Mechanical Engineering (Department of Scientific and Industrial Research,
  Transl. 337), 1949.

\bibitem{Dowson1995}
\BIBentryALTinterwordspacing
D.~Dowson, ``Elastohydrodynamic and micro-elastohydrodynamic lubrication,''
  \emph{Wear}, vol. 190, no.~2, pp. 125--138, Dec. 1995. [Online]. Available:
  \url{https://doi.org/10.1016/0043-1648(95)06660-8}
\BIBentrySTDinterwordspacing

\bibitem{Moes1992}
\BIBentryALTinterwordspacing
H.~Moes, ``Optimum similarity analysis with applications to elastohydrodynamic
  lubrication,'' \emph{Wear}, vol. 159, no.~1, pp. 57--66, Nov. 1992. [Online].
  Available: \url{https://doi.org/10.1016/0043-1648(92)90286-h}
\BIBentrySTDinterwordspacing

\bibitem{Hamrock2004}
\BIBentryALTinterwordspacing
B.~J. Hamrock, S.~R. Schmid, and B.~O. Jacobson, \emph{Fundamentals of fluid
  film lubrication}.\hskip 1em plus 0.5em minus 0.4em\relax Marcel Dekker, Mar.
  2004. [Online]. Available: \url{https://doi.org/10.1201/9780203021187}
\BIBentrySTDinterwordspacing

\bibitem{Stachowiak2004}
G.~Stachowiak and A.~W. Batchelor, \emph{Experimental Methods in Tribology},
  ser. Tribology and Interface Engineering.\hskip 1em plus 0.5em minus
  0.4em\relax Elsevier Science, 2004, vol.~44.

\bibitem{Manjunath2024}
\BIBentryALTinterwordspacing
M.~Manjunath, S.~Hausner, A.~Heine, P.~De~Baets, and D.~Fauconnier,
  ``Electrical impedance spectroscopy for precise film thickness assessment in
  line contacts,'' \emph{Lubricants}, vol.~12, no.~2, p.~51, Feb. 2024.
  [Online]. Available: \url{https://doi.org/10.3390/lubricants12020051}
\BIBentrySTDinterwordspacing

\bibitem{Jablonka2012}
\BIBentryALTinterwordspacing
K.~Jablonka, R.~Glovnea, and J.~Bongaerts, ``Evaluation of {EHD} films by
  electrical capacitance,'' \emph{Journal of Physics D: Applied Physics},
  vol.~45, no.~38, p. 385301, Sep. 2012. [Online]. Available:
  \url{https://doi.org/10.1088/0022-3727/45/38/385301}
\BIBentrySTDinterwordspacing

\bibitem{Scurria2021}
\BIBentryALTinterwordspacing
L.~Scurria, T.~Tamarozzi, O.~Voronkov, and D.~Fauconnier, ``Quantitative
  analysis of {R}eynolds and {N}avier{\textendash}{S}tokes based modeling
  approaches for isothermal {N}ewtonian elastohydrodynamic lubrication,''
  \emph{Journal of Tribology}, vol. 143, no.~12, Mar. 2021. [Online].
  Available: \url{https://doi.org/10.1115/1.4050272}
\BIBentrySTDinterwordspacing

\bibitem{Hajishafiee2017}
\BIBentryALTinterwordspacing
A.~Hajishafiee, A.~Kadiric, S.~Ioannides, and D.~Dini, ``A coupled
  finite-volume {CFD} solver for two-dimensional elasto-hydrodynamic
  lubrication problems with particular application to rolling element
  bearings,'' \emph{Tribology International}, vol. 109, pp. 258--273, May 2017.
  [Online]. Available: \url{https://doi.org/10.1016/j.triboint.2016.12.046}
\BIBentrySTDinterwordspacing

\bibitem{Singh2021}
\BIBentryALTinterwordspacing
K.~Singh, F.~Sadeghi, T.~Russell, S.~J. Lorenz, W.~Peterson, J.~Villarreal, and
  T.~Jinmon, ``Fluid{\textendash}structure interaction modeling of
  elastohydrodynamically lubricated line contacts,'' \emph{Journal of
  Tribology}, vol. 143, no.~9, Jan. 2021. [Online]. Available:
  \url{https://doi.org/10.1115/1.4049260}
\BIBentrySTDinterwordspacing

\bibitem{Greenshields2022}
\BIBentryALTinterwordspacing
C.~J. Greenshields and H.~G. Weller, \emph{Notes on Computational Fluid
  Dynamics: General Principles}.\hskip 1em plus 0.5em minus 0.4em\relax
  Reading, United Kingdom: CFD Direct Ltd, 2022. [Online]. Available:
  \url{https://doc.cfd.direct/notes/cfd-general-principles/}
\BIBentrySTDinterwordspacing

\bibitem{Stewart1984}
\BIBentryALTinterwordspacing
H.~B. Stewart and B.~Wendroff, ``Two-phase flow: Models and methods,''
  \emph{Journal of Computational Physics}, vol.~56, no.~3, pp. 363--409, Dec.
  1984. [Online]. Available: \url{https://doi.org/10.1016/0021-9991(84)90103-7}
\BIBentrySTDinterwordspacing

\bibitem{Karrholm2007}
\BIBentryALTinterwordspacing
F.~P. K\"{a}rrholm, H.~Weller, and N.~Nordin, ``Modelling injector flow
  including cavitation effects for diesel applications,'' in
  \emph{5\textsuperscript{th}\ Joint ASME/JSME Fluids Engineering
  Conference}.\hskip 1em plus 0.5em minus 0.4em\relax San Diego: {ASMEDC}, Jan.
  2007, pp. 1--10. [Online]. Available:
  \url{https://doi.org/10.1115/fedsm2007-37518}
\BIBentrySTDinterwordspacing

\bibitem{Hartinger2008}
\BIBentryALTinterwordspacing
M.~Hartinger, M.-L. Dumont, S.~Ioannides, D.~Gosman, and H.~Spikes, ``{CFD}
  modeling of a thermal and shear-thinning elastohydrodynamic line contact,''
  \emph{Journal of Tribology}, vol. 130, no.~4, Aug. 2008. [Online]. Available:
  \url{https://doi.org/10.1115/1.2958077}
\BIBentrySTDinterwordspacing

\bibitem{Almqvist2002}
\BIBentryALTinterwordspacing
T.~Almqvist and R.~Larsson, ``The {N}avier-{S}tokes approach for thermal {EHL}
  line contact solutions,'' \emph{Tribology International}, vol.~35, no.~3, pp.
  163--170, Mar. 2002. [Online]. Available:
  \url{https://doi.org/10.1016/s0301-679x(01)00112-8}
\BIBentrySTDinterwordspacing

\bibitem{Moran2006}
M.~J. Moran and H.~N. Shapiro, \emph{Fundamentals of Engineering
  Thermodynamics}.\hskip 1em plus 0.5em minus 0.4em\relax John Wiley {\&} Sons,
  2006.

\bibitem{Bair1993}
\BIBentryALTinterwordspacing
S.~Bair, ``A note on the use of {R}oelands equation to describe viscosity for
  {EHD} hertzian zone calculations,'' \emph{Journal of Tribology}, vol. 115,
  no.~2, pp. 333--334, Apr. 1993. [Online]. Available:
  \url{https://doi.org/10.1115/1.2921011}
\BIBentrySTDinterwordspacing

\bibitem{Bair2000}
\BIBentryALTinterwordspacing
------, ``The variation of viscosity with temperature and pressure for various
  real lubricants,'' \emph{Journal of Tribology}, vol. 123, no.~2, pp.
  433--436, Jun. 2000. [Online]. Available:
  \url{https://doi.org/10.1115/1.1308024}
\BIBentrySTDinterwordspacing

\bibitem{Tosic2019}
\BIBentryALTinterwordspacing
M.~To{\v{s}}i{\'{c}}, R.~Larsson, J.~Jovanovi{\'{c}}, T.~Lohner, M.~Björling,
  and K.~Stahl, ``A computational fluid dynamics study on shearing mechanisms
  in thermal elastohydrodynamic line contacts,'' \emph{Lubricants}, vol.~7,
  no.~8, p.~69, Aug. 2019. [Online]. Available:
  \url{https://doi.org/10.3390/lubricants7080069}
\BIBentrySTDinterwordspacing

\bibitem{Havaej2023}
\BIBentryALTinterwordspacing
P.~Havaej, J.~Degroote, and D.~Fauconnier, ``Sensitivity of {TEHL} simulations
  to the use of different models for the constitutive behaviour of
  lubricants,'' \emph{Lubricants}, vol.~11, no.~3, p. 151, Mar. 2023. [Online].
  Available: \url{https://doi.org/10.3390/lubricants11030151}
\BIBentrySTDinterwordspacing

\bibitem{Dowson1977}
\BIBentryALTinterwordspacing
D.~Dowson and G.~R. Higginson, \emph{Elasto-hydrodynamic lubrication}.\hskip
  1em plus 0.5em minus 0.4em\relax Pergamon Press, 1977. [Online]. Available:
  \url{https://doi.org/10.1016/c2013-0-05764-7}
\BIBentrySTDinterwordspacing

\bibitem{Roelands1966}
\BIBentryALTinterwordspacing
C.~J.~A. Roelands, ``Correlational aspects of the
  viscosity-temperature-pressure relationship of lubricating oils,'' Ph.D.
  dissertation, Techinical University of Delft, Delft, the Netherlands, 1966.
  [Online]. Available: \url{https://doi.org/10.1115/1.3451519}
\BIBentrySTDinterwordspacing

\bibitem{Houpert1985}
\BIBentryALTinterwordspacing
L.~Houpert, ``New results of traction force calculations in elastohydrodynamic
  contacts,'' \emph{Journal of Tribology}, vol. 107, no.~2, pp. 241--245, Apr.
  1985. [Online]. Available: \url{https://doi.org/10.1115/1.3261033}
\BIBentrySTDinterwordspacing

\bibitem{Evans1986}
\BIBentryALTinterwordspacing
C.~R. Evans and K.~L. Johnson, ``The rheological properties of
  elastohydrodynamic lubricants,'' \emph{Proceedings of the Institution of
  Mechanical Engineers, Part C: Journal of Mechanical Engineering Science},
  vol. 200, no.~5, pp. 303--312, Sep. 1986. [Online]. Available:
  \url{https://doi.org/10.1243/pime_proc_1986_200_134_02}
\BIBentrySTDinterwordspacing

\bibitem{Bair2019}
\BIBentryALTinterwordspacing
S.~Bair, ``Correlations for the temperature and pressure and composition
  dependence of low-shear viscosity,'' in \emph{High Pressure Rheology for
  Quantitative Elastohydrodynamics}.\hskip 1em plus 0.5em minus 0.4em\relax
  Elsevier, 2019, pp. 135--182. [Online]. Available:
  \url{https://doi.org/10.1016/b978-0-444-64156-4.00006-4}
\BIBentrySTDinterwordspacing

\bibitem{Zolper2020}
\BIBentryALTinterwordspacing
T.~J. Zolper, S.~Bair, and K.~Horne, ``Revisiting the {ASME} pressure-viscosity
  report using the tait-doolittle correlations,'' \emph{Journal of Tribology},
  vol. 143, no.~6, Oct. 2020. [Online]. Available:
  \url{https://doi.org/10.1115/1.4048605}
\BIBentrySTDinterwordspacing

\bibitem{Bjoerling2014}
\BIBentryALTinterwordspacing
M.~Björling, W.~Habchi, S.~Bair, R.~Larsson, and P.~Marklund, ``Friction
  reduction in elastohydrodynamic contacts by thin-layer thermal insulation,''
  \emph{Tribology Letters}, vol.~53, no.~2, pp. 477--486, Jan. 2014. [Online].
  Available: \url{https://doi.org/10.1007/s11249-013-0286-8}
\BIBentrySTDinterwordspacing

\bibitem{Bair2002}
\BIBentryALTinterwordspacing
S.~Bair, C.~McCabe, and P.~T. Cummings, ``Calculation of viscous {EHL} traction
  for squalane using molecular simulation and rheometry,'' \emph{Tribology
  Letters}, vol.~13, no.~4, pp. 251--254, 2002. [Online]. Available:
  \url{https://doi.org/10.1023/a:1021011225316}
\BIBentrySTDinterwordspacing

\bibitem{Baur2007}
\BIBentryALTinterwordspacing
O.~Baur, N.~Sneeuw, and E.~W. Grafarend, ``Methodology and use of tensor
  invariants for satellite gravity gradiometry,'' \emph{Journal of Geodesy},
  vol.~82, no. 4-5, pp. 279--293, Aug. 2007. [Online]. Available:
  \url{https://doi.org/10.1007/s00190-007-0178-5}
\BIBentrySTDinterwordspacing

\bibitem{Gurevich1964}
G.~B. Gurevich, J.~R.~M. Radok, and A.~J.~M. Spencer, \emph{Foundations of the
  theory of algebraic invariants}.\hskip 1em plus 0.5em minus 0.4em\relax
  Groningen, the Netherlands: Noordhoff, 1964.

\bibitem{Gray1998}
\BIBentryALTinterwordspacing
R.~A. Gray, S.~Chynoweth, Y.~Michopoulos, and G.~S. Pawley, ``Shear
  localisation in simulated fluid,'' \emph{Europhysics Letters}, vol.~43,
  no.~5, pp. 491--496, Sep. 1998. [Online]. Available:
  \url{https://doi.org/10.1209/epl/i1998-00387-4}
\BIBentrySTDinterwordspacing

\bibitem{Habchi2010}
\BIBentryALTinterwordspacing
W.~Habchi, P.~Vergne, S.~Bair, O.~Andersson, D.~Eyheramendy, and G.~E.
  Morales-Espejel, ``Influence of pressure and temperature dependence of
  thermal properties of a lubricant on the behaviour of circular {TEHD}
  contacts,'' \emph{Tribology International}, vol.~43, no.~10, pp. 1842--1850,
  Oct. 2010. [Online]. Available:
  \url{https://doi.org/10.1016/j.triboint.2009.10.002}
\BIBentrySTDinterwordspacing

\bibitem{Hartinger2007}
M.~Hartinger, ``{CFD} modelling of elastohydrodynamic lubrication,'' Ph.D.
  dissertation, University of London, London, United Kingdom, 2007.

\bibitem{Havaej2023a}
\BIBentryALTinterwordspacing
P.~Havaej, J.~Degroote, and D.~Fauconnier, ``A quantitative analysis of
  double-sided surface waviness on {TEHL} line contacts,'' \emph{Tribology
  International}, vol. 183, p. 108389, May 2023. [Online]. Available:
  \url{https://doi.org/10.1016/j.triboint.2023.108389}
\BIBentrySTDinterwordspacing

\bibitem{Ferziger2002}
J.~H. Ferziger and M.~Peri\'{c}, \emph{Computational Methods for Fluid
  Dynamics}.\hskip 1em plus 0.5em minus 0.4em\relax Springer Berlin Heidelberg,
  2002.

\bibitem{Weller1998}
H.~G. Weller, G.~Tabor, H.~Jasak, and C.~Fureby, ``A tensorial approach to
  computational continuum mechanics using object orientated techniques,''
  \emph{Computers in Physics}, vol.~12, pp. 620--631, 1998.

\bibitem{Greenshields2020}
\BIBentryALTinterwordspacing
C.~Greenshields, \emph{{OpenFOAM} v8 User Guide}.\hskip 1em plus 0.5em minus
  0.4em\relax London, United Kingdom: The OpenFOAM Foundation, 2020. [Online].
  Available: \url{https://doc.cfd.direct/openfoam/user-guide-v8}
\BIBentrySTDinterwordspacing

\bibitem{Karrholm2008}
F.~P. K\"{a}rrholm, ``Numerical modelling of diesel spray injection, turbulence
  interaction and combustion,'' Ph.D. dissertation, Chalmers University of
  Technology, Goteborg, Sweden, 2008.

\bibitem{Dadvand2010}
\BIBentryALTinterwordspacing
P.~Dadvand, R.~Rossi, and E.~O\~{n}ate, ``An object-oriented environment for
  developing finite element codes for multi-disciplinary applications,''
  \emph{Archives of Computational Methods in Engineering}, vol.~17, no.~3, pp.
  253--297, Jul. 2010. [Online]. Available:
  \url{https://doi.org/10.1007/s11831-010-9045-2}
\BIBentrySTDinterwordspacing

\bibitem{Delaisse2023}
\BIBentryALTinterwordspacing
N.~Delaiss\'{e}, T.~Demeester, R.~Haelterman, and J.~Degroote, ``Quasi-{N}ewton
  methods for partitioned simulation of fluid-structure interaction reviewed in
  the generalized {B}royden framework,'' \emph{Archives of Computational
  Methods in Engineering}, vol.~30, no.~5, pp. 3271--3300, Apr. 2023. [Online].
  Available: \url{https://doi.org/10.1007/s11831-023-09907-y}
\BIBentrySTDinterwordspacing

\bibitem{Delaisse2022}
\BIBentryALTinterwordspacing
N.~Delaiss{\'{e}}, T.~Demeester, D.~Fauconnier, and J.~Degroote,
  ``Surrogate-based acceleration of quasi-{N}ewton techniques for
  fluid-structure interaction simulations,'' \emph{Computers \&\ Structures},
  vol. 260, p. 106720, Feb. 2022. [Online]. Available:
  \url{https://doi.org/10.1016/j.compstruc.2021.106720}
\BIBentrySTDinterwordspacing

\bibitem{Bair2006}
\BIBentryALTinterwordspacing
S.~Bair, ``Reference liquids for quantitative elastohydrodynamics: selection
  and rheological characterization,'' \emph{Tribology Letters}, vol.~22, no.~2,
  pp. 197--206, May 2006. [Online]. Available:
  \url{https://doi.org/10.1007/s11249-006-9083-y}
\BIBentrySTDinterwordspacing

\bibitem{Kim2001a}
\BIBentryALTinterwordspacing
H.~J. Kim, P.~Ehret, D.~Dowson, and C.~M. Taylor, ``Thermal elastohydrodynamic
  analysis of circular contacts {P}art 1: {N}ewtonian model,''
  \emph{Proceedings of the Institution of Mechanical Engineers, Part J: Journal
  of Engineering Tribology}, vol. 215, no.~4, pp. 339--352, Apr. 2001.
  [Online]. Available: \url{https://doi.org/10.1243/1350650011543583}
\BIBentrySTDinterwordspacing

\bibitem{Kim2001b}
\BIBentryALTinterwordspacing
------, ``Thermal elastohydrodynamic analysis of circular contacts {P}art 2:
  Non-{N}ewtonian model,'' \emph{Proceedings of the Institution of Mechanical
  Engineers, Part J: Journal of Engineering Tribology}, vol. 215, no.~4, pp.
  353--362, Apr. 2001. [Online]. Available:
  \url{https://doi.org/10.1243/1350650011543592}
\BIBentrySTDinterwordspacing

\bibitem{Srirattayawong2014}
S.~Srirattayawong, ``{CFD} study of surface roughness effects on the
  thermo-elastohydrodynamic lubrication line contact problem,'' Ph.D.
  dissertation, University of Leicester, Leicester, United Kingdom, 2014.

\end{thebibliography}

\end{document}